\documentclass[twocolumn]{aastex63}
\usepackage[utf8]{inputenc}
\usepackage{amsmath}
\usepackage{natbib}
\usepackage{epstopdf}
\usepackage{gensymb}
\shorttitle{PAH Bands versus 33 GHz}
\shortauthors{Whitcomb et al.}

\newcommand{\sumneon}{$\Sigma\rm$Ne}
\newcommand{\nethree}{15.56~\micron\ [Ne~III]}
\newcommand{\netwo}{12.81~\micron\ [Ne~II]}
\newcommand{\Zoff}{$\Omega_Z$}
\newcommand{\rhos}{$\rho_s$}
\newcommand{\dellam}{$\Delta\lambda$}
\newcommand{\tildel}{$\sigma$}
\newcommand{\lammax}{$\lambda_{max}$}
\newcommand{\halpha}{H$\alpha$}
\newcommand{\unit}{\rm10^{-4}~erg~s^{-1}~cm^{-2}~sr^{-1}\rm}
\newcommand{\sfrsig}{$\Sigma_{SFR}$}
\newcommand{\twenmic}{24~\micron}
\begin{document}

%%%%%%%%%%%%%%
% TITLE PAGE %
%%%%%%%%%%%%%%

\title{A Comparative Study of Mid-Infrared Star-Formation Rate Tracers and Their Metallicity Dependence}

\author[0000-0003-2093-4452]{C. M. Whitcomb}
\author[0000-0002-4378-8534]{K. Sandstrom}
\affil{Center for Astrophysics and Space Sciences, Department of Physics, University of California, San Diego, \\ 9500 Gilman Drive, La Jolla, CA 92093, USA}
\correspondingauthor{Cory Whitcomb}
\email{coryw777@gmail.com}

\author[0000-0001-7089-7325]{E. J. Murphy}
\affil{National Radio Astronomy Observatory, 520 Edgemont Road, Charlottesville, VA 22903, USA}

\author[0000-0002-1000-6081]{S. Linden}
\affil{Department of Astronomy, University of Virginia, 530 McCormick Road, Charlottesville, VA 22904, USA}

%%%%%%%%%%%%
% ABSTRACT %
%%%%%%%%%%%%

\begin{abstract}
We present a comparative study of a set of star-formation rate tracers based on mid-infrared emission in the \netwo\ line, the \nethree\ line, and emission features from polycyclic aromatic hydrocarbons (PAHs) between 5.2 and 14.7~\micron. We calibrate our tracers with the thermal component of the radio continuum emission at 33 GHz from 33 extranuclear star-forming regions observed in nearby galaxies. Correlations between mid-IR emission features and thermal 33 GHz star-formation rates (SFR) show significant metallicity-dependent scatter and offsets. We find similar metallicity-dependent trends in commonly used SFR tracers such as \halpha\ and \twenmic. As seen in previous studies, PAH emission alone is a poor SFR tracer due to a strong metallicity dependence: lower metallicity regions show decreased PAH emission relative to their SFR compared to higher metallicity regions. We construct combinations of PAH bands, neon emission lines, and their respective ratios to minimize metallicity trends. The calibrations that most accurately trace SFR with minimal metallicity dependence involve the sum of the integrated intensities of the \netwo\ line and the \nethree\ line combined with any major PAH feature normalized by dust continuum emission. This mid-IR calibration is a useful tool for measuring SFR as it is minimally sensitive to variations in metallicity and it is composed of bright, ubiquitous emission features. The Mid-Infrared Instrument (MIRI) on the James Webb Space Telescope will detect these features from galaxies as far as redshift z~$\sim$~1. We also investigate the behavior of the PAH band ratios and find that subtracting the local background surrounding a star-forming region decreases the ratio of PAH 11.3~\micron\ to 7.7~\micron\ emission. This implies PAHs are more ionized in star-forming regions relative to their surroundings.

\end{abstract}


%%%%%%%%
% BODY %
%%%%%%%%

\section{Introduction}\label{sec:intro}
The rate at which new stars are formed is a fundamental property that influences how a galaxy evolves. Many methods to measure the star-formation rate (SFR) in galaxies at low and high redshifts have been employed with various complications present in each \citep[for a review, see][]{KennEvans+12}. One of the main complications is interstellar dust that pervades star-forming regions. This dust protects molecular gas from ionizing or photodissociating radiation, but it also prevents such radiation from leaving the cloud. Absorbed ultraviolet and optical light is re-emitted in the infrared by a complex population of dust grains, revealing their physics and that of their environment \citep[e.g.][and references therein]{draine03,tielens+08}. This often necessitates hybrid, multi-wavelength SFR tracers that include infrared emission from dust in addition to the more direct UV and optical tracers of massive star formation~\citep[e.g.][]{buat92,meurer95,calzetti,kennicutt07,cortese08,kennicutt09,murphy+11,leroy12}. Hybrid tracers that use both UV/optical and infrared emission such as \halpha\ and \twenmic\ require observations on multiple telescopes. It is therefore of significant interest to find a tracer or multiple tracers that can predict SFR from a single spectral observation; for this purpose, emission features present in the mid-IR may be ideal.

The mid-infrared spectrum hosts many strong emission lines and broad emission bands that have the potential to be SFR tracers. Among the brightest of these features are the forbidden lines [Ne~II] at 12.81~\micron\ and [Ne~III] at 15.56~\micron. These lines contribute a significant portion of the cooling in H~II regions as relatively abundant ionized species with lines that can be easily collisionally excited \citep{burbidge63,osterbrock65}. Both emission lines have a low energy spacing and a high critical density so their emissivity varies little in typical H~II regions. The ionization potentials of Ne~I$\rightarrow$Ne~II and Ne~II$\rightarrow$Ne~III are greater than that of hydrogen at 21.56~eV and 40.96~eV, respectively, so both [Ne~II] and [Ne~III] emission trace highly ionized gas. Ne~III can be the dominant ion at low densities and in harder radiation fields \citep[e.g.][]{giveon02}.Ne~IV is not abundant in H~II regions since it requires more energetic photons than O and B stars can produce. Combining [Ne~II] and [Ne~III] emission from H~II regions thus traces the total emission from ionized neon, which traces some fraction of the overall cooling luminosity. Given a steady-state balance between heating and cooling, the neon emission should therefore trace SFR to some degree. The efficacy of SFR tracers based on ionized neon emission has been studied in several previous publications \citep{keto07,treyer10,zho_ne19}.

Also seen in the mid-IR spectra of galaxies are broad features which originate from a class of molecules called polycyclic aromatic hydrocarbons \citep[PAHs, e.g.][]{roche91,helou00,peeters02,lu03,smith07}. PAHs emit strong features relative to the infrared thermal continuum between 3.3 and 17~\micron\ due to vibrations in their carbon and hydrogen bonds when exposed to UV/optical light \citep{sellgren84,leger84,allamandola85,draine85,allaman+89,schutte93,draine01}. The efficiency with which PAHs absorb and re-radiate UV/optical light suggests that they are ideal candidates for components of a SFR tracer, and many such calibrations exist \citep[e.g.][]{roussel01,forster04,peeters04,wu05,shipley,cluver17,xie19}. 

However, PAH emission traces only the light that has been absorbed, so a PAH-based tracer would not trace the ionized gas and young stars directly, and would include potential biases if the attenuation of UV/optical light varies with environment. In particular, it has been demonstrated that the properties of PAH emission are dependent on several environmental factors such as metallicity and radiation field hardness \citep{engel05,madden+06,ohalloran06,engel08} which must be accounted for to produce a reliable SFR tracer \citep[e.g.][]{calzetti,treyer10,wen14,shipley,mosdef,xie19}. In accounting for environmental variations, ratios of PAH bands and PAH-to-continuum ratios may be useful tools. Ratios of PAH bands such as 11.3~\micron\ to 7.7~\micron\ reveal the properties of the local PAH population such as size and ionization \citep[e.g.][]{bakes01,peeters02,peeters04,bauschlicher08}. Similarly, ratios of PAH bands to mid-IR dust continuum measure the amount of emission from PAHs relative to that of the entire dust population, tracing their abundance. For instance, photometry at 8~\micron\ (dominated by the PAH feature) and the continuum measured at 24~\micron\ has often been used to trace PAH abundance \citep{engel05,engel08,draine07}.

To fully escape the complications related to dust attenuation in a SFR tracer, one must move to long wavelengths. Thermal 33 GHz radio continuum (T33) originates from free-free emission and is directly proportional to the amount of ionizing radiation from young stars \citep[e.g.][]{condon92,murphy+11,KennEvans+12}. This combination of independence from the presence of interstellar dust and direct correlation with radiation from young stars makes T33 emission an excellent standard of reference for SFR. Emission at 33 GHz, however, is faint even in local sources and difficult to detect in distant galaxies. Systematic observations of the 3 - 33 GHz continuum have been made as part of the Star Formation in Radio Survey \citep[SFRS;][]{sfrs,linden} for 50 nearby galaxies in the {\it Spitzer} Infrared Nearby Galaxies Survey \citep[SINGS;][]{sings} and Key Insights on Nearby Galaxies: A Far-Infrared Survey with Herschel \citep[KINGFISH;][]{KINGFISH}.

In the following, we use T33 as a standard of reference to calibrate SFR tracers with emission in the mid-infrared from neon and PAHs. We investigate the correlation between the strengths of these mid-IR features and T33 to find a tracer insensitive to absorption by interstellar dust (in normal galaxies with low mid-IR optical depth) that will be easily detected with the instruments on the James Webb Space Telescope (JWST). The \nethree\ line has the longest wavelength of the mid-IR features considered in this work. Thus the Mid-Infrared Instrument (MIRI) on JWST will detect the same emission lines as the SINGS spectra described in Section~\ref{sec:sings_spec} in galaxies out to redshift z~$\approx$~0.85. Restricting the maximum wavelength further to include only emission below the \netwo\ line allows observations at z~$\approx$~1.25 with MIRI.

A SFR tracer that is insensitive to metallicity is crucial for application to galaxies at higher redshifts, as the average metallicity of galaxies changes with redshift \citep{gallazzi08,madau14}. In addition, the optical diagnostic lines needed to measure metallicity may or may not be available for higher redshift targets. Therefore to construct a SFR calibration applicable to a range of redshifts observed with JWST, we aim to ensure that any metallicity dependence is accounted for. To this end, we directly study the metallicity dependence of each SFR calibration using local measurements of the metallicity. The local metallicity of each region is determined from metallicity gradients as a function of galactocentric radius measured from H~II region spectra \citep[e.g.][in our case]{moustakas}.

The paper is organized as follows: Section~\ref{sec:obs} outlines the SINGS, SFRS, and metallicity data used in this work and the methods of our analysis. Sections \ref{sec:T33_corr_1} and \ref{sec:T33_corr_2} present the results of our correlations between the SINGS mid-IR features and SFRS T33 measurements. Section~\ref{sec:SFR_plots} presents a series of mid-infrared star-formation rate tracers and comparisons of our methods against established calibrations, including the T33 tracer and the \halpha-\twenmic\ tracer from \cite{murphy+11}, the PAH tracer from \cite{shipley}, and the neon-metallicity tracer from \cite{zho_ne19}. Section~\ref{sec:disc} describes the implications of these results and novel patterns seen therein. Section~\ref{sec:concl} summarizes our key conclusions and explores potential directions for future research.

\section{Observations \& Analysis}\label{sec:obs}

\subsection{Sample}\label{sec:sample}
The basis of our work consists of 33 GHz measurements from the National Science Foundation's Karl G. Jansky Very Large Array (VLA)\footnote{The National Radio Astronomy Observatory is a facility of the National Science Foundation operated under cooperative agreement by Associated Universities, Inc.} as part of SFRS \citep{sfrs,linden}. Our full sample is the overlap between SFRS 7$\arcsec$ diameter aperture 33 GHz observations and regions observed in SINGS SL and SH spectral maps. We find 56 regions from SFRS have full 7$\arcsec$ aperture coverage in SINGS SL data and 52 of these were found to have full-aperture coverage in SH data. 

Our key goals in this study require knowledge of the thermal component of the 33 GHz emission and a local measurement of the metallicity. There are 43 of the 56 regions in the SFRS-SINGS overlap that have measured thermal fractions. Of those, 33 have local metallicity measurements (see Section~\ref{sec:Z_data}). These 33 regions form the basis of our SFR calibrations in Sections \ref{sec:T33_corr_1} and \ref{sec:T33_corr_2}. In the comparisons to literature SFR calibrations in Section \ref{sec:SFR_plots}, we use the largest possible subset of the data which will be noted in the text. In Section~\ref{sec:bandratios} we investigate the PAH band ratios separately from the SFR calibrations, and therefore use a subset of 49 regions on which the only requirement is full 11\arcsec\ aperture coverage in SINGS SL mapping.

\begin{deluxetable*}{lcccccr}
\tabletypesize{\scriptsize}
\tablecaption{7$\arcsec$ Region Information from \citet{sfrs} \label{tab:murphy}}
\tablewidth{1pt}
\tablehead{
\colhead{Region ID} &
\colhead{RA} &
\colhead{Dec} &
\colhead{T33} &
\colhead{\halpha/10$^{4}$} &
\colhead{{\bf $\nu I_{\nu}$} (24\micron)~/10$^{4}$} &
\colhead{r$_G$} \\ [-3mm]
\colhead{ } &
\colhead{(J2000)} &
\colhead{(J2000)} &
\colhead{(mJy)} &
\colhead{(erg s$^{-1}$cm$^{-2}$sr$^{-1}$)} &
\colhead{(erg s$^{-1}$cm$^{-2}$sr$^{-1}$)} &
\colhead{(kpc)}}
\startdata
NGC 0628 Enuc.1 & 01 36 45.27 & 15 47 48.3 & 0.28 $\pm$~ 0.04 &
0.79 $\pm$~ 0.16 & 109 $\pm$~ 5 & 2.478 \\
NGC 0628 Enuc.2 & 01 36 37.65 & 15 45 07.2 & 0.17 $\pm$~ 0.03 &
0.48 $\pm$~ 0.10 & 58.2 $\pm$~ 2.9 & 4.468 \\
NGC 0628 Enuc.3 & 01 36 38.78 & 15 44 23.2 & 0.18 $\pm$~ 0.05 &
0.86 $\pm$~ 0.18 & 43.0 $\pm$~ 2.1 & 5.720 \\
NGC 0628 Enuc.4 & 01 36 35.72 & 15 50 07.2 & 0.12 $\pm$~ 0.03 &
0.54 $\pm$~ 0.11 & 14.2 $\pm$~ 0.7 & 7.608 \\
NGC 1097 Enuc.1c & 02 46 24.06 & -30 17 50.9 & \nodata &
0.01 $\pm$~ 0.00 & 11.8 $\pm$~ 0.6 & 7.392 \\
NGC 1097 Enuc.2 & 02 46 14.40 & -30 15 04.0 & \nodata &
0.01 $\pm$~ 0.00 & 4.85 $\pm$~ 0.3 & 7.435 \\
NGC 2403 Enuc.1b & 07 36 45.50 & 65 37 00.9 & 0.55 $\pm$~ 0.03 &
3.10 $\pm$~ 0.62 & 121 $\pm$~ 6 & 1.192 \\
NGC 2403 Enuc.2b & 07 36 52.36 & 65 36 46.9 & 0.48 $\pm$~ 0.03 &
2.34 $\pm$~ 0.46 & 84.5 $\pm$~ 4.2 & 1.249 \\
NGC 2403 Enuc.3 & 07 37 06.95 & 65 36 39.0 & 1.62 $\pm$~ 0.03 &
5.44 $\pm$~ 1.08 & 408 $\pm$~ 20 & 2.811 \\
NGC 2403 Enuc.4 & 07 37 18.19 & 65 33 48.1 & 0.33 $\pm$~ 0.02 &
0.82 $\pm$~ 0.17 & 24.6 $\pm$~ 1.2 & 3.455 \\
NGC 2403 Enuc.5 & 07 36 19.84 & 65 37 05.5 & 0.56 $\pm$~ 0.03 &
2.44 $\pm$~ 0.49 & 75.7 $\pm$~ 3.8 & 3.464 \\
NGC 2403 Enuc.6 & 07 36 28.69 & 65 33 49.4 & 0.30 $\pm$~ 0.02 &
1.26 $\pm$~ 0.25 & 12.9 $\pm$~ 0.7 & 5.380 \\
Holmberg II 0 & 08 19 13.06 & 70 43 08.0 & 0.21 $\pm$~ 0.03 &
1.65 $\pm$~ 0.33 & 23.1 $\pm$~ 1.2 & 0.738 \\
NGC 2976 Enuc.1b & 09 47 07.64 & 67 55 54.7 & 0.85 $\pm$~ 0.04 &
3.42 $\pm$~ 0.69 & 185 $\pm$~ 9 & 1.201 \\
NGC 2976 Enuc.2a & 09 47 23.83 & 67 53 54.9 & 0.45 $\pm$~ 0.04 &
2.10 $\pm$~ 0.42 & 76.2 $\pm$~ 3.8 & 1.394 \\
NGC 2976 Enuc.2b & 09 47 23.94 & 67 54 02.1 & 0.20 $\pm$~ 0.04 &
1.21 $\pm$~ 0.24 & 32.1 $\pm$~ 1.6 & 1.310 \\
IC 2574 b & 10 28 48.40 & 68 28 03.5 & 0.24 $\pm$~ 0.03 &
1.92 $\pm$~ 0.39 & 28.8 $\pm$~ 1.4 & 5.254 \\
NGC 3521 Enuc.1 & 11 05 46.30 & 00 04 09.0 & \nodata &
0.13 $\pm$~ 0.02 & 10.1 $\pm$~ 0.5 & 9.929 \\
NGC 3521 Enuc.2b & 11 05 49.94 & 00 03 55.9 & \nodata &
0.06 $\pm$~ 0.01& 4.1 $\pm$~ 0.2 & 6.044 \\
NGC 3521 Enuc.3 & 11 05 47.60 & 00 00 33.0 & \nodata &
0.11 $\pm$~ 0.02 & 10.3 $\pm$~ 0.5 & 9.509 \\
NGC 3627 Enuc.1 & 11 20 16.32 & 12 57 49.2 & 0.83 $\pm$~ 0.03 &
0.43 $\pm$~ 0.09 & 245 $\pm$~ 12 & 4.712 \\
NGC 3627 Enuc.2 & 11 20 16.46 & 12 58 43.4 & 1.64 $\pm$~ 0.03 &
1.02 $\pm$~ 0.20 & 686 $\pm$~ 34 & 2.746 \\
NGC 3938 Enuc.2a & 11 53 00.06 & 44 08 00.0 & \nodata &
0.32 $\pm$~ 0.07 & 11.0 $\pm$~ 0.6 & 11.16 \\
NGC 3938 Enuc.2b & 11 53 00.19 & 44 07 48.3 & 0.11 $\pm$~ 0.04 &
0.38 $\pm$~ 0.08 & 30.5 $\pm$~ 1.5 & 11.05 \\
NGC 4254 Enuc.1a & 12 18 49.20 & 14 23 57.9 & 0.11 $\pm$~ 0.03 &
0.25 $\pm$~ 0.06 & 31.7 $\pm$~ 1.6 & 4.428 \\
NGC 4321 Enuc.1 & 12 22 58.90 & 15 49 35.0 & \nodata &
0.11 $\pm$~ 0.02 & 6.7 $\pm$~ 0.3 & 4.520 \\
NGC 4321 Enuc.2b & 12 22 49.90 & 15 50 27.8 & \nodata &
0.17 $\pm$~ 0.03 & 9.0 $\pm$~ 0.4 & 7.979 \\
NGC 4631 Enuc.1 & 12 41 40.47 & 32 31 49.1 & 0.14 $\pm$~ 0.02 &
0.91 $\pm$~ 0.18 & 10.6 $\pm$~ 0.5 & 13.76 \\
NGC 4631 Enuc.2a & 12 42 21.42 & 32 33 06.3 & 0.17 $\pm$~ 0.02 &
0.76 $\pm$~ 0.16 & 27.1 $\pm$~ 1.4 & 9.974 \\
NGC 4736 Enuc.1a & 12 50 56.41 & 41 07 14.3 & 0.31 $\pm$~ 0.03 &
0.99 $\pm$~ 0.20 & 129 $\pm$~ 6 & 0.864 \\
NGC 5055 Enuc.1 & 13 15 58.32 & 42 00 27.4 & 0.15 $\pm$~ 0.03 &
0.71 $\pm$~ 0.14 & 32.3 $\pm$~ 1.6 & 5.630 \\
NGC 5194 Enuc.10b & 13 29 56.52 & 47 10 46.9 & 0.13 $\pm$~ 0.03 &
0.43 $\pm$~ 0.09 & 54.8 $\pm$~ 2.7 & 2.723 \\
NGC 5194 Enuc.11d & 13 29 49.58 & 47 13 28.7 & \nodata &
0.13 $\pm$~ 0.02 & 21.0 $\pm$~ 1.1 & 4.078 \\
NGC 5194 Enuc.11e & 13 29 50.64 & 47 13 44.9 & 0.12 $\pm$~ 0.02 &
0.29 $\pm$~ 0.06 & 33.1 $\pm$~ 1.7 & 4.633 \\
NGC 5194 Enuc.1c & 13 29 53.13 & 47 12 39.4 & 0.14 $\pm$~ 0.03 &
0.74 $\pm$~ 0.14 & 45.8 $\pm$~ 2.3 & 2.323 \\
NGC 5194 Enuc.2 & 13 29 44.10 & 47 10 23.4 & 0.36 $\pm$~ 0.02 &
1.55 $\pm$~ 0.31 & 142 $\pm$~ 7 & 6.834 \\
NGC 5194 Enuc.3 & 13 29 45.13 & 47 09 57.4 & 0.26 $\pm$~ 0.02 &
0.74 $\pm$~ 0.14 & 90.8 $\pm$~ 4.5 & 7.048 \\
NGC 5194 Enuc.4b & 13 29 55.49 & 47 14 01.6 & 0.18 $\pm$~ 0.02 &
0.02 $\pm$~ 0.00 & 48.1 $\pm$~ 2.4 & 6.193 \\
NGC 5194 Enuc.5 & 13 29 59.60 & 47 13 59.8 & 0.16 $\pm$~ 0.03 &
0.02 $\pm$~ 0.00 & 46.1 $\pm$~ 2.3 & 7.742 \\
NGC 5194 Enuc.7b & 13 30 02.38 & 47 09 48.7 & 0.16 $\pm$~ 0.03 &
1.15 $\pm$~ 0.23 & 72.5 $\pm$~ 3.6 & 6.329 \\
NGC 5194 Enuc.8 & 13 30 01.48 & 47 12 51.7 & 0.29 $\pm$~ 0.03 &
0.69 $\pm$~ 0.13 & 157 $\pm$~ 8 & 6.650 \\
NGC 5194 Enuc.9 & 13 29 59.78 & 47 11 12.3 & 0.21 $\pm$~ 0.04 &
0.43 $\pm$~ 0.09 & 73.6 $\pm$~ 3.7 & 4.131 \\
NGC 5713 Enuc.2a & 14 40 10.80 & 00 17 35.5 & 0.20 $\pm$~ 0.03 &
0.61 $\pm$~ 0.12 & 83.5 $\pm$~ 4.2 & 1.984 \\
NGC 6946 Enuc.1 & 20 35 16.80 & 60 11 00.0 & 0.34 $\pm$~ 0.03 &
3.02 $\pm$~ 0.61 & 67.0 $\pm$~ 3.3 & 6.989 \\
NGC 6946 Enuc.2b & 20 35 25.38 & 60 09 58.8 & 0.67 $\pm$~ 0.03 &
8.23 $\pm$~ 1.65 & 97.6 $\pm$~ 4.9 & 8.595 \\
NGC 6946 Enuc.3a & 20 34 49.86 & 60 12 40.7 & 0.10 $\pm$~ 0.02 &
1.28 $\pm$~ 0.25 & 7.4 $\pm$~ 0.4 & 7.742 \\
NGC 6946 Enuc.3b & 20 34 52.24 & 60 12 43.7 & 0.17 $\pm$~ 0.02 &
2.63 $\pm$~ 0.53 & 31.1 $\pm$~ 1.6 & 7.729 \\
NGC 6946 Enuc.4a & 20 34 19.84 & 60 10 06.6 & 0.83 $\pm$~ 0.02 &
0.01 $\pm$~ 0.00 & 23.9 $\pm$~ 1.2 & 9.044 \\
NGC 6946 Enuc.5b & 20 34 39.36 & 60 04 52.4 & 0.16 $\pm$~ 0.02 &
1.44 $\pm$~ 0.29 & 15.1 $\pm$~ 0.8 & 9.704 \\
NGC 6946 Enuc.6a & 20 35 06.08 & 60 10 58.5 & 0.45 $\pm$~ 0.03 &
3.25 $\pm$~ 0.65 & 147 $\pm$~ 7 & 4.855 \\
NGC 6946 Enuc.7 & 20 35 12.97 & 60 08 50.5 & \nodata &
1.32 $\pm$~ 0.27 & 102 $\pm$~ 5 & 5.637 \\
NGC 6946 Enuc.8 & 20 34 32.28 & 60 10 19.3 & 0.48 $\pm$~ 0.03 &
1.52 $\pm$~ 0.30 & 113 $\pm$~ 6 & 6.122 \\
NGC 6946 Enuc.9 & 20 35 11.09 & 60 08 57.5 & \nodata &
3.64 $\pm$~ 0.73 & 165 $\pm$~ 8 & 5.071 \\
NGC 7793 Enuc.1 & 23 57 48.80 & -32 36 58.0 & \nodata &
0.69 $\pm$~ 0.13 & 7.1 $\pm$~ 0.4 & 2.574 \\
NGC 7793 Enuc.2 & 23 57 56.10 & -32 35 40.0 & \nodata &
0.24 $\pm$~ 0.04 & 8.0 $\pm$~ 0.4 & 1.526 \\
NGC 7793 Enuc.3 & 23 57 48.80 & -32 34 52.0 & \nodata &
0.86 $\pm$~ 0.18 & 41.4 $\pm$~ 2.1 & 1.016 \\
\enddata
\end{deluxetable*}

\subsection{SFRS Photometry}\label{sec:sfrsphoto}
SFRS observed 33 GHz radio continuum emission with the VLA at $\sim$2$\arcsec$ resolution in 50 local galaxies ($<$ 30 Mpc) from the SINGS \citep{sings} and KINGFISH \citep{KINGFISH} surveys. The resulting maps were then convolved to 7$\arcsec$ Gaussian resolution to match the resolution of SINGS \twenmic\ images. 33 GHz measurements were extracted from 7$\arcsec$ diameter apertures in these convolved maps and  are listed in Table~\ref{tab:murphy}. This Table also lists the location of each aperture, its galactocentric radius, and emission from \halpha\ and \twenmic\ we adopt from \cite{sfrs}. \halpha\ and \twenmic\ emission are converted to $\unit$\ assuming a 7\arcsec\ aperture in order to facilitate comparison with mid-IR emission results described in Section~\ref{sec:pahfit}. We convert \twenmic\ continuum to $\nu I_{\nu}$ units by multiplying by 12.5 THz (i.e. \twenmic\ in terms of frequency) for ease of comparison with line integrated intensities. The thermal component of 33 GHz emission (T33) for each region was obtained from \cite{linden} where fractions were determined by fitting radio spectra with a sychrotron emission component (non-thermal) and a free-free emission component (thermal). There are 43 of the 56 regions in our SFRS-SINGS overlap sample that have thermal fractions at 33 GHz available. We convert the T33 flux densities into surface brightness units by dividing by the solid angle of the 7$\arcsec$ diameter aperture.

\subsection{SINGS Spectroscopy}\label{sec:sings_spec}
The mid-infrared data used in this work is from the fifth release of the {\it Spitzer} Infrared Nearby Galaxies Survey \citep[SINGS, DR5][]{sings,dale06} which mapped nuclear and extranuclear regions of nearby galaxies with the Infrared Spectrometer \citep[IRS][]{houck04} on the {\it Spitzer} Space Telescope \citep{werner04}. The SINGS DR5 spectral flux and uncertainty FITS files for extranuclear regions were downloaded from the Infrared Science Archive (IRSA) hosted by the Infrared Processing and Analysis Center (IPAC). Spectra in each common aperture between the SINGS and SFRS data were extracted from the short-low SL1 and SL2 (5.2 - 14.7~\micron) modules. Additional spectra were extracted from SINGS data in the IRS short-high (SH) module (9.9 - 19.4~\micron) for the 52 regions in our sample where observations are available. The extracted spectra are shown in Figure~\ref{fig:spectra} ordered by their \netwo\ line strength.

\begin{deluxetable*}{lccccccrc}[t]
\tabletypesize{\scriptsize}
\tablecaption{Galaxy Information \label{tab:galaxy}}
\tablewidth{0pt}
\tablehead{
\colhead{Galaxy} &
\colhead{12 + log[O/H]$_{KK04}$} &
\colhead{12 + log[O/H]$_{PT05}$} &
\colhead{Gradient$_{KK04}$} &
\colhead{Gradient$_{PT05}$} &
\colhead{$R_{25}$} &
\colhead{Distance} &
\colhead{Velocity} &
\colhead{Mass}\\
\colhead{} &
\colhead{at R = 0} &
\colhead{at R = 0} &
\colhead{(dex~$R_{25}^{-1}$)} &
\colhead{(dex~$R_{25}^{-1}$)} &
\colhead{(arcmin)} &
\colhead{(Mpc)} &
\colhead{(km~s$^{-1}$)}&
\colhead{(log(M$_{\odot}$))}
}
\startdata
NGC 0628 &9.19 $\pm$~0.02 &8.43 $\pm$~0.02 &-0.57 $\pm$~0.04 &-0.27 $\pm$~0.05 &5.24  &7.20 &657 $\pm$~1 & 9.56\\
NGC 1097 &9.17 $\pm$~0.01 &8.57 $\pm$~0.01 &-0.29 $\pm$~0.09 &-0.37 $\pm$~0.13 &4.67  &14.2 &1271 $\pm$~3 & 10.48\\
NGC 2403\tablenotemark{c} &8.89 $\pm$~0.01&8.42 $\pm$~0.01 &-0.26 $\pm$ 0.03 &-0.32 $\pm$~0.03 &11.9  &3.22 & 133 $\pm$~1 & 9.57\\
Holmberg II\tablenotemark{a} &8.13 $\pm$~0.11 &7.72 $\pm$~0.14 &\nodata &\nodata &3.97  &3.05 &142 $\pm$~1 & 7.59 \\
IC 2574\tablenotemark{a} &8.24 $\pm$~0.11 &7.85 $\pm$~0.14 &\nodata &\nodata &6.59 &3.79 &57 $\pm$~2 & 8.20 \\
NGC 2976\tablenotemark{a} &8.98 $\pm$~0.03&8.36 $\pm$~0.06&\nodata &\nodata &2.94 &3.55 &3 $\pm$~0  & 8.96 \\
NGC 3521 &9.00 $\pm$~0.02&8.44 $\pm$~0.05 &-0.69 $\pm$~0.20 &-0.16 $\pm$~0.33 &5.48  &11.2 &801 $\pm$~3 & 10.69 \\
NGC 3627 &8.99 $\pm$~0.10 &8.34 $\pm$~0.24&\nodata &\nodata &4.56 &9.38 &727 $\pm$~3 & 10.49 \\
NGC 3938\tablenotemark{b} & 9.06 & 8.42  &\nodata &\nodata &2.69 &17.9 &810 $\pm$~4 & 9.46 \\
NGC 4254 &9.14 $\pm$~0.01&8.56 $\pm$~0.02 &-0.42 $\pm$~0.06 &-0.37 $\pm$~0.08 &2.69 &14.4 &2407 $\pm$~3 & 9.56 \\
NGC 4321 &9.18 $\pm$~0.01 &8.50 $\pm$~0.03 &-0.35 $\pm$~0.13 &-0.38 $\pm$~0.21 &3.71 &14.3 &1571 $\pm$~1 & 10.30 \\
NGC 4631\tablenotemark{a} &8.75 $\pm$~0.09 &8.12 $\pm$~0.11 &\nodata &\nodata &7.74 &7.62 &606 $\pm$~3 & 9.76 \\
NGC 4736 &9.01 $\pm$~0.03&8.40 $\pm$~0.01 &-0.11 $\pm$~0.15 &-0.33 $\pm$~0.18 &5.61 &4.66 &308 $\pm$~2 & 10.34 \\
NGC 5055 &9.14 $\pm$~0.02&8.59 $\pm$~0.07 &-0.54 $\pm$~0.18 &-0.63 $\pm$~0.29 &6.30 &7.94 &484 $\pm$~1 & 10.55 \\
NGC 5194\tablenotemark{c} &9.18 $\pm$~0.01&8.64 $\pm$~0.01 &-0.50 $\pm$~0.05 &-0.31 $\pm$~0.06 &5.61 &7.62 &463 $\pm$~3 & 10.73 \\
NGC 5713\tablenotemark{a} &9.03 $\pm$~0.03&8.24 $\pm$~0.06&\nodata &\nodata &1.38 &21.4 &1899 $\pm$~7 & 10.07 \\
NGC 6946 &9.05 $\pm$~0.02&8.45 $\pm$~0.06 &-0.28 $\pm$~0.10 &-0.17 $\pm$~0.15 &5.74 &6.80 &40 $\pm$~2 & 9.96 \\
NGC 7793 &8.87 $\pm$~0.01&8.34 $\pm$~0.02 &-0.36 $\pm$~0.07 &-0.10 $\pm$~0.08 &4.67 &3.91 &230 $\pm$~4 & 9.00 
\enddata
\tablerefs{Columns 2 to 6: \cite{moustakas}, Distances: \cite{sfrs}, Velocities: NED\tablenotemark{d}, Masses: \cite{KINGFISH} and \cite{leroy19}}
\tablenotemark{a}{Characteristic abundance values where no value for R = 0 is given.}\\
\tablenotemark{b}{L-Z determined values where no characteristic or R = 0 value is given.}\\
\tablenotemark{c}{Stellar mass values from \cite{leroy19}\\
\tablenotemark{d}{The NASA/IPAC Extragalactic Database (NED) is funded by the National Aeronautics and Space Administration and operated by the California Institute of Technology}}
\end{deluxetable*}

\subsection{Ancillary Data}\label{sec:Z_data}
Metallicities for the regions in this sample were determined using data from \cite{moustakas}. This work presents metallicity gradients and central values in both the KK04 \citep{kk04} and PT05 \citep{pt05} strong-line calibrations for galaxies in SINGS. These values are shown in Table~\ref{tab:galaxy} for the galaxies in our sample. Our investigations into the metallicity-dependence of mid-IR emission features rely on having a local measurement of the metallicity at the galactocentric radius of each star-forming region. For this purpose, we select the galaxies that have a measured H~II region metallicity gradient or a reported average metallicity if the galaxy is a dwarf ($\leq$ 10$^9$ M$_{\odot}$). Of the 56 SFRS-SINGS overlap regions, 49 have local metallicities. For dwarf galaxies, which are expected to have small or negligible gradients, we adopt the characteristic oxygen abundance if no gradient is measured, and for those without either a gradient or characteristic abundance we use values calculated from the luminosity-metallicity (L - Z) relation. These regions are used in our investigation into metallicity correlations in Appendix~\ref{sec:Z_corr}. In general, we use the KK04 calibration throughout; changing to the PT05 calibration minimally affects the values of some of our fits and does not alter the qualitative results of our study. Table~\ref{tab:galaxy} also lists the 25th-magnitude isophotal radii (R$_{25}$) from \cite{moustakas}, distances from \cite{sfrs}, heliocentric velocities from NED\footnote{The NASA/IPAC Extragalactic Database (NED) is funded by the National Aeronautics and Space Administration and operated by the California Institute of Technology}, and masses from \cite{KINGFISH} or other sources that we adopt for these galaxies.

\subsection{Matched Aperture Extraction}\label{sec:spec_extract}
Spectra from the SINGS SL1, SL2, and SH spectral cubes were extracted in matched 7$\arcsec$ diameter apertures identical to those used for the 33 GHz measurements \citep{sfrs}. We remove regions where the 7$\arcsec$ aperture does not completely fall within the coverage of the SINGS data. The \texttt{aperture\char`_photometry} module of the Python \texttt{astropy} \texttt{photutils} package was used at each wavelength of the Spitzer SL and SH cubes to extract mid-IR spectra for each region \citep{astropy,astropy18}. This function measures the flux density within the complete circular aperture by using exact fractions for partially enclosed pixels and we repeat this for each wavelength in the data cube. We then divide by the total solid angle of the aperture to obtain spectra in units of MJy/sr. Spectral uncertainties are also propagated by the \texttt{aperture\char`_photometry} function from the corresponding SINGS uncertainty cube. 

We do not perform any PSF matching across wavelength in the IRS cubes prior to aperture photometry.  In the wavelength range of interest for our study, the PSF of IRS spectral mapping observations is found to oscillate around $\sim3-4''$ FWHM for SL and $\sim4-5''$ FWHM for SH \citep{pereira-santaella}. The FWHM of the PSF is always smaller than our extraction aperture. Given the small observed variations in FWHM as a function of wavelength, PSF matching would not greatly affect our measurements and given the uncertainties in the convolution kernels for IRS, we chose to proceed without performing any PSF matching.

\begin{figure*}
\centering
\includegraphics[width=3.02in]{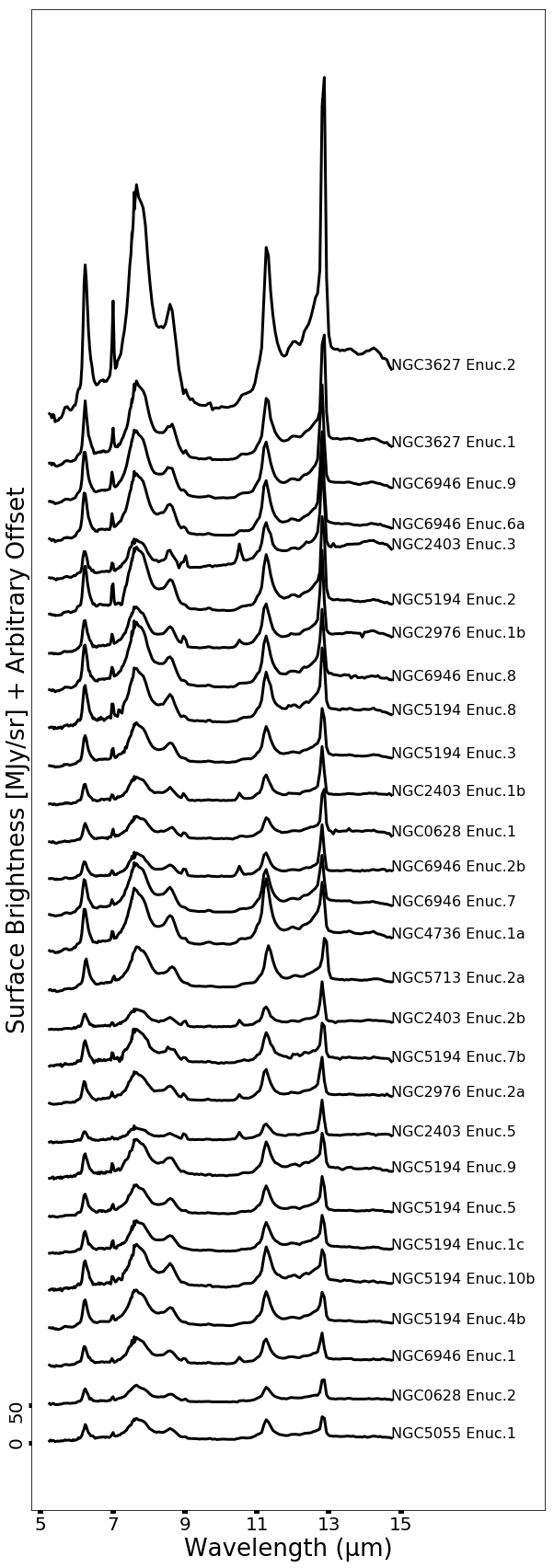}
\includegraphics[width=3in]{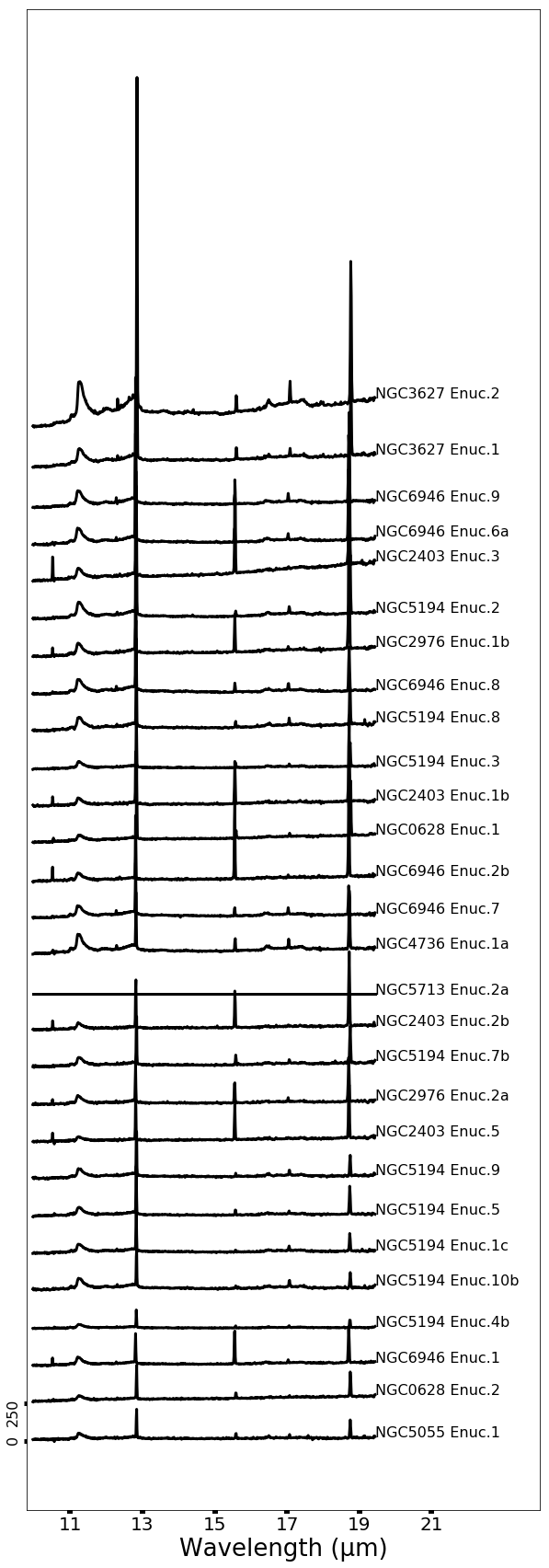}
\caption{Extracted spectra from SL wavelengths 5.2 to 14.7~\micron\ (left) and SH wavelengths 9.9 to 19.4~\micron\ (right). Shown are the 28 brightest of the 56 spectra in order of \netwo\ emission measurements from PAHFIT on the SL spectra. Regions with no SH data are shown as a horizontal line in the right panel.}

\label{fig:spectra}
\end{figure*}
\begin{figure*}
\centering
\includegraphics[width=3.02in]{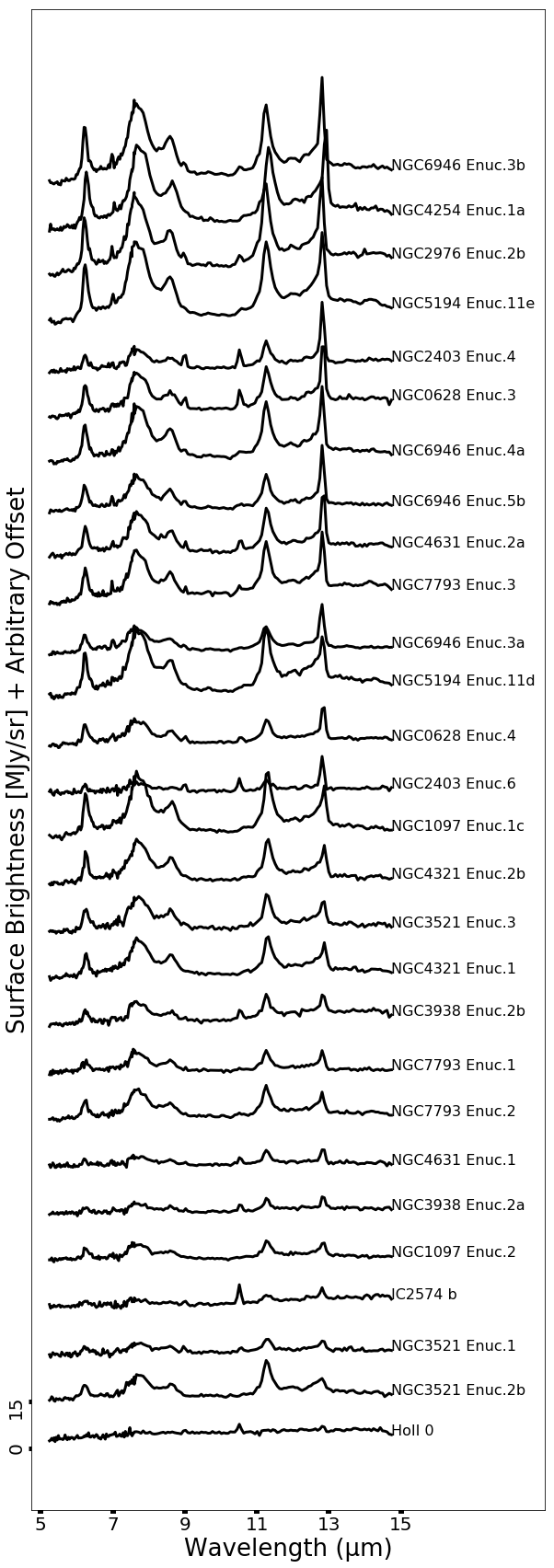}
\includegraphics[width=3in]{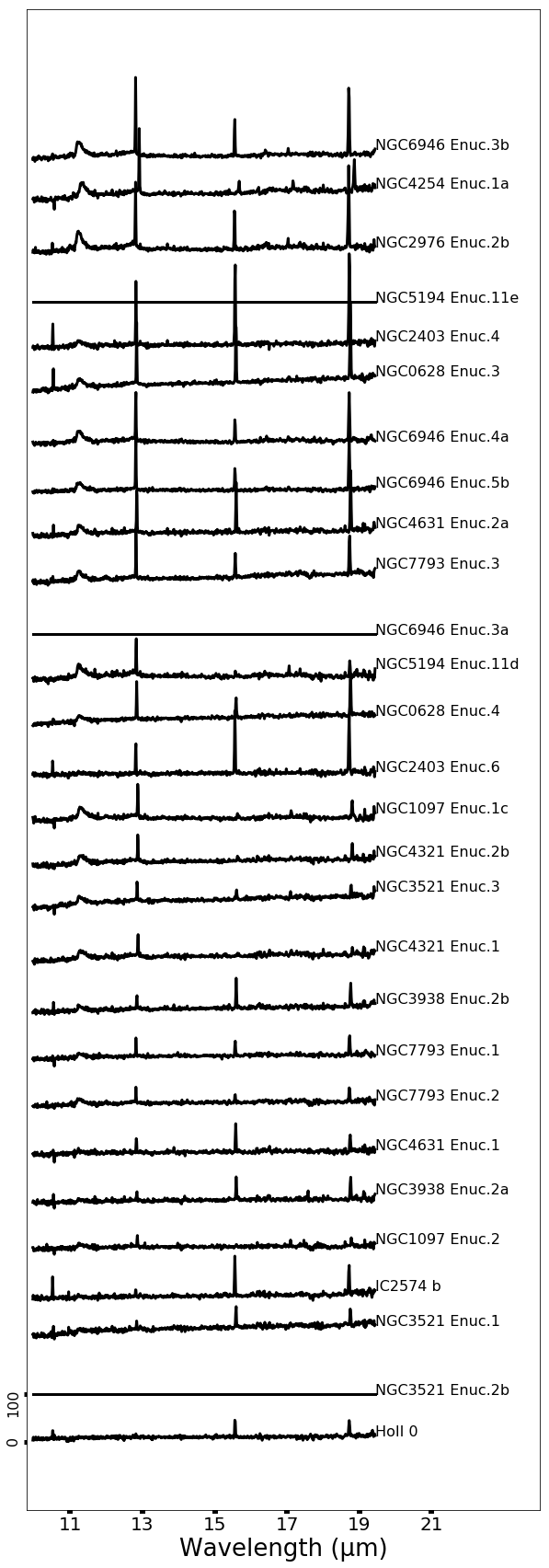}
\caption{Extracted spectra from SL wavelengths 5.2 to 14.7~\micron\ (left) and SH wavelengths 9.9 to 19.4~\micron\ (right). Shown are the 28 dimmest of the 56 spectra in order of \netwo\ emission measurements from PAHFIT on the SL spectra. Regions with no SH data are shown as a horizontal line in the right panel.}
\label{fig:lower_spectra}
\end{figure*}

We produced a second SL dataset with the local background removed by subtracting the average background level determined in an annulus surrounding the circular aperture. At aperture scales of radius 3.5\arcsec~these regions are $<$ 360 pc in physical radius and the majority of the emission from each star-forming region is enclosed. We investigate the effect of local background subtraction for these regions using the surface brightness in an annulus between radius 3.5 and 5.5\arcsec\ surrounding the original 7\arcsec~diameter aperture. The surface brightness at each wavelength in the background annulus is then averaged and subtracted from that of the inner aperture. We exclude regions that do not have full SL data coverage in the background annulus, leaving 49 of the 56 regions in the background-subtracted dataset. In Section~\ref{sec:bandratios} we use this additional dataset to study the effect of local background subtraction on PAH band ratios.

\subsection{Measuring Integrated Intensity of PAH Bands and Emission Lines}\label{sec:pahfit}

\begin{deluxetable*}{lrrrrrcrc}
\tabletypesize{\scriptsize}
\tablecaption{Mid-IR Measurement Results [10$^{-4}$~erg~s$^{-1}$~cm$^{-2}$~sr$^{-1}$] \label{tab:pahfit}}
\tablewidth{0pt}
\tablehead{
\colhead{Region ID} &
\colhead{PAH 6.2\micron} &
\colhead{PAH 7.7\micron} &
\colhead{PAH 11.3\micron} &
\colhead{$\Sigma$PAH} &
\colhead{[NeII] 12.81\micron} &
\colhead{[NeIII] 15.56\micron} &
\colhead{{\bf $\nu I_{\nu}$} (10\micron)}}
\startdata
NGC0628 Enuc.1\tablenotemark{$\star$} & 5.37 $\pm$~ 0.17 & 17.2 $\pm$~ 0.18 & 
2.70 $\pm$~ 0.06 & 33.21 $\pm$~ 0.28 & 1.28 $\pm$~ 0.01 & 
0.38 $\pm$~ 0.01 & 21.47 $\pm$~ 1.06 \\
NGC0628 Enuc.2\tablenotemark{$\star$} & 4.26 $\pm$~ 0.12 & 12.3 $\pm$~ 0.19 & 
1.84 $\pm$~ 0.04 & 23.02 $\pm$~ 0.25 & 0.59 $\pm$~ 0.01 & 
0.27 $\pm$~ 0.01 & 13.59 $\pm$~ 1.01 \\
NGC0628 Enuc.3\tablenotemark{$\star$} & 2.25 $\pm$~ 0.07 & 6.48 $\pm$~ 0.24 & 
1.20 $\pm$~ 0.03 & 12.72 $\pm$~ 0.27 & 0.39 $\pm$~ 0.01 & 
0.88 $\pm$~ 0.01 & 11.27 $\pm$~ 1.31 \\
NGC0628 Enuc.4\tablenotemark{$\star$} & 1.48 $\pm$~ 0.05 & 4.34 $\pm$~ 0.15 & 
0.69 $\pm$~ 0.02 & 8.09 $\pm$~ 0.18 & 0.23 $\pm$~ 0.01& 
0.31 $\pm$~ 0.02 & 5.10 $\pm$~ 0.87 \\
NGC1097 Enuc.1c & 2.96 $\pm$~ 0.11 & 8.95 $\pm$~ 0.30 & 
1.89 $\pm$~ 0.03 & 17.9 $\pm$~ 0.33 & 0.22 $\pm$~ 0.01 & 
0.06 $\pm$~ 0.03 & 9.69 $\pm$~ 1.18 \\
NGC1097 Enuc.2 & 0.84 $\pm$~ 0.07 & 2.41 $\pm$~ 0.28 & 
0.52 $\pm$~ 0.03 & 5.07 $\pm$~ 0.59 & 0.07 $\pm$~ 0.01 & 
0.04 $\pm$~ 0.02 & 2.51 $\pm$~ 1.16 \\
NGC2403 Enuc.1b\tablenotemark{$\star$} & 5.85 $\pm$~ 0.15 & 17.4 $\pm$~ 0.20 & 
3.18 $\pm$~ 0.05 & 34.17 $\pm$~ 0.28 & 1.28 $\pm$~ 0.01 & 
2.13 $\pm$~ 0.11 & 24.44 $\pm$~ 1.07 \\
NGC2403 Enuc.2b\tablenotemark{$\star$} & 4.47 $\pm$~ 0.13 & 13.5 $\pm$~ 0.21 & 
2.38 $\pm$~ 0.05 & 26.06 $\pm$~ 0.27 & 1.07 $\pm$~ 0.01 & 
1.85 $\pm$~ 0.01 & 17.88 $\pm$~ 1.03 \\
NGC2403 Enuc.3\tablenotemark{$\star$} & 8.08 $\pm$~ 0.23 & 23.5 $\pm$~ 0.26 & 
4.98 $\pm$~ 0.10 & 48.15 $\pm$~ 0.54 & 2.35 $\pm$~ 0.01 & 
4.73 $\pm$~ 0.02 & 67.3 $\pm$~ 1.14 \\
NGC2403 Enuc.4\tablenotemark{$\star$} & 1.05 $\pm$~ 0.08 & 2.91 $\pm$~ 0.27 & 
0.79 $\pm$~ 0.03 & 6.44 $\pm$~ 0.31 & 0.39 $\pm$~ 0.01 & 
1.27 $\pm$~ 0.01 & 5.86 $\pm$~ 1.09 \\
NGC2403 Enuc.5\tablenotemark{$\star$} & 2.96 $\pm$~ 0.10 & 9.08 $\pm$~ 0.27 & 
1.98 $\pm$~ 0.03 & 17.99 $\pm$~ 0.31 & 0.95 $\pm$~ 0.01 & 
1.77 $\pm$~ 0.09 & 14.69 $\pm$~ 1.27 \\
NGC2403 Enuc.6\tablenotemark{$\star$} & 0.50 $\pm$~ 0.09 & 1.54 $\pm$~ 0.43 & 
0.34 $\pm$~ 0.03 & 3.09 $\pm$~ 0.45 & 0.22 $\pm$~ 0.01 & 
0.99 $\pm$~ 0.01 & 3.22 $\pm$~ 1.15 \\
Holmberg II 0\tablenotemark{$\star$} & 0.16 $\pm$~ 0.06 & 0.53 $\pm$~ 0.28 & 0.07 $\pm$~ 0.02 & 
1.28 $\pm$~ 0.30 & 0.03 $\pm$~ 0.01 & 0.27 $\pm$~ 0.01 & 
14.81 $\pm$~ 1.03 \\
NGC2976 Enuc.1b\tablenotemark{$\star$} & 9.66 $\pm$~ 0.27 & 30.7 $\pm$~ 0.30 & 
5.41 $\pm$~ 0.09 & 60.06 $\pm$~ 0.43 & 2.21 $\pm$~ 0.01 & 
1.42 $\pm$~ 0.48 & 36.72 $\pm$~ 1.14 \\
NGC2976 Enuc.2a\tablenotemark{$\star$} & 6.37 $\pm$~ 0.16 & 20.2 $\pm$~ 0.29 & 
3.86 $\pm$~ 0.06 & 39.12 $\pm$~ 0.35 & 0.96 $\pm$~ 0.01 & 
1.01 $\pm$~ 0.01 & 22.01 $\pm$~ 1.13 \\
NGC2976 Enuc.2b\tablenotemark{$\star$} & 3.89 $\pm$~ 0.11 & 12.0 $\pm$~ 0.20 & 
2.47 $\pm$~ 0.05 & 23.4 $\pm$~ 0.25 & 0.46 $\pm$~ 0.01 & 
0.60 $\pm$~ 0.01 & 13.47 $\pm$~ 1.02 \\
IC2574 b\tablenotemark{$\star$} & 0.41 $\pm$~ 0.07 & 0.81 $\pm$~ 0.09 & 
0.25 $\pm$~ 0.02 & 2.08 $\pm$~ 0.15 & 0.07 $\pm$~ 0.01 & 
0.34 $\pm$~ 0.08 & 3.98 $\pm$~ 1.06 \\
NGC3521 Enuc.1 & 0.50 $\pm$~ 0.07 & 1.60 $\pm$~ 0.22 & 
0.44 $\pm$~ 0.02 & 3.46 $\pm$~ 0.25 & 0.05 $\pm$~ 0.01 & 
0.32 $\pm$~ 0.01 & 1.63 $\pm$~ 1.26 \\
NGC3521 Enuc.2b\tablenotemark{$\Delta$} & 1.02 $\pm$~ 0.08 & 3.75 $\pm$~ 0.26 & 
1.13 $\pm$~ 0.03 & 8.01 $\pm$~ 0.62 & 0.05 $\pm$~ 0.01 & 
\nodata & 4.93 $\pm$~ 1.15 \\
NGC3521 Enuc.3 & 1.66 $\pm$~ 0.07 & 5.51 $\pm$~ 0.29 & 
1.10 $\pm$~ 0.03 & 10.83 $\pm$~ 0.32 & 0.14 $\pm$~ 0.01 & 
0.16 $\pm$~ 0.03 & 5.82 $\pm$~ 1.30 \\
NGC3627 Enuc.1 & 18.9 $\pm$~ 0.55 & 57.3 $\pm$~ 0.37 & 
9.86 $\pm$~ 0.16 & 110.97 $\pm$~ 0.72 & 3.47 $\pm$~ 0.02 & 
0.59 $\pm$~ 0.01 & 40.59 $\pm$~ 1.24 \\
NGC3627 Enuc.2 & 51.6 $\pm$~ 1.25 & 169.0 $\pm$~ 0.67 & 
29.1 $\pm$~ 0.33 & 322.47 $\pm$~ 1.55 & 9.23 $\pm$~ 0.02 & 
0.80 $\pm$~ 0.01 & 100.56 $\pm$~ 1.48 \\
NGC3938 Enuc.2a & 0.50 $\pm$~ 0.08 & 1.44 $\pm$~ 0.22 & 
0.40 $\pm$~ 0.06 & 2.95 $\pm$~ 0.28 & 0.09 $\pm$~ 0.01 & 
0.31 $\pm$~ 0.08 & 2.67 $\pm$~ 1.11 \\
NGC3938 Enuc.2b & 1.11 $\pm$~ 0.08 & 2.93 $\pm$~ 0.25 & 
0.73 $\pm$~ 0.03 & 6.15 $\pm$~ 0.28 & 0.13 $\pm$~ 0.01 & 
0.41 $\pm$~ 0.10 & 7.03 $\pm$~ 1.14 \\
NGC4254 Enuc.1a\tablenotemark{$\star$} & 3.81 $\pm$~ 0.11 & 12.6 $\pm$~ 0.28 & 
2.32 $\pm$~ 0.06 & 24.59 $\pm$~ 0.33 & 0.47 $\pm$~ 0.01 & 
0.18 $\pm$~ 0.02 & 11.3 $\pm$~ 1.26 \\
NGC4321 Enuc.1 & 1.57 $\pm$~ 0.07 & 6.12 $\pm$~ 0.30 & 
1.21 $\pm$~ 0.03 & 11.64 $\pm$~ 0.33 & 0.14 $\pm$~ 0.01 & 
0.05 $\pm$~ 0.02 & 6.52 $\pm$~ 1.19 \\
NGC4321 Enuc.2b & 2.03 $\pm$~ 0.08 & 6.61 $\pm$~ 0.28 & 
1.29 $\pm$~ 0.03 & 12.87 $\pm$~ 0.31 & 0.16 $\pm$~ 0.01 & 
0.06 $\pm$~ 0.03 & 7.55 $\pm$~ 1.29 \\
NGC4631 Enuc.1\tablenotemark{$\Delta$} & 0.46 $\pm$~ 0.07 & 1.42 $\pm$~ 0.28 & 
0.40 $\pm$~ 0.02 & 3.02 $\pm$~ 0.31 & 0.10 $\pm$~ 0.01 & 
0.40 $\pm$~ 0.16 & 2.81 $\pm$~ 1.08 \\
NGC4631 Enuc.2a & 2.02 $\pm$~ 0.08 & 6.49 $\pm$~ 0.28 & 
1.36 $\pm$~ 0.03 & 12.98 $\pm$~ 0.31 & 0.36 $\pm$~ 0.01 & 
0.40 $\pm$~ 0.03 & 7.80 $\pm$~ 1.12 \\
NGC4736 Enuc.1a\tablenotemark{$\star$}\tablenotemark{$\Delta$} & 12.3 $\pm$~ 0.28 & 39.9 $\pm$~ 0.28 & 
8.32 $\pm$~ 0.22 & 78.98 $\pm$~ 0.49 & 1.20 $\pm$~ 0.01 & 
0.61 $\pm$~ 0.01 & 50.41 $\pm$~ 1.25 \\
NGC5055 Enuc.1\tablenotemark{$\star$} & 4.41 $\pm$~ 0.17 & 14.6 $\pm$~ 0.27 & 
2.61 $\pm$~ 0.05 & 27.73 $\pm$~ 0.34 & 0.55 $\pm$~ 0.01 & 
0.25 $\pm$~ 0.01 & 13.76 $\pm$~ 1.14 \\
NGC5194 Enuc.10b\tablenotemark{$\star$} & 8.49 $\pm$~ 0.30 & 27.5 $\pm$~ 0.35 & 
5.33 $\pm$~ 0.09 & 54.42 $\pm$~ 0.51 & 0.80 $\pm$~ 0.01 & 
0.13 $\pm$~ 0.02 & 24.67 $\pm$~ 1.71 \\
NGC5194 Enuc.11d & 3.01 $\pm$~ 0.08 & 10.8 $\pm$~ 0.21 & 
1.96 $\pm$~ 0.05 & 20.84 $\pm$~ 0.25 & 0.23 $\pm$~ 0.01 & 
0.07 $\pm$~ 0.03 & 9.26 $\pm$~ 1.12 \\
NGC5194 Enuc.11e\tablenotemark{$\Delta$} & 3.98 $\pm$~ 0.10 & 13.2 $\pm$~ 0.14 & 
2.35 $\pm$~ 0.06 & 25.47 $\pm$~ 0.48 & 0.40 $\pm$~ 0.01& 
\nodata & 11.49 $\pm$~ 0.79 \\
NGC5194 Enuc.1c\tablenotemark{$\star$} & 6.20 $\pm$~ 0.19 & 19.9 $\pm$~ 0.20 & 
3.77 $\pm$~ 0.07 & 39.50 $\pm$~ 0.30 & 0.80 $\pm$~ 0.01 & 
0.09 $\pm$~ 0.02 & 18.23 $\pm$~ 1.06 \\
NGC5194 Enuc.2\tablenotemark{$\star$} & 14.3 $\pm$~ 0.39 & 39.6 $\pm$~ 0.44 & 
7.23 $\pm$~ 0.14 & 79.88 $\pm$~ 0.63 & 2.29 $\pm$~ 0.01 & 
0.27 $\pm$~ 0.06 & 36.54 $\pm$~ 1.13 \\
NGC5194 Enuc.3\tablenotemark{$\star$} & 8.90 $\pm$~ 0.26 & 27.5 $\pm$~ 0.17 & 
4.61 $\pm$~ 0.09 & 52.98 $\pm$~ 0.34 & 1.29 $\pm$~ 0.01 & 
0.23 $\pm$~ 0.03 & 28.25 $\pm$~ 1.05 \\
NGC5194 Enuc.4b\tablenotemark{$\star$} & 8.16 $\pm$~ 0.18 & 25.2 $\pm$~ 0.16 & 
4.33 $\pm$~ 0.10 & 48.01 $\pm$~ 0.28 & 0.74 $\pm$~ 0.01 & 
0.10 $\pm$~ 0.01 & 20.79 $\pm$~ 0.97 \\
NGC5194 Enuc.5\tablenotemark{$\star$} & 6.18 $\pm$~ 0.19 & 19.6 $\pm$~ 0.28 & 
3.53 $\pm$~ 0.10 & 37.34 $\pm$~ 0.38 & 0.86 $\pm$~ 0.01 & 
0.25 $\pm$~ 0.01 & 18.37 $\pm$~ 1.16 \\
NGC5194 Enuc.7b\tablenotemark{$\star$} & 7.13 $\pm$~ 0.22 & 21.1 $\pm$~ 0.36 & 
3.96 $\pm$~ 0.07 & 41.26 $\pm$~ 0.48 & 1.01 $\pm$~ 0.01 & 
0.48 $\pm$~ 0.01 & 21.37 $\pm$~ 1.55 \\
NGC5194 Enuc.8\tablenotemark{$\star$} & 12.5 $\pm$~ 0.35 & 38.1 $\pm$~ 0.31 & 
6.57 $\pm$~ 0.14 & 74.48 $\pm$~ 0.53 & 1.66 $\pm$~ 0.01 & 
0.29 $\pm$~ 0.01 & 39.31 $\pm$~ 1.47 \\
NGC5194 Enuc.9\tablenotemark{$\star$} & 7.52 $\pm$~ 0.23 & 24.5 $\pm$~ 0.44 & 
4.98 $\pm$~ 0.17 & 48.89 $\pm$~ 0.57 & 0.93 $\pm$~ 0.01 & 
0.13 $\pm$~ 0.04 & 21.74 $\pm$~ 1.67 \\
NGC5713 Enuc.2a & 8.92 $\pm$~ 0.21 & 28.7 $\pm$~ 0.34 & 
5.56 $\pm$~ 0.10 & 55.76 $\pm$~ 0.99 & 1.19 $\pm$~ 0.01 & 
\nodata & 26.48 $\pm$~ 1.06 \\
NGC6946 Enuc.1\tablenotemark{$\star$}\tablenotemark{$\Delta$} & 5.72 $\pm$~ 0.14 & 16.7 $\pm$~ 0.31 & 
3.08 $\pm$~ 0.04 & 32.89 $\pm$~ 0.36 & 0.66 $\pm$~ 0.01 & 
1.67 $\pm$~ 0.01 & 17.54 $\pm$~ 1.03 \\
NGC6946 Enuc.2b\tablenotemark{$\star$}\tablenotemark{$\Delta$} & 5.23 $\pm$~ 0.10 & 15.9 $\pm$~ 0.30 & 
2.90 $\pm$~ 0.05 & 30.81 $\pm$~ 0.33 & 1.24 $\pm$~ 0.01 & 
3.86 $\pm$~ 0.02 & 20.3 $\pm$~ 1.04 \\
NGC6946 Enuc.3a & 1.19 $\pm$~ 0.04 & 3.97 $\pm$~ 0.12 & 
0.75 $\pm$~ 0.02 & 7.55 $\pm$~ 0.32 & 0.28 $\pm$~ 0.01& 
\nodata & 3.56 $\pm$~ 0.75 \\
NGC6946 Enuc.3b\tablenotemark{$\star$}\tablenotemark{$\Delta$} & 3.74 $\pm$~ 0.10 & 12.6 $\pm$~ 0.15 & 
2.13 $\pm$~ 0.03 & 23.49 $\pm$~ 0.21 & 0.54 $\pm$~ 0.01 & 
0.58 $\pm$~ 0.01 & 12.33 $\pm$~ 1.25 \\
NGC6946 Enuc.4a & 2.68 $\pm$~ 0.07 & 8.00 $\pm$~ 0.33 & 
1.74 $\pm$~ 0.03 & 16.33 $\pm$~ 0.35 & 0.38 $\pm$~ 0.01& 
0.35 $\pm$~ 0.01 & 9.99 $\pm$~ 0.97 \\
NGC6946 Enuc.5b\tablenotemark{$\star$} & 1.75 $\pm$~ 0.07 & 5.27 $\pm$~ 0.34 & 
1.04 $\pm$~ 0.03 & 10.42 $\pm$~ 0.36 & 0.37 $\pm$~ 0.01 & 
0.35 $\pm$~ 0.01 & 5.36 $\pm$~ 1.11 \\
NGC6946 Enuc.6a\tablenotemark{$\star$} & 14.1 $\pm$~ 0.32 & 44.1 $\pm$~ 0.20 & 
7.07 $\pm$~ 0.12 & 84.12 $\pm$~ 0.43 & 2.38 $\pm$~ 0.01 & 
0.70 $\pm$~ 0.01 & 37.77 $\pm$~ 1.14 \\
NGC6946 Enuc.7 & 11.2 $\pm$~ 0.22 & 35.5 $\pm$~ 0.24 & 
6.33 $\pm$~ 0.10 & 68.61 $\pm$~ 0.38 & 1.20 $\pm$~ 0.01 & 
0.40 $\pm$~ 0.01 & 27.53 $\pm$~ 1.50 \\
NGC6946 Enuc.8\tablenotemark{$\star$}  & 14.0 $\pm$~ 0.34 & 46.7 $\pm$~ 0.32 & 
7.47 $\pm$~ 0.12 & 87.16 $\pm$~ 0.52 & 1.73 $\pm$~ 0.01 & 
0.45 $\pm$~ 0.01 & 29.84 $\pm$~ 1.19 \\
NGC6946 Enuc.9 & 16.1 $\pm$~ 0.37 & 49.7 $\pm$~ 0.26 & 
8.33 $\pm$~ 0.13 & 95.01 $\pm$~ 0.53 & 2.73 $\pm$~ 0.01 & 
0.47 $\pm$~ 0.01 & 37.54 $\pm$~ 1.49 \\
NGC7793 Enuc.1 & 0.72 $\pm$~ 0.07 & 2.74 $\pm$~ 0.22 & 
0.64 $\pm$~ 0.02 & 5.53 $\pm$~ 0.25 & 0.11 $\pm$~ 0.01 & 
0.23 $\pm$~ 0.01 & 3.15 $\pm$~ 1.11 \\
NGC7793 Enuc.2 & 1.35 $\pm$~ 0.07 & 4.51 $\pm$~ 0.28 & 
0.95 $\pm$~ 0.03 & 8.85 $\pm$~ 0.31 & 0.11 $\pm$~ 0.01 & 
0.12 $\pm$~ 0.01 & 4.95 $\pm$~ 1.23 \\
NGC7793 Enuc.3 & 2.37 $\pm$~ 0.08 & 7.30 $\pm$~ 0.26 & 
1.57 $\pm$~ 0.03 & 14.63 $\pm$~ 0.29 & 0.33 $\pm$~ 0.01 & 
0.37 $\pm$~ 0.03 & 13.5 $\pm$~ 1.11 \\
\enddata
\tablenotemark{$\star$}{Region is part of calibration dataset as defined in Section~\ref{sec:sample}.}\\
\tablenotemark{$\Delta$}{Corresponding SL data does not have full coverage in 11$\arcsec$ aperture, region is not included in background-subtracted subset used in Section~\ref{sec:bandratios}.}\\
\end{deluxetable*}

We analyzed the SINGS SL spectra using the IDL program PAHFIT to determine the strengths and uncertainties of emission features from PAHs and ions \citep{smith07}. Spectral lines from ions such as \netwo\ and pure rotational lines from molecular hydrogen such as 12.28~\micron\ H$_2$\it{S}\rm(2) were fit with Gaussian profiles while PAH features were modelled with the Drude profile of a classical damped harmonic oscillator. We input the redshift of the galaxies to obtain the most accurate PAHFIT measurements using values listed in Table~\ref{tab:galaxy}. We also allowed a fit to the optical depth of the silicate absorption feature at 9.7~\micron\ but find no significant detection in any region. A selection of PAHFIT results are shown in Table~\ref{tab:pahfit}. We obtained measurements for some emission lines and PAH features that are not included in this Table, but we do not consider them for our SFR comparisons because a significant portion of the regions are non-detections. These include the weak PAH feature at 14.2~\micron\ and the rotational lines of hydrogen H$_2$\it{S}\rm(2) and \it{S}\rm(3) at 12.28 and 9.67~\micron, respectively. We also omit the forbidden line emission from [Ar~II] at 6.99~\micron\ and [Ar~III] at  8.99~\micron\ since they are non-detections in many regions (see Figure~\ref{fig:spectra} and~\ref{fig:lower_spectra}). The emission from [S~IV] at 10.51~\micron\ is similarly too weak to be measured at the resolution of SL spectra. Several other interesting emission lines (e.g. [S~III] 33~\micron) lie outside the wavelength range of our SL and SH spectra, but may be an interesting topic for future studies.

SINGS SH spectra were analyzed with the \texttt{specutils} package of \texttt{astropy} to calculate the integrated line emission of \nethree. We extracted a spectral region containing the emission feature between 14.9 and 16.3~\micron\ to include $\sim$0.4~\micron\ of continuum around the emission line. The continuum that underlies the emission feature was fit by taking the average surface brightness between 14.9 and 15.3~\micron\ and another average between 15.9 and 16.3~\micron. We determined the linear fit that connects these two averages to model the continuum, then subtracted this continuum from the extracted spectrum. 

We set the maximum surface brightness measurement between 14.9 and 16.3~\micron\ as an initial estimate of the peak of the \nethree\ emission line. We then fit a Gaussian to the continuum-subtracted data using the emission line parameter modelling functions of \texttt{specutils} to determine the integrated intensity. Uncertainties were propagated by the Monte Carlo method where spectral data were randomly offset according to their corresponding uncertainty in each wavelength bin and the Gaussian was re-fit. We repeated 500 trials in this way, using the standard deviation of the results to calculate the uncertainty on the integrated intensity of the \nethree\ emission. The mean and standard deviation of these 500 trials are our [Ne~III] line strength and its respective uncertainty. These measurements for regions with SH spectra are shown in Table~\ref{tab:pahfit}.

Table~\ref{tab:pahfit} also shows the surface brightness of each H~II region at 10~\micron\ in the SL spectra in $\nu I_{\nu}$ units. We calculate this value as the average over three wavelength bins at $\sim$9.9, 10.0, and 10.1~\micron\ to obtain a measure of the mid-IR continuum. This average is then multiplied by 30 THz (i.e. 10~\micron\ in terms of frequency). This unit conversion ensures ratios of [Ne~III] emission and PAHFIT results with 10~\micron\ continuum values are unitless. We note that 10 \micron\ emission will be a poor tracer of the underlying dust emission in the case where there is strong silicate absorption. We caution the reader against using any SFR calibrations that involve 10 \micron\ continuum in that case.

\subsection{Correlation Methodology}\label{sec:corr_methods}
We use T33 surface brightness as the reference SFR indicator for our calibrations. Converting T33 to SFR surface density \sfrsig\ by Eq.~11 from \cite{murphy+11} requires a multiplicative factor, which we omit from the calibrations for convenience, but is reproduced in Equation~\ref{eq:sigsfr}. This relation also has a dependence on the local electron temperature and we assume a constant value of T$_{e}$~=~10$^{4}$K that is typical for H~II regions in nearby galaxies. Inherent in the multiplicative factor from \citet{murphy+11} is the assumption of the IMF, which other SFR calibrations may not share. The SFR calibration for neon and metallicity from \cite{zho_ne19}, for example, assumes a Salpeter IMF which differs from the others which assume a Kroupa IMF. As shown in \cite{murphy+11}, calibrations based on the Salpeter IMF result in SFR about 50$\%$ greater than those that use the Kroupa IMF. To compare each model equally in Section~\ref{sec:SFR_plots}, we take this offset into account.

We assume all emission features can be treated in the same manner and that the functional form of the T33 calibration is a power law with the general form: 

\begin{equation}\label{eq:triple}
\begin{split}
    \log_{10}\left[\frac{T\textit{33}}{\rm mJy~sr^{-1}}\right] &~=~\alpha \log_{10}\left[\frac{X}{\unit}\right] \\ 
    & + \beta \log_{10}\left[Y\right] + \phi \log_{10}\left[Z\right] + \gamma
\end{split}
\end{equation}

where \it{X}\rm, \it{Y}\rm, and \it{Z}\rm\  are any emission features and \it{$\alpha$}\rm, \it{$\beta$}\rm, \it{$\phi$}\rm, and \it{$\gamma$}\rm\ are constants we fit by ordinary least squares. Parameters \textit{Y} and \textit{Z} are typically unitless band ratios or metallicity values (converted to [O/H]) but the integrated intensity of emission lines or PAH features have the units of \textit{X}.
We first investigate single-feature models where \it{$\beta$}\rm\ and \it{$\phi$}\rm\ in Eq.~\ref{eq:triple} are zero. All emission features in Table~\ref{tab:pahfit} have been converted to $\unit$ to simplify use in Eq.~\ref{eq:triple}. In Section~\ref{sec:SFR_plots} we convert to SFR using distances listed in Table~\ref{tab:galaxy}.

We also use Eq.~11 from \cite{murphy+11} to write our Eq.~\ref{eq:triple} as a relation for determining \sfrsig:

\begin{equation}\label{eq:sigsfr}
\begin{split}
    \log_{10}\left(\frac{\Sigma_{SFR}}{\rm M_{\odot} yr^{-1} kpc^{-2}}\right)~=\\
    \alpha~\log_{10}\left[\frac{X}{\unit}\right]~+~\gamma - 1.08
\end{split}
\end{equation}

where notation is preserved from Eq.~\ref{eq:triple} and parameters log$_{10}$[$Y$] and log$_{10}$[$Z$] are not shown for brevity but can be included following the functional form of Eq.~\ref{eq:triple}. We attempted to further improve our single-feature model by including a second feature \it{Y}\rm, setting only $\phi$ to zero. These regressions are run independently of the single-feature correlations so that the constants \it{$\alpha$}\rm\ and \it{$\gamma$}\rm\ in Eq.~\ref{eq:triple} are not necessarily equal for a given feature \it{X}\rm\ after including another feature \it{Y}\rm. If KK04 or PT05 metallicity values are used for log$_{10}$[\it{Y}\rm], we set \textit{Y} in terms of [O/H]. Band ratios and 10~\micron-normalized bands used for \it{Y}\rm\ are unitless. Finally, we attempted to improve our double-feature models by including a third feature \it{Z}\rm. As before, these trials are run independently of the previous correlations so that no constants are necessarily preserved from single- or double-feature calibrations.

In our calibrations to Eq.~\ref{eq:triple}, we employed a Monte Carlo approach to propagate uncertainties in both the independent and dependent variables. In each trial, random offsets were added to each point according to their corresponding measurement uncertainties, assuming a Gaussian distribution. We also quantified the dependence of our results on the specific regions considered using the bootstrap technique. This approach measures the uncertainty that results from the small sample size and non-Gaussian scatter of our dataset. From our set of 33 regions, we create a new set of the same size where each value is randomly selected from the 33 regions. This allows some regions to contribute to the fit more than once in a given trial, or not at all. The resulting set is then used to fit the constants of Eq.~\ref{eq:triple} by the ordinary least-squares method.

We measured the quality of single- and multi-feature models found by Eq.~\ref{eq:triple} using the Spearman rank correlation coefficient between the predicted T33 and the observed values. In this way we uniformly quantify the strength of each correlation regardless of how many parameters are used to predict T33 by Eq.~\ref{eq:triple}. We then determined the residuals as the difference between the predicted and observed T33 values for each region. From the residuals we calculated the standard deviation \tildel\ and quantified the presence of any metallicity-dependent offset, as follows. The set of regions was divided into two groups based on the median metallicity value (12 + log[O/H] = 8.98$_{KK04}$ = 8.36$_{PT05}$). We calculated the median of the residuals in the two metallicity bins, and the difference between the high- and low-metallicity\footnote{We find the separation based on metallicity is the same in KK04 and PT05, except for two regions, which makes a negligible difference in the results.} residual medians is what we define as the metallicity offset \Zoff. 

We perform 10$^3$ bootstrap/Monte Carlo trials and calculate the mean and standard deviation of all values to quantify the uncertainty. We find the uncertainty from bootstrap variation exceeds that returned by the Monte Carlo method. We apply both methods in each trial: measurement uncertainties are propagated via the Monte Carlo technique, then sampling uncertainties are included by applying the bootstrap method.

We confirm the significance of our \Zoff\ parameter using the 2-sample Kolmogorov-Smirnov test (K-S test) and the 2-sample Anderson-Darling test (A-D test). Each of these tests yields a probability that the two samples are drawn from the same distribution. For each correlation, we divide the regions based on their metallicity as described previously for determining \Zoff, then we run the K-S and A-D test on these two portions of the sample. We find both tests return similar probabilities. We consider \Zoff\ values significant when these tests find $>90\%$ confidence that the two metallicity subsets are not drawn from the same distribution. We note that the magnitude of \Zoff\ does not necessarily indicate a statistically significant difference between the two metallicity subsets. In some cases, a given tracer has a poor correlation with T33 for reasons not tied to metallicity, leading to a large scatter in the residuals. In such a case, a relatively large offset between the median residual in the upper and lower metallicity bin may be statistically insignificant. On the opposite extreme, for tracers that are well correlated with T33, a small \Zoff\ can be significant when the intrinsic scatter in the relationship is small.

A perfect correlation would have \rhos~=~1, \Zoff~=~0, and \tildel~=~0. However, our T33 data have non-negligible measurement uncertainties which limit the practical maximum value of \rhos, and minimum values of \Zoff\ and \tildel\ when used in our combined bootstrap and Monte Carlo perturbation method \citep{MCspearman}. We attempted to quantify this limit by using T33 and its associated uncertainty as feature \it{X}\rm\ in bootstrap Monte Carlo trials using Eq.~\ref{eq:triple} with $\beta$ and $\phi$ set to zero. We expect the mean of these trials to show a perfect correlation if the random offsets in each trial were negligible, but we find a practical maximum \rhos\ of 0.92, and no minimum for \Zoff\ or \tildel.

On the other end of the scale, we investigated the null hypothesis for \rhos\ where two datasets are considered statistically indistinguishable from random, uncorrelated datasets. Correlations with a Spearman coefficient less than or equal to a critical value $\rho_c$ are consistent with this null hypothesis. This $\rho_c$ is determined by producing two random sets of 33 points to match the size of our calibration set. In each trial, two sets of 33 random points are created and \rhos\ is calculated between them. We repeat 10$^4$ such trials and find the \rhos\ values converge on zero with a standard deviation that defines the 1$\sigma$ null range: $\rho_c$~=~0.18. Datasets that have $|$\rhos$|$\ within this range are considered uncorrelated. We also note that due to the properties of our sample, metallicity is correlated with T33 with \rhos~=~0.3. This correlation likely results from a sparse sampling of low-metallicity regions below about 12 + log[O/H]$_{KK04}$ $\sim8.8$, with exception of two dwarf galaxy regions at about 8.2. We anticipate that this could introduce spurious correlations with T33 at the level of \rhos~=~0.3 for highly metallicity-dependent emission features.

\begin{figure*}
\centering
\includegraphics[width=3.5in]{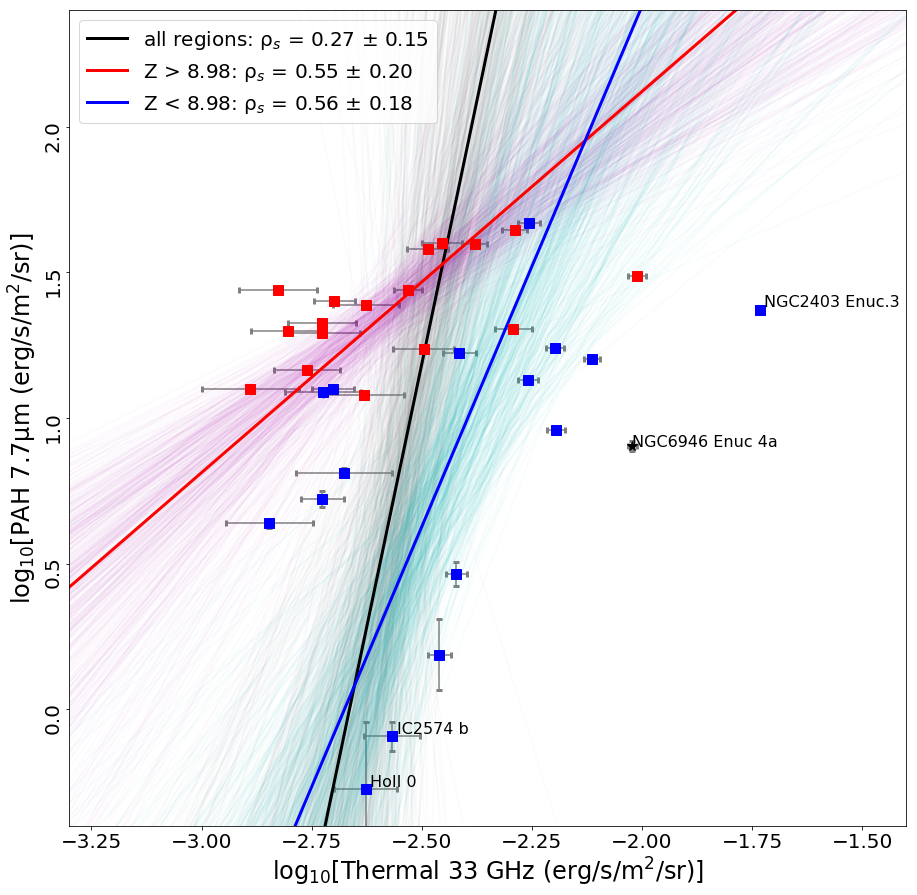}
\includegraphics[width=3.5in]{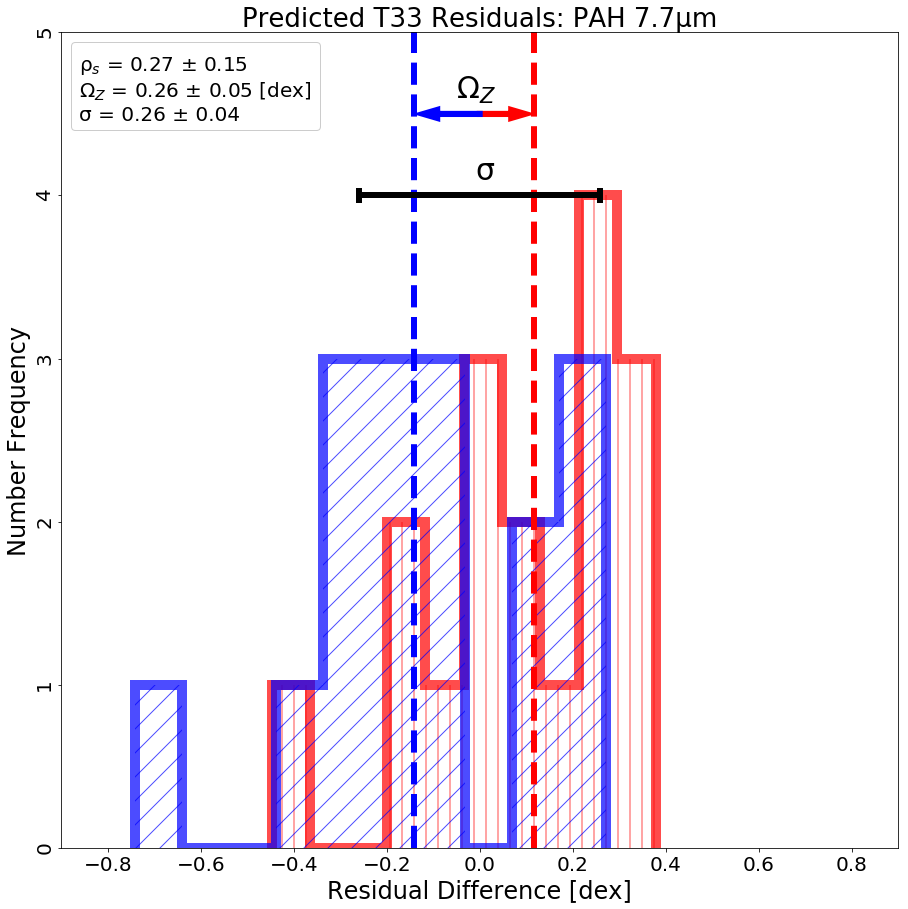}
\caption{(a) Observed T33 in $\nu I_{\nu}$ units as a function of PAH 7.7~\micron\ brightness with Monte Carlo fit lines to Eq.~\ref{eq:triple} (with $\beta$~=~$\phi$~=~0) for high- (red) and low-metallicity (blue) points separately and all points (black).
(b) Residual difference between observed T33 and predicted T33 based on PAH 7.7~\micron. A horizontal line annotates the standard deviation of the residuals $\pm$\tildel. Vectors indicating the metallicity offset \Zoff\ as defined using the high- and low-metallicity median of residuals are shown with color labels preserved from (a). We subtract the red value by the blue value to determine \Zoff; in this example we find a positive \Zoff\ from the difference between the positive high-metallicity median in red and negative low-metallicity median in blue.}
\label{fig:PAH77}
\end{figure*}

Figure~\ref{fig:PAH77} shows an example of a single-feature correlation with T33 using Eq.~\ref{eq:triple}, specifically PAH 7.7~\micron\ versus T33. To emphasize the relative strength of mid-IR bands, we have converted T33 to $\nu I_{\nu}$ units by multiplying the observed T33 by its frequency to compare with the PAH 7.7~\micron\ feature only in Figure~\ref{fig:PAH77}a. This figure also shows the fit from each regression trial and their mean in black. We also repeat this for the high- and low-metallicity groups individually, shown in red and blue lines, respectively. Figure~\ref{fig:PAH77}b summarizes our statistics on the residuals: \Zoff\ and \tildel. In this case, the correlation of T33 with PAH 7.7~\micron\ is weak, with \rhos\ $=$ 0.27, significant scatter \tildel\ $=$ 0.26, and a large metallicity offset \Zoff\ $=$ 0.26 (with $<10\%$ chance that the metallicity subsets of the residuals come from the same distribution). In Figure~\ref{fig:PAH77}b we illustrate our \Zoff\ statistic as the difference between the red and blue vertical lines, and $\pm$ the standard deviation~\tildel\ is shown as the length of the horizontal black line. Table~\ref{tab:model_quality} lists \rhos, \Zoff, and \tildel\ for the correlations discussed in Section~\ref{sec:results} and Appendix~\ref{sec:hists} contains histograms in the form of Figure~\ref{fig:PAH77}b for these correlations.

There are a few measurements with notable characteristics indicated in Figure~\ref{fig:PAH77}a. These are the dwarf galaxy regions Holmberg II-0 and IC 2574-b (the two lowest metallicity points in our sample) and the known anomalous microwave emission (AME) source designated NGC6946 Enuc.4a \citep{murphy+10}. The AME source is plotted but not included in our calculations of \rhos, \Zoff, and \tildel\ for T33 correlations due to the observed excess 33 GHz emission relative to its SFR. We also indicate the region NGC2403 Enuc.3 which is a more significant outlier than the known AME source. This region has recently been proposed as a potential AME source as well \citep{linden} which likely explains the significant excess of T33 relative to PAH 7.7~\micron\ emission observed.

\section{Results}\label{sec:results}
\begin{figure*}
\includegraphics[width=6.5in]{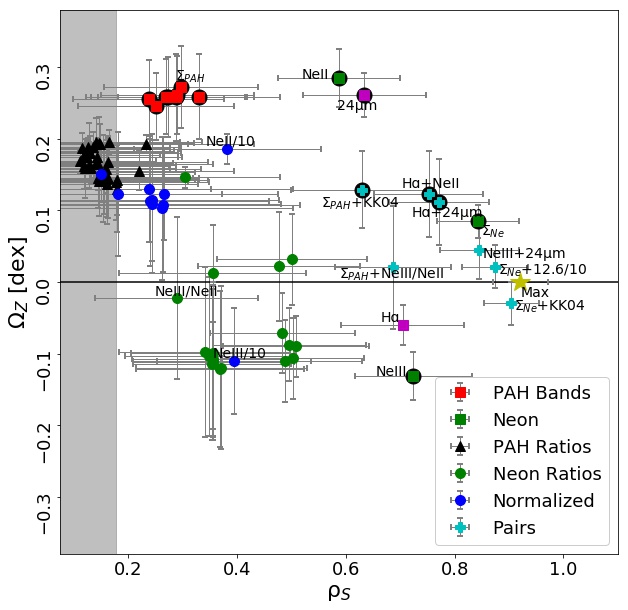}
\caption{Metallicity offset \Zoff\ as a function of Spearman correlation coefficient with T33 \rhos\ for individual mid-IR bands, their ratios and continuum-normalized values, and several pair tracers of interest. The `Max' value indicated by the $\star$ is the practical maximum \rhos\ defined by the noise in our T33 measurements (see end of Section~\ref{sec:corr_methods})}
\label{fig:offsetspear}
\end{figure*}

We measured the properties of various correlations of mid-IR emission features with T33 and their metallicity dependence. For each correlation we measured \rhos, \Zoff, and \tildel, and their uncertainties. We focused primarily on maximizing \rhos\ in the following sections since this generally minimizes \tildel\ as well. In cases of similar \rhos\ we then looked for minimum \Zoff\ to find tracers that are more independent of metallicity. These parameters describing the quality of our results are summarized in Figure~\ref{fig:offsetspear} where metallicity offset \Zoff\ is plotted against Spearman rank correlation coefficient \rhos. In this figure the features that have a statistically significant \Zoff, as defined in Section~\ref{sec:corr_methods}, are indicated by a black circle surrounding the symbol. Table~\ref{tab:model_quality} lists our main results for the following model quality statistics for several tracers: Spearman rank correlation coefficient \rhos, metallicity offset \Zoff, standard deviation of residuals \tildel, wavelength range \dellam, and maximum wavelength \lammax. Wavelength range \dellam\ and maxima \lammax\ for these tracers are calculated from the peak location of constituent emission features but additional spectral coverage will be necessary to separate most PAH and neon features from continuum. The following sections present details on trends and features of interest in this Table and Figure~\ref{fig:offsetspear}. In the Appendix, we also describe the optimal correlations obtained in restricted wavelength ranges, which may be of interest for higher redshift observations.

\subsection{Correlations of Individual Mid-IR Emission Features with Thermal 33 GHz Emission}\label{sec:T33_corr_1}

In this section, we present single-feature fit results for Eq.~\ref{eq:triple} (with $\beta$~=~$\phi$~=~0). The best-correlated mid-IR observable with T33 is the \nethree\ line with \rhos~=~0.72. This emission line is the only single feature considered that has a statistically significant negative metallicity offset (\Zoff~=~$-0.13$ dex). Low-metallicity H~II regions show increased \nethree\ emission relative to their SFR compared to higher-metallicity regions. On average, a low-metallicity region is found to have about 35$\%$ more [Ne~III] emission than a typical high-metallicity region of the same T33. After [Ne~III] the next best individual feature is the \netwo\ line which has a strong positive metallicity offset (\rhos~=~0.59 and \Zoff~=~0.28 dex). In comparison to [Ne~III], an average high-metallicity region is found to emit about 90$\%$ more [Ne~II] emission than a typical low-metallicity region of the same T33.

All PAH bands in this study show weak correlation with T33, characterized by \rhos~$\sim~0.3$. We find significant positive metallicity offsets of \Zoff~$\sim~0.25$ dex for each PAH band correlation with T33. Adding the PAH band emission into a total $\Sigma$PAH does not improve T33 correlations or reduce the metallicity offset. We find that our high- and low-metallicity sub-samples are better correlated separately than combined because of these metallicity-dependent offsets as seen in Figure~\ref{fig:PAH77}a. Ratios of PAH bands are uncorrelated (i.e. \rhos$~<~\rho_{C}$). If we normalize the PAH bands by 10~\micron\ continuum, the magnitude of \rhos\ is typically unaffected, but the metallicity offset is reduced and the two metallicity samples become consistent with being drawn from the same distribution.

Following the work of \cite{keto07}, we combine the \netwo\ and \nethree\ emission line integrated intensities into a new `single' feature: \sumneon. We refer to \sumneon\ as an individual parameter when used in Eq.~\ref{eq:triple} despite being a combination of two distinct emission features. This combined feature is our best single parameter \textit{X} for predicting T33 by Eq.~\ref{eq:triple}, with \rhos~=~0.84 and a small, but statistically significant \Zoff~=~0.08 dex (see Appendix~\ref{sec:hists} for histogram). We find that adding the [Ne~II] and [Ne~III] emission lines to create \sumneon\ results in a similar \rhos\ within about 1$\sigma$ to that obtained by using them separately as parameters \textit{X} and \textit{Y} (\rhos~=~0.80~$\pm~$0.09). We discuss this observation in Section~\ref{sec:disc}.

\subsection{Correlations of Multiple Observables with Thermal 33 GHz Emission}\label{sec:T33_corr_2}

\begin{deluxetable*}{cccccccc}
\tabletypesize{\scriptsize}
\tablecaption{T33 Correlation Statistics \label{tab:model_quality}} 
\tablewidth{0pt}
\tablehead{
\colhead{ } &
\colhead{\textbf{Feature}} &
\colhead{ } &
\colhead{Correlation Coefficient} &
\colhead{Metallicity Offset} &
\colhead{Standard Deviation}&
\colhead{Wavelength Range}&
\colhead{Max Wavelength}\\ [-3mm]
\colhead{\textit{X}} &
\colhead{\textit{Y}} &
\colhead{\textit{Z}} &
\colhead{\rhos} &
\colhead{\Zoff\ [dex]} &
\colhead{\tildel\ [dex]} &
\colhead{\dellam\ [\micron]}&
\colhead{$\lambda_{max}$\ [\micron]}
}
\startdata
\sumneon & KK04  & \nodata & 0.91 $\pm~$ 0.05 & -0.03 $\pm~$ 0.03 & 0.09 $\pm~$ 0.01 & \nodata & \nodata \\
\sumneon & 12.6/10  & \nodata & 0.87 $\pm~$ 0.06 & 0.02 $\pm~$ 0.03 & 0.11 $\pm~$ 0.02 & 5.6 & 15.6   \\
$[$Ne III$]$ & 24\micron & \nodata & 0.85 $\pm~$ 0.07 & 0.04 $\pm~$ 0.04 & 0.11 $\pm~$ 0.01 & 8.4 & 24.0  \\
\sumneon & \nodata  & \nodata & 0.84 $\pm~$ 0.08 & 0.08 $\pm~$ 0.02\tablenotemark{$\star$} & 0.14 $\pm~$ 0.03 & 2.8 & 15.6  \\
$[$Ne III$]$ & 13.6 & \nodata & 0.80 $\pm~$ 0.09 & -0.01 $\pm~$ 0.05 & 0.13 $\pm~$ 0.01 & 2.0 & 15.6 \\
$[$Ne  II$]$ & 12.6  & 12.6/10 & 0.78 $\pm~$ 0.10 & 0.07 $\pm~$ 0.07\tablenotemark{$\star$} & 0.16 $\pm~$ 0.03 & 2.8 & 12.8 \\
H$\alpha$ & 24\micron & \nodata & 0.77 $\pm~$ 0.09 & 0.11 $\pm~$ 0.06\tablenotemark{$\star$} & 0.16 $\pm~$ 0.02 & \nodata & \nodata  \\
$[$Ne III$]$ & \nodata & \nodata & 0.72 $\pm~$ 0.11 & -0.13 $\pm~$ 0.03\tablenotemark{$\star$} & 0.17 $\pm~$ 0.03 & 0.0 & 15.6  \\
$[$Ne II$]$ & 12.6  & \nodata & 0.72 $\pm~$ 0.11 & 0.11 $\pm~$ 0.07 & 0.22 $\pm~$ 0.03 & 0.2 & 12.8 \\
$[$Ne II$]$ & \nodata & \nodata & 0.59 $\pm~$ 0.11 & 0.28 $\pm~$ 0.04\tablenotemark{$\star$} & 0.23 $\pm~$ 0.03  & 0.0 & 12.8 \\
PAH 11.3\micron & 12.0/10 & \nodata & 0.55 $\pm~$ 0.15 & 0.24 $\pm~$ 0.05\tablenotemark{$\star$} & 0.23 $\pm~$ 0.02  & 2.0 & 12.0 \\
PAH 7.7\micron & 7.7/10 & \nodata & 0.54 $\pm~$ 0.15 & 0.26 $\pm~$ 0.07\tablenotemark{$\star$} & 0.26 $\pm~$ 0.04 & 2.3 & 10.0 \\
\enddata
\tablenotemark{$\star$}{Metallicity offset is statistically significant as defined in Section~\ref{sec:corr_methods}.}\\
\end{deluxetable*}

\begin{deluxetable*}{ccccccr}
\tabletypesize{\scriptsize}
\tablecaption{T33 and SFR Tracer Constants for Eqs.~\ref{eq:triple}\tablenotemark{$\star$} and \ref{eq:SFR_power} \label{tab:sfr_constants}}
\tablewidth{0pt}
\tablehead{
\colhead{ } &
\colhead{\textbf{Feature}} &
\colhead{ } &
\colhead{ } &
\colhead{\textbf{Constants}} &
\colhead{ } &
\colhead{ }\\ [-3mm]
\colhead{\textit{X}} &
\colhead{\textit{Y}} &
\colhead{\textit{Z}} &
\colhead{$\alpha$} &
\colhead{$\beta$} &
\colhead{$\phi$}&
\colhead{$\gamma$}
}
\startdata
\sumneon & KK04  & \nodata & 0.87 $\pm~$ 0.09 & -0.48 $\pm~$ 0.14 & \nodata & -2.17 $\pm~$ 0.42 \\
\sumneon & 12.6/10  & \nodata & 0.84 $\pm~$ 0.10 & -0.40 $\pm~$ 0.18 & \nodata & -1.14 $\pm~$ 0.19  \\
$[$Ne III] & 24\micron & \nodata & 0.38 $\pm~$ 0.06 & 0.38 $\pm~$ 0.09 & \nodata & -1.14 $\pm~$ 0.17 \\
\sumneon & \nodata  & \nodata & 0.76 $\pm~$ 0.15 & \nodata & \nodata & -0.69 $\pm~$ 0.05 \\
$[$Ne III] & 13.6 & \nodata & 0.48 $\pm~$ 0.06 & 0.26 $\pm~$ 0.09 & \nodata & -0.29 $\pm~$ 0.06\\
$[$Ne II] & 12.6  & 12.6/10 & -0.47 $\pm~$ 0.24 & 1.10 $\pm~$ 0.25 & -0.86 $\pm~$ 0.35 & -1.27 $\pm~$ 0.41\\
H$\alpha$ & 24\micron & \nodata & 0.25 $\pm~$ 0.10 & 0.43 $\pm~$ 0.11 & \nodata & -1.34 $\pm~$ 0.20 \\
$[$Ne III] & \nodata & \nodata & 0.48 $\pm~$ 0.08 & \nodata & \nodata & -0.43 $\pm~$ 0.04\\
$[$Ne II] & 12.6  & \nodata & 1.09 $\pm~$ 0.35 & -0.76 $\pm~$ 0.25 & \nodata & -0.30 $\pm~$ 0.09\\
$[$Ne II] & \nodata & \nodata & 0.36 $\pm~$ 0.16 & \nodata & \nodata  & -0.52 $\pm~$ 0.05\\
PAH 11.3\micron & 12.0/10 & \nodata & 0.44 $\pm~$ 0.21 & -0.84 $\pm~$ 0.45 & \nodata  & -1.95 $\pm~$ 0.70\\
PAH 7.7\micron & 7.7/10 & \nodata & 0.48 $\pm~$ 0.19 & -0.89 $\pm~$ 0.50 &\nodata & 1.20 $\pm~$ 0.26 \\
\enddata
\tablenotemark{$\star$}[Constant $\gamma$ as listed gives T33 in mJy; units of Eq.~\ref{eq:triple}\\(mJy/sr) are obtained by adding a constant 9.54 to $\gamma$.]
\end{deluxetable*}

\subsubsection{Improvements to \textit{$\Sigma$Ne} Correlations with T33}\label{sec:totne}
In the previous section, we showed that single-feature correlations with T33 have metallicity-dependent residuals that can decrease the correlation coefficient. To improve the correlations, we therefore attempt to add an additional feature to the model that acts to remove metallicity offsets. Section~\ref{sec:T33_corr_1} shows using \sumneon\ results in the highest \rhos\ and lowest \Zoff\ values of the single mid-IR feature models. We first explicitly include metallicity as the second parameter to check that the correlation improves and to what degree. Indeed, Table~\ref{tab:model_quality} shows that using the KK04 metallicity in combination with \sumneon\ results in \rhos~=~0.91 which is consistent within 1$\sigma$ with the maximum~0.92 noted in Section~\ref{sec:corr_methods}. We also find an equivalent improvement using the PT05 metallicity values.

The KK04 and PT05 metallicity calibrations require optical emission lines and observations may not be available for all galaxies, particularly at high redshift. To avoid requiring this additional information, we searched for observables in the mid-IR that could improve the \sumneon\ correlation with T33 by acting as a proxy for metallicity. The ratio of \nethree\ and \netwo\ is expected and observed to correlate well with metallicity, however we find including it results in no significant improvement in \rhos\ for \sumneon. We discuss this observation further in Section~\ref{sec:disc}.

Another potential proxy for metallicity is our 10~\micron\ continuum-normalized PAH bands, which should trace the abundance of PAHs relative to dust and correlate well with metallicity. We find combining \sumneon\ with the strongest PAH bands normalized by 10~\micron\ continuum results in \rhos~=~0.87 and a statistically insignificant \Zoff. These values are within 1$\sigma$ of those obtained from combining \sumneon\ with KK04 or PT05 metallicities. Table~\ref{tab:model_quality} shows that the 12.6~\micron\ PAH feature normalized by continuum at 10~\micron\ can be used to remove the metallicity offset and improve the \sumneon-T33 correlation in place of KK04 or PT05 metallicity. We find that 10~\micron\ continuum-normalized PAH bands improve \rhos\ and \Zoff\ for \sumneon\ better than ratios of two PAH bands for every band considered. In Appendix~\ref{sec:Z_corr} we investigate the correlation between continuum-normalized PAH bands and metallicity. In theory, adding another feature to fit could further improve the correlation, however, we find including a third mid-IR observable results in no significant improvement to \rhos\ for the \sumneon-12.6/10~\micron\ tracer.

\subsubsection{Improvements to Other Correlations with T33}\label{sec:others}
We next investigated two-feature correlations with T33 that do not involve \sumneon. No pair of two features was found that predicts T33 measurements better than \sumneon\ paired with metallicity or a normalized PAH band. In Section~\ref{sec:T33_corr_1} we showed that the individual emission feature that best correlates with T33 is the \nethree\ line. We find that this [Ne~III] correlation is improved best by including \twenmic\ continuum as parameter \textit{Y} or with [Ne~II] to form \sumneon; each combination results in the same \rhos\ of 0.85 within 1$\sigma$. Figure~\ref{fig:offsetspear} summarizes this similarity between the [Ne~III]-24~\micron\ tracer and the \sumneon\ tracer.

A commonly used hybrid SFR tracer involves linearly combining \halpha\ emission from ionized gas and \twenmic\ continuum from dust \citep[e.g.][]{calzetti}. Although we use a different functional form (power-law rather than a linear combination), we tested the quality of the combination of \halpha\ and \twenmic\ by using the \cite{sfrs} measurements listed in Table~\ref{tab:murphy} as parameters in Eq.~\ref{eq:triple}. As expected, the correlation is strong with \rhos~=~0.77 as shown in Table~\ref{tab:model_quality} but with a statistically significant metallicity offset \Zoff~=~0.11 dex. 

We then searched for individual features that could replace either \halpha\ or \twenmic\ to give an improved correlation with T33 based on a larger \rhos, and smaller \Zoff\ and \tildel. As mentioned previously, pairing \twenmic\ with \nethree\ emission results in \rhos~=~0.85 and \Zoff~=~0.04~$\pm~$0.04 with no statistical difference between the two metallicity bins, which is a significant improvement over pairing \twenmic\ with \halpha\ in our power-law model Eq.~\ref{eq:triple}. We find no better feature to pair with \halpha\ emission than \twenmic, however pairing \halpha\ with the \netwo\ line results in an equivalent \rhos\ and \Zoff. These different combinations of \halpha, \nethree, \twenmic, and \netwo\ are further discussed in Section~\ref{sec:results_2_summ}. We also directly compare our power-law \halpha\ and \twenmic\ model with the commonly used linear hybrid \halpha\ and \twenmic\ tracer from \cite{calzetti} in Section~\ref{sec:SFR_plots}.

Despite their weak correlation with T33 individually, we investigated the full potential of PAH emission as a SFR tracer by including metallicity information. As before, the correlations are similar for each PAH band with \rhos~$\approx$~0.6 after KK04 metallicity is included. We also attempted using the ratio of \nethree\ to \netwo\ emission again as a mid-IR proxy for metallicity. Figure~\ref{fig:PAH_Z} in Appendix~\ref{sec:hists} shows combining $\Sigma$PAH with this neon line ratio results in a stronger correlation with T33 than if $\Sigma$PAH is combined with KK04 metallicity (\rhos~=~0.69 and statistically insignificant \Zoff~=~0.02 dex compared to \rhos~=~0.63 and \Zoff~=~0.13 dex). These results imply emission from PAHs is a poor option for a SFR indicator since even the metallicity-corrected PAH correlation with T33 has a \rhos\ similar to some single-feature tracers such as \nethree\ emission.

\subsubsection{Summary}\label{sec:results_2_summ}
\begin{figure}
\centering
\includegraphics[width=3.5in]{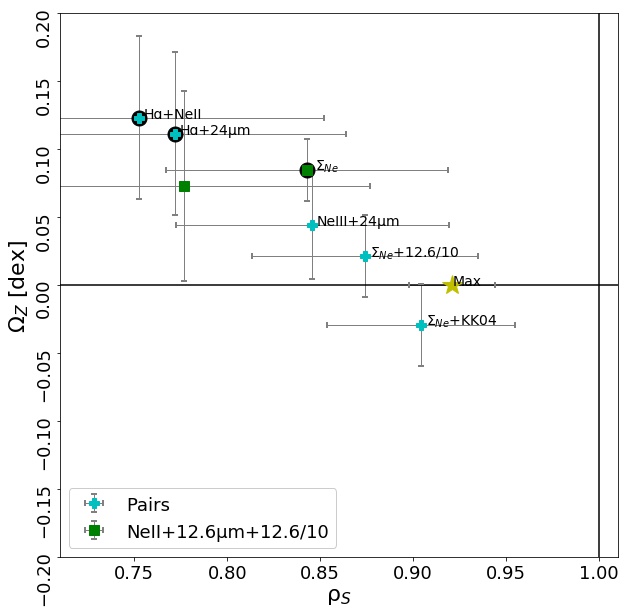}
\caption{\Zoff\ as a function of \rhos, zoomed on the best pairs and one of the best three-feature tracers.}
\label{fig:offsetspear_2}
\end{figure}

We summarize our results for mid-IR/T33 correlations by plotting their \Zoff\ as a function of \rhos\ in Figure~\ref{fig:offsetspear}. We have shown in Section~\ref{sec:corr_methods} that the T33 measurement uncertainties limit the maximum value of \rhos\ for any correlation to $\sim$0.92 which is indicated by the $\star$ in this figure. Figure~\ref{fig:offsetspear} also shows the null range for \rhos\ with dataset size of 33 points as a gray band around zero as defined in Section~\ref{sec:corr_methods}. In Figure~\ref{fig:offsetspear_2} we focus on some of the best-correlated tracers from Table~\ref{tab:model_quality}, including some that are not shown in Figure~\ref{fig:offsetspear} for clarity. Figure~\ref{fig:offsetspear_2} shows \rhos\ for the tracer using \netwo, PAH 12.6~\micron, and 10~\micron\ continuum is equivalent to the \halpha\ and \twenmic\ tracer but with a smaller \Zoff. The \sumneon\ and PAH 12.6/10~\micron\ tracer has a statistically insignificant \Zoff\ similar to the \sumneon\ and KK04 metallicity tracer and with \rhos\ within the 1$\sigma$ uncertainty.

In addition to each individual feature, we have included several pair tracers of interest in Figure~\ref{fig:offsetspear}. The strong similarities between the \sumneon\ tracer and the [Ne~III]-\twenmic\ pair tracer noted in Section~\ref{sec:others} are evident in their near overlap on this plot. Figure~\ref{fig:offsetspear} reveals a trend that combining two emission features with opposite metallicity dependence results in a stronger correlation with T33. For example, combining \nethree\ (negative \Zoff) with any PAH feature (positive \Zoff) results in a strong correlation with T33 and negligible \Zoff. The frequently used hybrid tracer based on \halpha\ and \twenmic\ is found to follow a similar pattern, however, the small, negative \Zoff\ determined for \halpha\ emission is not statistically significant. 

This trend is further supported by considering our previous observation that pairing \twenmic\ with [Ne~III] performs better than pairing with \halpha\ since the value of \Zoff\ for [Ne~III] is statistically significant while \halpha\ has a negligible \Zoff. Similarly, pairing [Ne~II] with \halpha\ performs as well as \twenmic\ with \halpha. We also find that pairing features with \Zoff\ of the same sign such as [Ne~II] with \twenmic\ or [Ne~III] with \halpha\ does not improve \rhos, \Zoff, or \tildel, based on calculated uncertainties.

\subsection{Comparisons with Previous SFR Calibrations}\label{sec:SFR_plots}

Our calibrations for \sfrsig~ tracers based on T33 were performed assuming that a power-law functional form (Eq.~\ref{eq:triple}) can be applied to all emission features considered. We compare our tracers built using this assumption against other calibrations from the literature. These other calibrations differ from that of this work in functional form and few publications have directly studied the metallicity dependence of SFR tracers using local metallicities. We used the largest applicable subset of the 56 regions for any given comparison.

Observed T33 surface brightnesses were used to derive SFR by Eq.~11 from \cite{murphy+11}. We refer to the SFR calculated using observed T33 as the reference model. We combined our Eq.~\ref{eq:triple} for predicting T33 with the SFR-T33 relation in Eq.~11 from \cite{murphy+11} such that:
\begin{equation}
\begin{split}
    \left(\frac{\rm SFR}{\rm M_{\odot} yr^{-1}}\right) &~=~7.8 \times 10^{\gamma -4}
    \left(\frac{T_e}{\rm10^{4}~K}\right)^{-0.45}
    \left(\frac{D}{\rm Mpc}\right)^2\\
    & \times \left(\frac{X}{\unit}\right)^{\alpha} \left(Y\right)^{\beta} \left(Z\right)^{\phi}
\end{split}
\label{eq:SFR_power}
\end{equation}

where notation is preserved from Eq.~\ref{eq:triple}, \textit{T}$_{e}$ is the local electron temperature, and \textit{D} represents the distance to the host galaxy. Units for \textit{Y}\rm\ and \textit{Z}\rm\ vary from ($\unit$) for emission features, to unitless [O/H] for KK04 and PT05 metallicities, while band ratios and 10~\micron-normalized bands are also unitless. SFRs derived using Eq.~\ref{eq:SFR_power} are referred to as the power-law model. The corresponding constants for use in Eq.~\ref{eq:SFR_power} are listed in Table~\ref{tab:sfr_constants}. As noted in this table these same constants can be used to model T33 in Eq.~\ref{eq:triple} if 9.04 (i.e. the logarithm of the reciprocal solid angle of a 7$"$ aperture in steradians) is added to the constant $\gamma$.

Lacking constraints on the electron temperature, we assume a value of 10$^4$ K which is typical of H~II regions. Changing this assumption by a factor of two results in a difference of $\sim37\%$. Converting T33 from Eq.~\ref{eq:triple} to SFR in Eq.~\ref{eq:SFR_power} requires distance as additional parameter. Therefore, we expect statistics such as the correlation coefficient or the standard deviation of residuals will not be preserved from Table~\ref{tab:model_quality}. The Spearman correlation coefficient between SFR derived using our power-law relation and a comparison tracer is referred to as $\rho_{SFR}$ to differentiate. We also calculate the median value of the residuals as $\tilde{\Delta}$ to measure any remaining offset between the two SFR tracers. We also list the standard deviation of the residuals when we report $\tilde{\Delta}$.

\begin{figure*}
\includegraphics[width=2.3in]{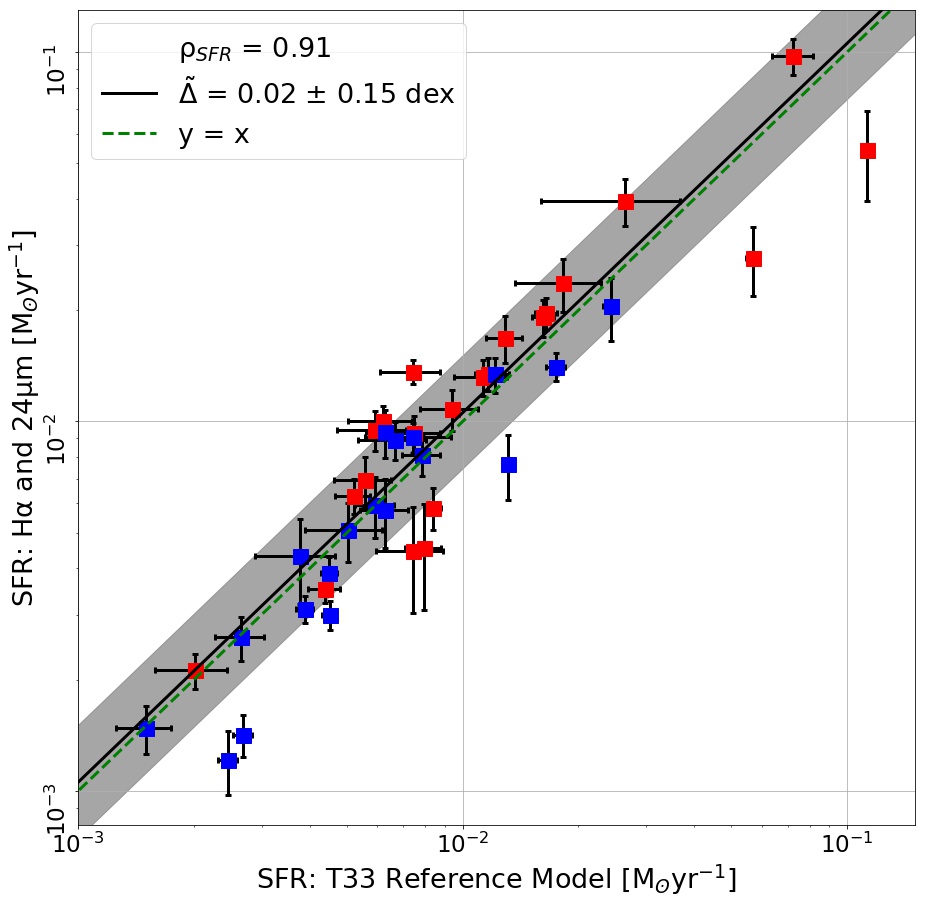}
\includegraphics[width=2.3in]{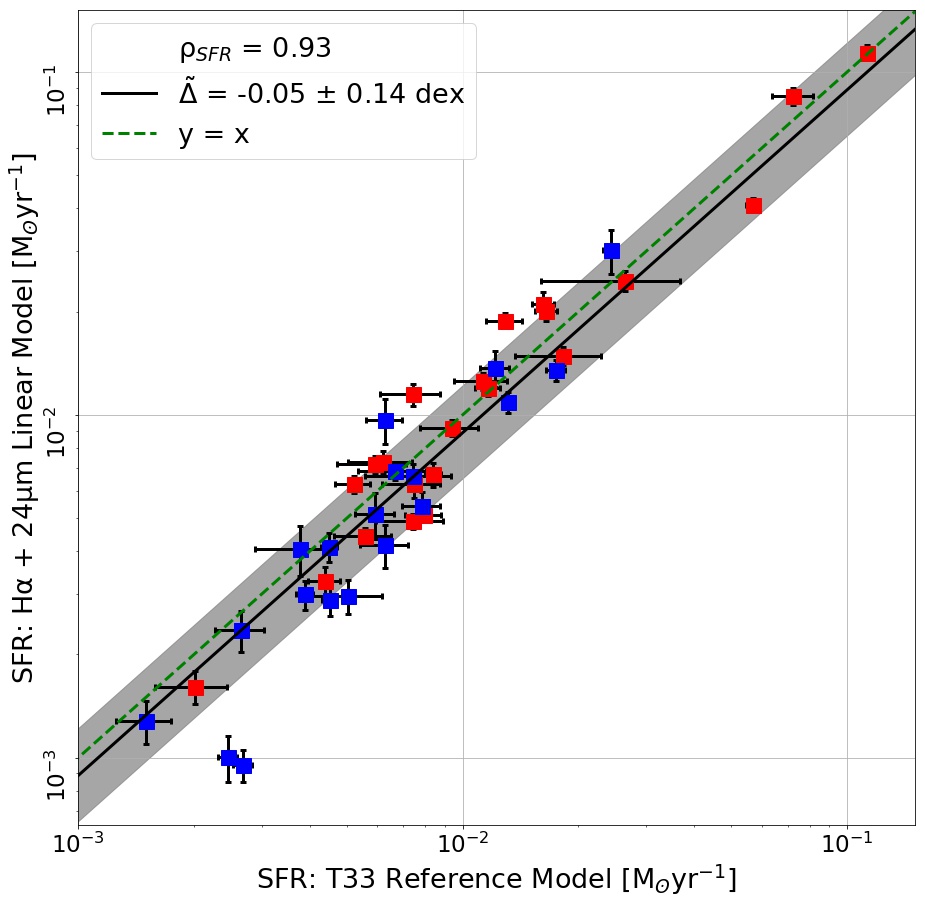}
\includegraphics[width=2.3in]{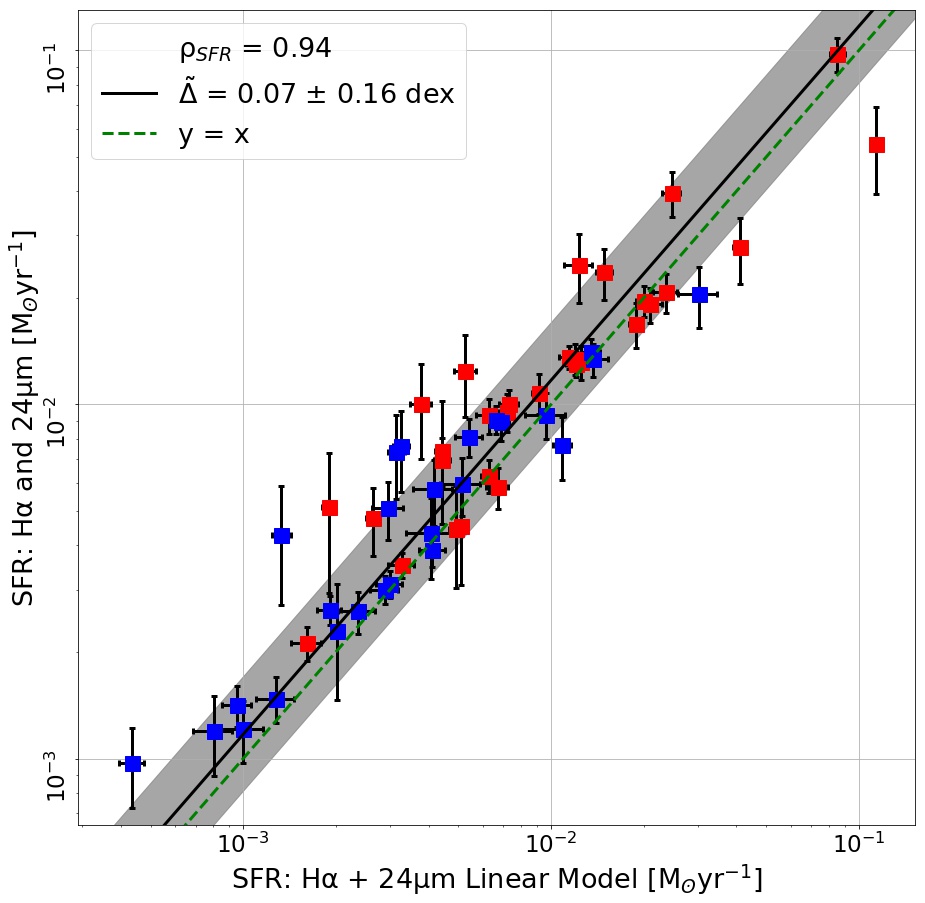}

\caption{(a) Star-formation rates calculated from \halpha~and \twenmic~emission using Eq.~\ref{eq:SFR_power} as a function of SFR derived from observed T33.
(b) SFR calculated using Eq.7 in \cite{murphy+11} as a function of T33 reference SFR.
(c) SFR calculated from \halpha\ and \twenmic~emission using Eq.~\ref{eq:SFR_power} as a function of SFR calculated using Eq.7 in \cite{murphy+11}. This is the only correlation that utilizes all 56 regions in our sample.}
\label{fig:sfr_Ha24_T33}
\end{figure*}

We compare our power-law model using \halpha\ and \twenmic\ emission in Eq.~\ref{eq:SFR_power} with Eq.~7 from \cite{murphy+11} which depends only on these same parameters. The constant in this relation is updated slightly from the primary source, \cite{calzetti}, but remains very similar. We refer to the relation from \cite{murphy+11} as the ``linear model'' due to its functional form. Neither model has an explicit metallicity, \nethree, or T33 dependence so the full dataset of 56 regions with \halpha\ and \twenmic\ emission is used in Figure~\ref{fig:sfr_Ha24_T33}c. Plots (a) and (b) of this figure show our power-law model predicts T33-derived SFR about as well as the commonly used linear model for \halpha\ and \twenmic\ emission. Figure~\ref{fig:sfr_Ha24_T33}c shows that our power-law model is also as accurate as the linear model for the 23 H~II regions outside the calibration subset of 33 regions.

\begin{figure*}
\centering
\includegraphics[width=2.3in]{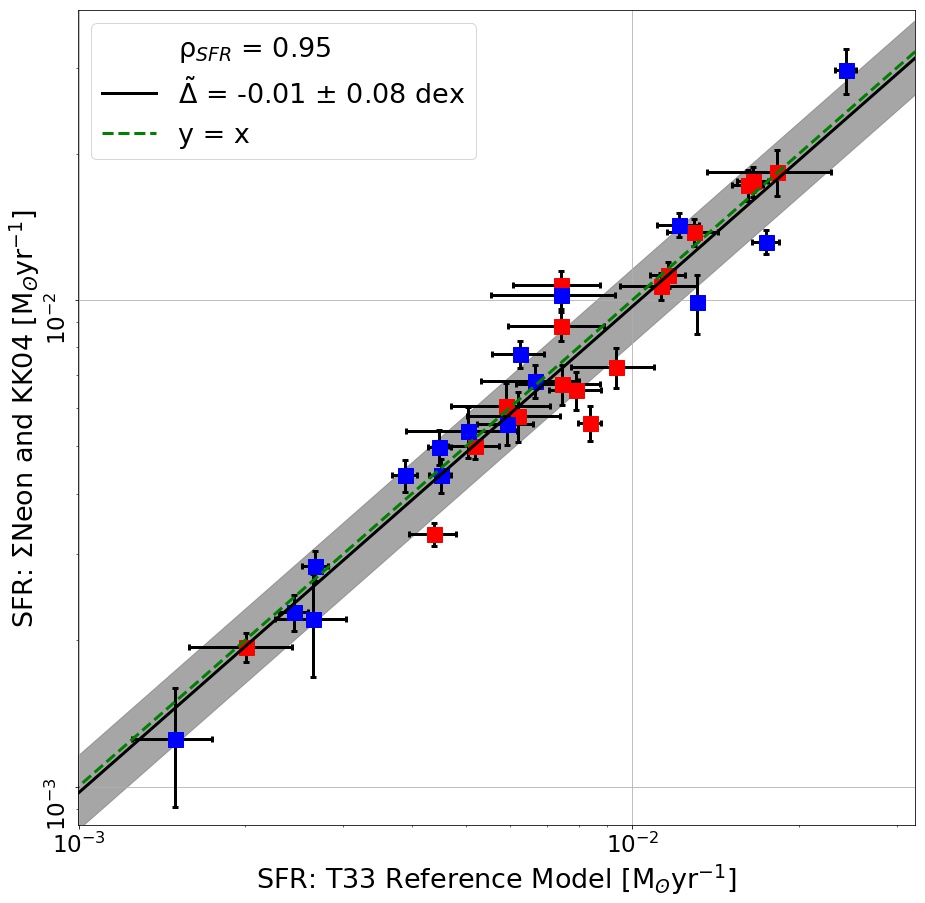}
\includegraphics[width=2.3in]{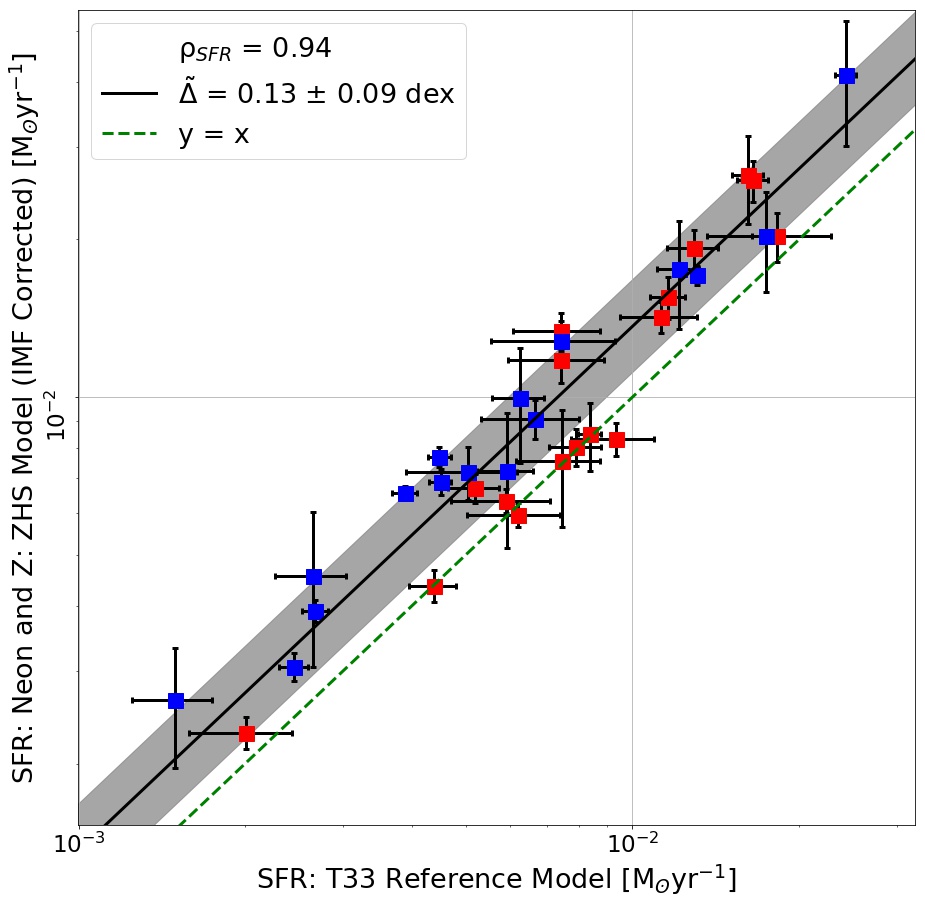}
\includegraphics[width=2.3in]{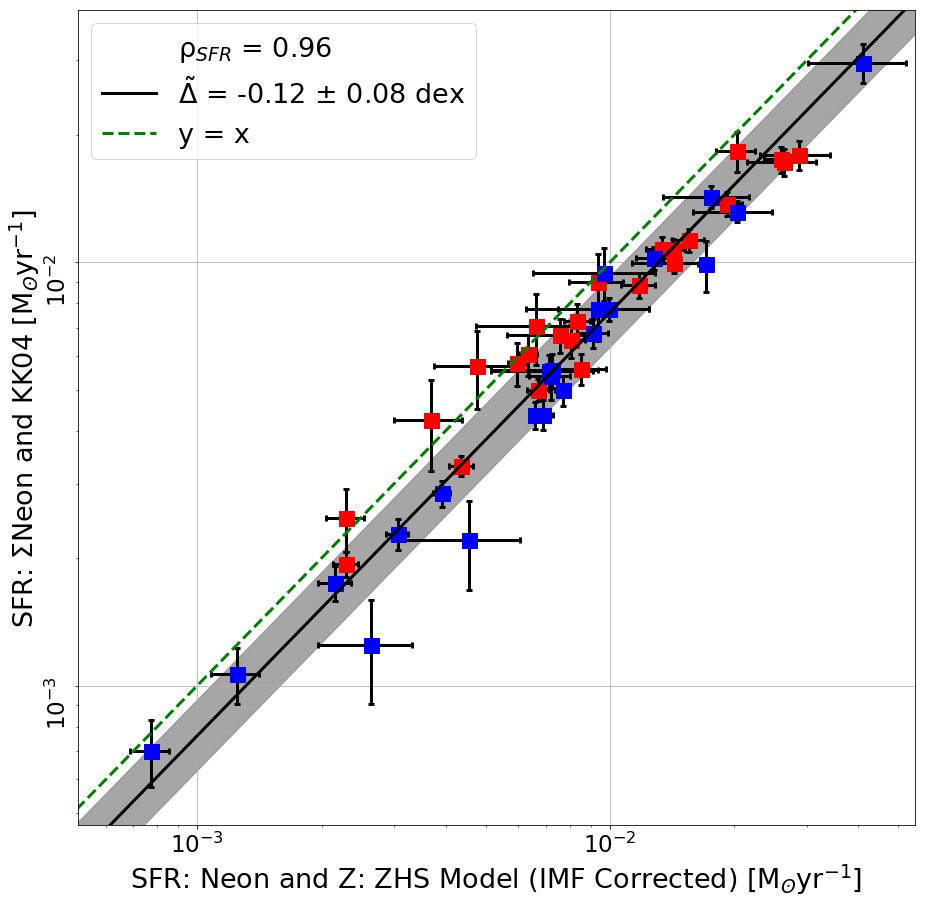}
\caption{(a) SFR calculated by Eq.~\ref{eq:SFR_power} using the sum of \netwo\ and \nethree\ emission paired with KK04 metallicity values as a function of T33-derived SFR.
(b) SFR calculated by Eq.6 from \cite{zho_ne19} using the same parameters as (a) with 0.18 dex subtracted to account for IMF assumption (see Section~\ref{sec:intro}), as a function of T33-derived SFR.
(c) SFR calculated by Eq.~\ref{eq:SFR_power} using \sumneon\ and KK04 metallicity as a function of using the model from \cite{zho_ne19}).
Regions are separated into high-metallicity(red) and low-metallicity(blue) groups according to the median value.}
\label{fig:sfr_NeZ_T33}
\end{figure*}

We also compare our \sumneon\ and metallicity power-law model to the model presented as Eq.~6 in \cite{zho_ne19} (hereafter ZHS model). Both models depend only on the metallicity and functions of the \netwo\ and \nethree\ bands.  As noted in Section~\ref{sec:corr_methods}, an offset is expected between the ZHS model based on the Salpeter IMF and the T33 model which is based on the Kroupa IMF. We corrected for this by including a factor of 1.5 decrease (-0.18 dex) to SFR from the ZHS model \citep{murphy+11}. We calculate SFR for the 46 regions with SH data and metallicity measurements when comparing our power-law model to the ZHS model in Figure~\ref{fig:sfr_NeZ_T33}c and the 33 regions in the calibration subset when comparing our power-law model or the ZHS model to the T33 reference model in Figure~\ref{fig:sfr_NeZ_T33}a and b. We set the solar metallicity in the KK04 calibration to be 12 + log$_{10}$[O/H]~=~8.8 for use in the ZHS model \citep[based on applying the KK04 calibration to the spectrum of Orion;][]{sandstrom13}.
We find SFR predicted by the ZHS model is well-correlated with T33-derived SFR, but with a significant average residual offset. Figure~\ref{fig:sfr_NeZ_T33}a shows the IMF-corrected SFR from the ZHS model are greater than SFR from T33 by about an average of 0.13 dex. This discrepancy will be further discussed in Section~\ref{sec:disc}. This observed offset is propagated into the comparison between our power-law model and the ZHS model in Figure~\ref{fig:sfr_NeZ_T33}c since our calibration is based on T33 emission, however the correlation is very strong as indicated by $\rho_{SFR}$. 

\begin{figure*}
\centering
\includegraphics[width=3.5in]{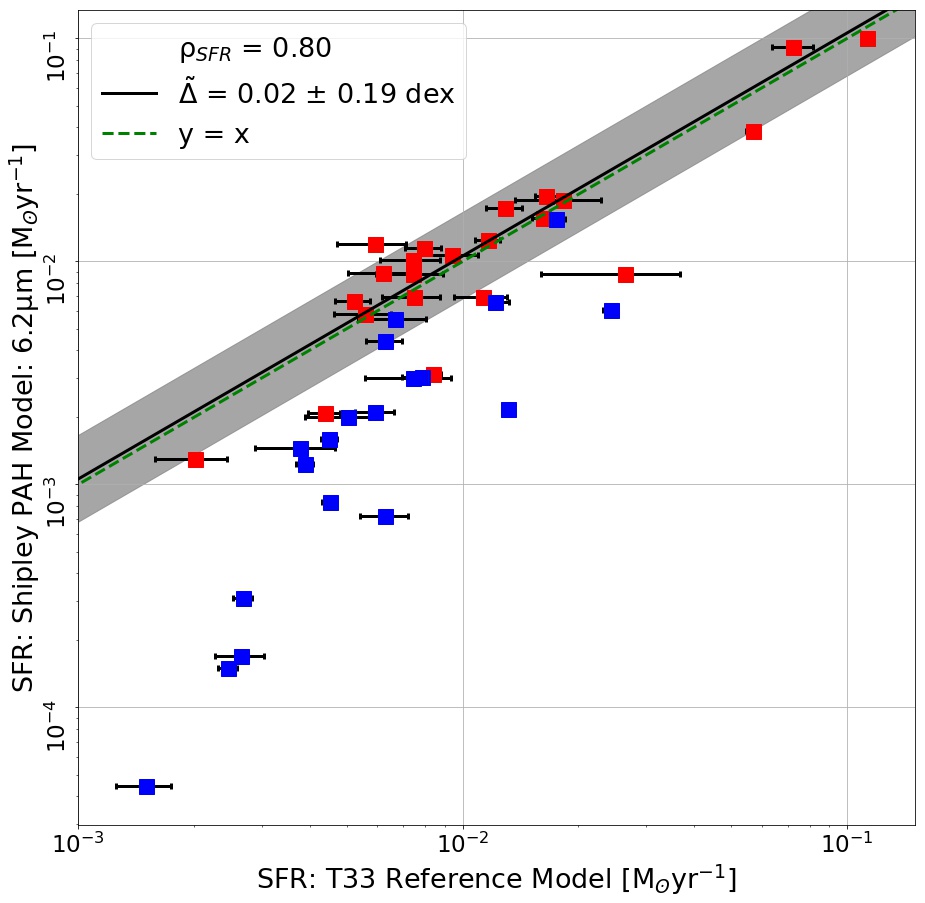}
\includegraphics[width=3.5in]{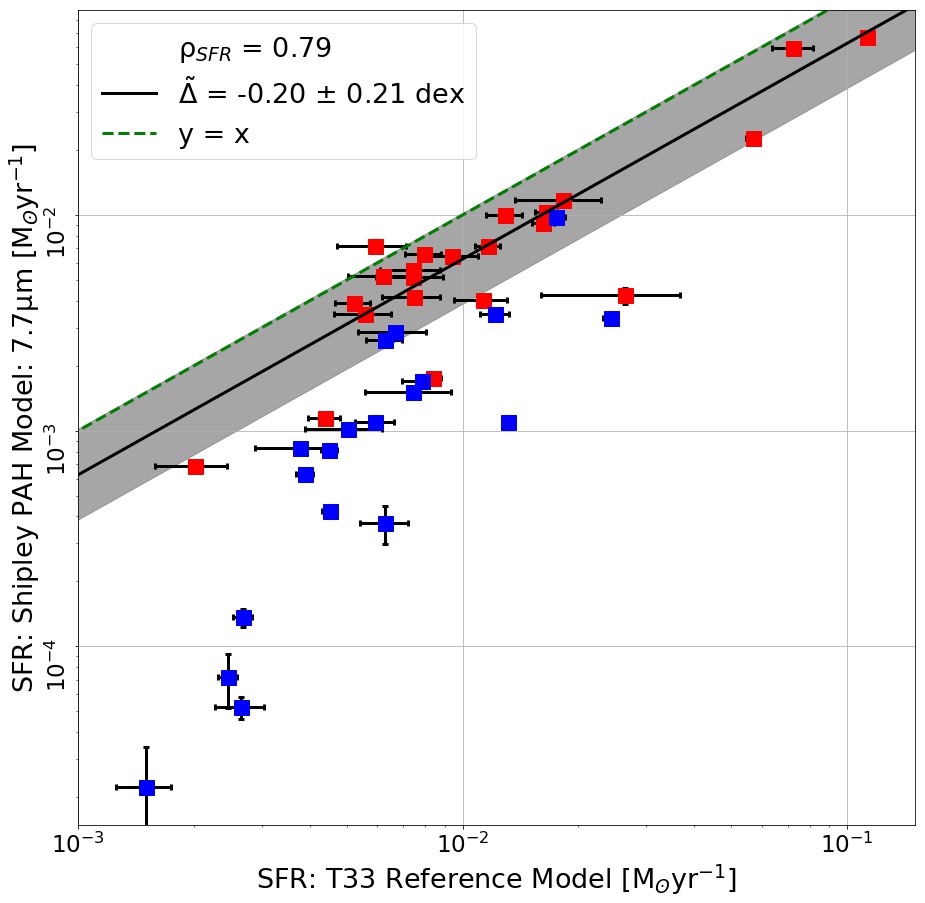}
\includegraphics[width=3.5in]{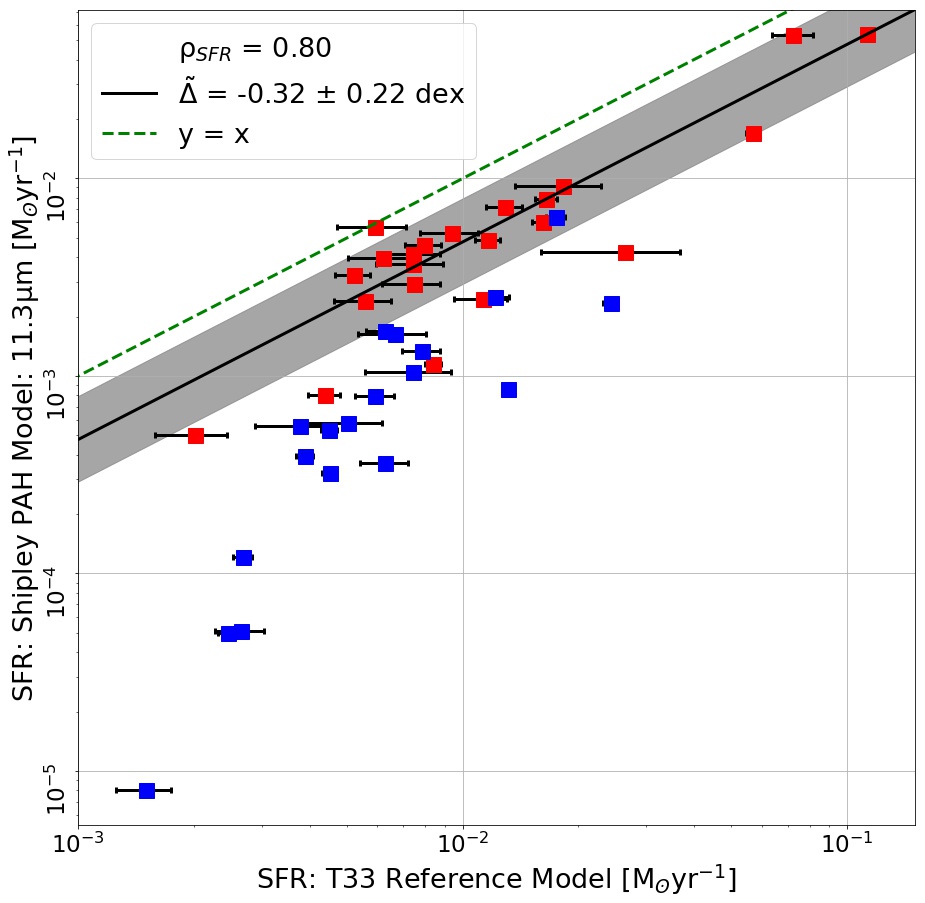}
\includegraphics[width=3.5in]{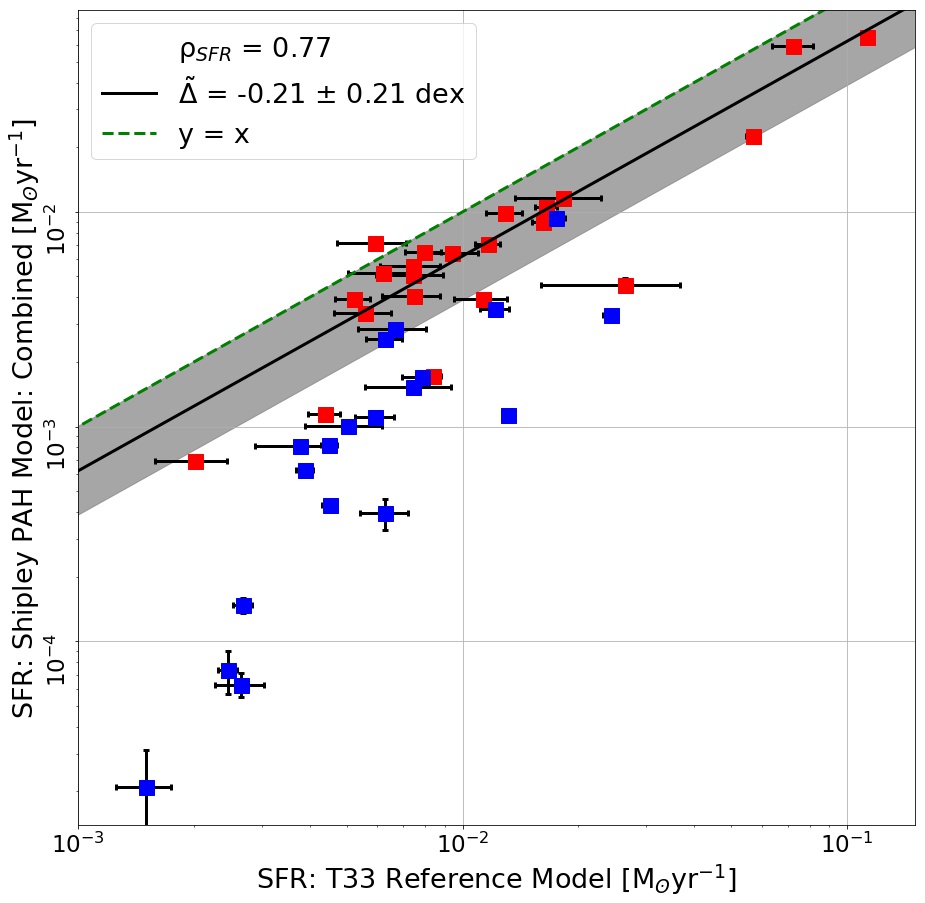}
\caption{SFR calculated by the Eqs. 8 and 12 from \cite{shipley} using the 43 7\arcsec\ regions in our sample with T33 measurements. Indicated statistics are calculated from high-metallicity points shown in red.
(a) PAH 6.2~\micron. 
(b) PAH 7.7~\micron.
(c) PAH 11.3~\micron.
(d) Combination of PAH 6.2~\micron, 7.7~\micron, and 11.3~\micron.}
\label{fig:sfr_ship}
\end{figure*}

Finally, we apply the PAH-based SFR tracers presented by \cite{shipley}. These tracers were calibrated for entire galaxies with a strict lower limit on metallicity; approximately equivalent to the median metallicity value used in this work. Therefore, we expect the tracers to be most accurate for the regions we have previously defined as high-metallicity. For fair comparison with the work of Shipley et al., we restrict our calculations of statistics such as the Spearman correlation coefficient $\rho_{Z}$ and the median of the residuals $\tilde{\Delta}$ to high-metallicity regions (12 + log[O/H]$_{KK04}~>~$8.98). The correlations with T33 are indeed much stronger among the high-metallicity points. We find each tracer is statistically equivalent in both $\rho_{SFR}$ and \tildel. Figure~\ref{fig:sfr_ship} shows a significant offset of about -0.2 dex in the PAH 7.7~\micron\ SFR as well as about -0.3 dex in the PAH 11.3~\micron\ SFR compared to that from T33. However, we find no significant offset between T33 SFR and the PAH 6.2~\micron\ SFR relation.

\subsection{PAH Band Ratios in H~II Regions}\label{sec:bandratios}
\begin{figure}
\includegraphics[width=3.3in]{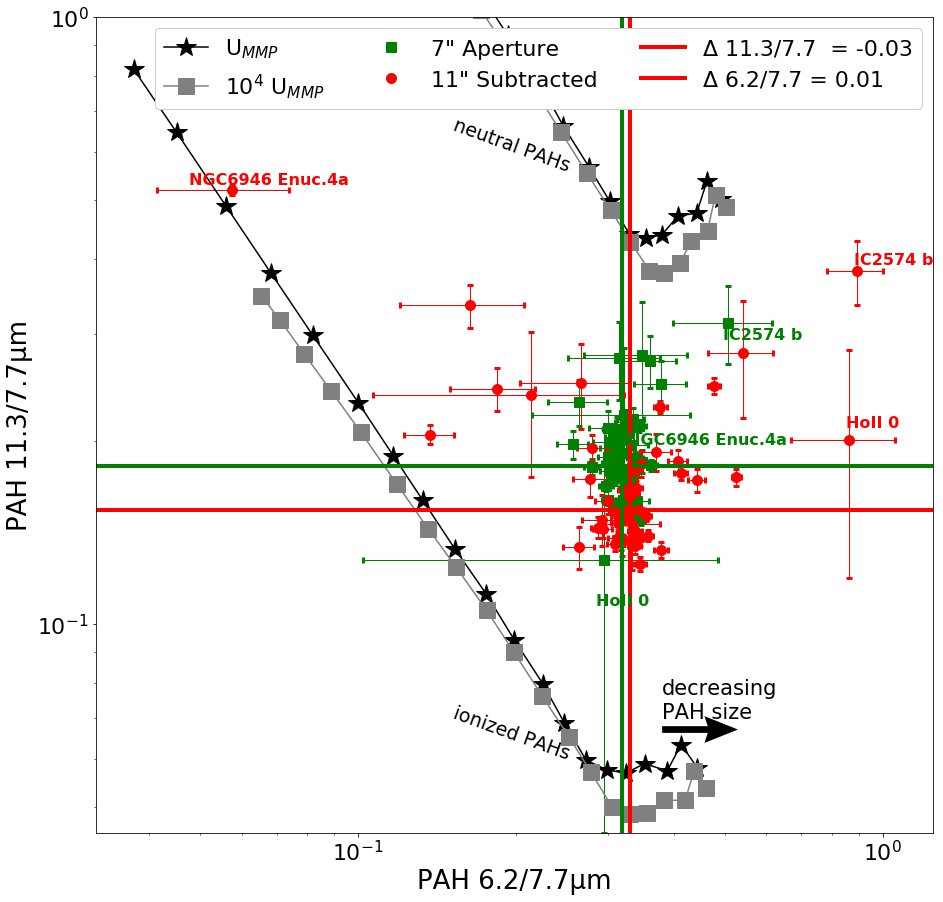}
\caption{Ratio of PAH features at 11.3~\micron\ and 7.7~\micron\ plotted against the ratio of PAH features at 6.2~\micron\ and 7.7~\micron\ for 49 extranuclear star-forming regions. Vertical and horizontal lines indicate the weighted mean and theoretical tracks obtained from \cite{DL01} depicted in gray squares and black stars represent expected PAH emission for interstellar radiation field strength U$_{\it{MMP}}$.}
\label{fig:DT}
\end{figure}

In Section~\ref{sec:spec_extract} we described our method for producing a set of background-subtracted star-forming region spectra. This dataset was used to compare changes in the PAH spectrum in star-forming regions after background subtraction; the ratio of PAH features at 11.3~\micron\ and 7.7~\micron\ is used as an indicator of the relative number of neutral to ionized PAH molecules, and the ratio of PAH features at 6.2~\micron\ and 7.7~\micron\ as an indicator of the relative number of small to large PAH molecules \citep[e.g.][]{DL01}.
We find a decrease in the weighted mean of the 11.3/7.7 ratio after the local background subtraction and a less significant increase in the weighted mean of the 6.2/7.7 ratio. Figure~\ref{fig:DT} shows the 6.2/7.7 PAH-size ratio increases by 3.6\% and the 11.3/7.7 PAH-ionization ratio decreases by 15.6\%. Also indicated in this figure are significant outliers such as the dwarf galaxy regions Holmberg II-0 and IC 2574-b and the known AME source designated NGC6946 Enuc. 4a \citep{murphy+10}. These outliers are found to be regions where the aperture is not centered on the peak of PAH emission or PAH emission is sufficiently extended such that subtracting the surrounding annulus removes a significant portion of the total aperture brightness. We find this form of annulus background subtraction results in no significant effect on T33 correlations. 

\section{Discussion}\label{sec:disc}

\subsection{Properties of Effective Mid-IR SF Tracers}

Our study has revealed a novel recipe for creating effective star-formation rate tracers by considering their dependence on metallicity. Common hybrid SFR tracers use UV/optical emission that originates from young stars and ionized gas, such as \halpha, and combine it with a correction factor for the light that was attenuated, which is assumed to be proportional to the \twenmic\ continuum \citep[see review by][]{KennEvans+12}. Mid-infrared emission is largely insensitive to dust attenuation and our study finds that the best SFR tracers based on these features instead pair an ionized gas tracer with a metallicity-dependent correction term. Our mid-IR tracers do not rely on two features to trace separate portions of emission from a star-forming region and instead employ a feature such as \sumneon\ to trace the emission from ionized gas and a second parameter which acts to further improve the correlation, typically by correcting for variations due to metallicity. 

In Section~\ref{sec:results_2_summ} we noted a pattern among SFR tracers where combining features with opposite metallicity offset \Zoff\ improves correlations with T33. For example, our \nethree-\twenmic\ tracer is constructed of \twenmic, which we found to have a positive metallicity offset, while [Ne~III] has a large negative metallicity offset. The direction of the [Ne~III] metallicity offset implies underestimated SFR in low-metallicity relative to high-metallicity regions for a given [Ne~III] surface brightness, and vice-versa for \twenmic\ emission. Pair tracers that combine features with these inverse metallicity dependencies in T33 correlations are found to best predict T33-based SFR.

We can also interpret the results for the often used \halpha-24~\micron\ calibration as following a similar trend of pairing strong SF tracers with opposite metallicity-dependence leading to the best correlation with SF. \halpha\ has a small negative metallicity offset, although the residuals in the two metallicity bins are consistent with being drawn from the same sample, and a moderate correlation with T33, while \twenmic\ has a strong positive metallicity offset and similar correlation with T33. The metallicity offset is expected for the dust attenuation-dependent emission from \twenmic, as found by \citet{relano07} in a similar study of star-forming regions in nearby galaxies. The weak metallicity dependence of \halpha\ with respect to T33 suggests attenuation geometry may dominate as the source of scatter between T33 and \halpha. Combining \halpha\ and \twenmic\ leads to an improvement in the correlation with T33 and a lower metallicity offset than the \twenmic\ has alone, reinforcing the idea that pairing strong SF tracers with opposite metallicity-dependence results in the best correlation with SF.

Interestingly, we found a SFR tracer using \nethree\ and \twenmic\ emission in Eq.~\ref{eq:SFR_power} outperforms a tracer using \halpha\ and \twenmic\ in all model quality parameters of Table~\ref{tab:model_quality}. [Ne~III] and \halpha\ each trace highly ionized gas. The negative metallicity dependence as defined in this work is found to be statistically significant for [Ne~III], but not for \halpha. This is likely because \halpha\ emission exclusively traces the unobscured portion of radiation from ionized hydrogen gas and thus depends less directly on metallicity than [Ne~III]. [Ne~III] emission has a complex dependence on both metallicity, density, and radiation field hardness, but a general trend of decreasing emission with respect to SFR as metallicity increases is expected and observed \citep[e.g.][]{giveon02,madden+06,zho_ne19}. Observations at 15.56~\micron\ are relatively unaffected by dust attenuation that hinders \halpha\ emission. These results imply that a [Ne~III] and \twenmic\ tracer outperforms the \halpha\ and \twenmic\ tracer because \halpha\ traces only the unobscured ionized gas while [Ne~III] emission has a dependence on all the ionized gas as well as the local metallicity. 

Generalizing this pattern we propose that a tracer is more effective if both component emission features have a strong correlation with star formation and the residual dependence on metallicity of one is opposite that of the other. The metallicity-dependence in the SFR correlation of each feature independently is mitigated after they are combined, strengthening the SFR dependence.

\subsection{SF Tracers From \sumneon}

In Section~\ref{sec:totne} we noted that combining \sumneon\ and KK04 metallicity in Eq.~\ref{eq:triple} results in a correlation coefficient consistent with the maximum allowed by the uncertainties of our T33 measurements. We also found that \sumneon\ alone better traces T33 than combining \halpha\ and \twenmic. The total emission from the mid-IR ionized neon lines depends on the total abundance of neon, so an ionized neon tracer could, in theory, have a complex dependence on metallicity and the radiation field hardness. These effects may cancel each other on average as high-metallicity regions will have more neon in total but less of it will be in an ionized state since the radiation field will be softer compared to lower metallicity regions. Similarly, the low-metallicity regions will have less neon in total but more of it will be in the Ne~II and Ne~III states due to the harder radiation field of stars at low metallicity. Because both are $\alpha$-elements \citep{woosley95}, the total neon abundance is expected and observed to correlate with oxygen abundance so regions of the same metallicity have similar neon abundance \citep[e.g.][]{bergIV}. These opposing effects may be mitigated on average in the \sumneon-T33 correlation, as seen in the significantly reduced metallicity offset compared with the correlation of either [Ne~II] or [Ne~III] individually. Residual scatter in the \sumneon-T33 correlation is found to be entirely dependent on metallicity since including metallicity as a second parameter in the fit results in an optimal correlation coefficient.

Another explanation for the success of the \sumneon\ calibration and its weak metallicity dependence may be related to the importance of the two mid-IR neon lines in H~II region cooling. Because Ne~II and Ne~III are relatively abundant and have low-lying fine structure levels that can be easily excited by collisions, the resulting \netwo\ and \nethree\ lines are important coolants in H~II region energy balance \citep[e.g.][]{burbidge63,osterbrock65}. In steady-state, the H~II region heating, which is dominated by photo-ionization of hydrogen, and the cooling, which is partially from the neon lines, will be equal. Therefore, \sumneon\ should trace the ionizing photon rate, modulo a potentially varying fraction of the cooling luminosity emerging in other emission lines.  The lack of substantial metallicity offset in \sumneon\ suggests that over the metallicity range of our study, the neon lines together carry a relatively constant fraction of the overall cooling luminosity of H~II regions.

In Section~\ref{sec:totne} we found that unlike local metallicity (12 + log$_{10}$[O/H]), the ratio of [Ne~III]/[Ne~II] emission does not improve \sumneon\ correlations with T33. However, we also found in Appendix~\ref{sec:Z_corr} that this ionized neon ratio has the strongest correlation with metallicity of all mid-IR ratios considered. The neon ratio illustrates an important point that a ratio of emission features can have a strong correlation with metallicity and not have a metallicity-dependent offset in T33 correlations. We find any other correlation that is improved by pairing with metallicity is similarly improved by pairing with the [Ne~III]/[Ne~II] ratio. This indicates that the \sumneon\ correlation with T33 is a special case that cannot be improved with the [Ne~III]/[Ne~II] ratio. It is possible that combining the \netwo\ and \nethree\ surface brightnesses into \sumneon\ includes all the relevant metallicity information about the relationship between [Ne~II] and [Ne~III] with respect to SFR. We find using metallicity itself or any other ratio that is well-correlated with metallicity (such as a normalized PAH band) improves the \sumneon\ correlation with T33 to approximately the maximum allowed.

\subsection{Metallicity Dependence of PAH-Based SF Tracers}

Individual PAH band correlations with T33 are found to all be similar within their measured uncertainties and weak: with Spearman coefficients of \rhos~$\sim$~0.3 (within about 1$\sigma$ of our calculated critical coefficient for the null hypothesis). We found this weak relation between PAH emission and T33 is due to a strong metallicity dependence. Combining emission from multiple PAH bands also does not improve the correlation, suggesting that band ratios do not vary substantially with SFR or metallicity. At a given T33 or SFR, we observed high-metallicity regions are approximately twice as bright in the 7.7~\micron\ PAH band as low-metallicity regions. As noted in Section~\ref{sec:corr_methods}, KK04 metallicity itself is found to have \rhos~=~0.3 in correlations with T33 due to the sparse sampling of low metallicites in our dataset. This fact combined with the strong metallicity-dependence seen in PAH emission implies the significance of PAH correlations with T33 may even be slightly exaggerated by our sample choice. The scatter in PAH-T33 correlations can, however, be improved significantly, to \rhos~$\sim~0.6$ by using a metallicity tracer as a second parameter. Therefore, we conclude the PAH correlations with T33-SFR are very weak due to strong metallicity-dependence but not uncorrelated.

In choosing the best second feature to include in a PAH-based tracer, we found pairing any PAH band with the [Ne~III]/[Ne~II] ratio improves T33 correlations and metallicity offsets. These improvements are greater than those obtained from pairing PAH bands with KK04 metallicity itself. The population of PAH molecules and their emission have a dependence on both metallicity and radiation field hardness \citep[e.g][]{madden+06,lebouteiller07,gordon08}, since PAHs can be destroyed by energetic photons. These dependencies are also expected in the ionized neon ratio [Ne~III]/[Ne~II] \citep{zho_ne19}. If the metallicity dependence of PAH emission is due in part to their destruction in harder radiation fields, the use of [Ne~III]/[Ne~II] as a second parameter in the T33 correlation may capture that effect better than metallicity alone.

\subsection{Comparison to Existing SFR Calibrations}

Our investigation provides an interesting comparison with existing SFR calibrations for several reasons. We calibrate our relationship using resolved measurements for individual star-forming regions, as was done in \citet{calzetti} and \citet{kennicutt09}, for instance. Our calibration is tied to thermal 33 GHz which has a relatively direct relationship with the ionizing flux from massive stars, subject to few of the systematic effects tied to UV/optical SF tracers \citep[e.g.][]{murphy+11}, infrared tracers \citep[e.g.][]{leroy12}, or SED-fitting tracers \citep[e.g.][]{leja}. In addition, our resolved measurements of SF regions minimize issues caused by diffuse emission unrelated to star formation (i.e. cirrus or diffuse ionized gas) that may contaminate integrated galaxy SF calibrations. 

One difference between our investigation and some previous studies is the assumption of a power-law form for the calibration, defined by Eq.~\ref{eq:SFR_power}. This assumption does not appear to affect the accuracy of our SF predictions, however. Section~\ref{sec:SFR_plots} shows our results are consistent with previous calibrations such as the linear hybrid tracers with \halpha\ and \twenmic\ \citep{murphy+11} and more complex SFR calibrations such as the \netwo, \nethree, and metallicity model from \cite{zho_ne19}.

In Section~\ref{sec:T33_corr_1} we found that combining the [Ne~II] with the [Ne~III] surface brightness linearly into a single feature \sumneon\ results in an equivalent correlation with T33 compared to including [Ne~II] and [Ne~III] in Eq.~\ref{eq:triple} as two separate parameters. This supports our assumption that all parameters can be treated similarly using a power law. As noted previously in Section~\ref{sec:intro}, the sum of [Ne~II] and [Ne~III] will trace the total emission from ionized neon but our study cannot definitively conclude if this is better modelled as a linear combination or a combination of two power laws.

We compared our \sumneon SFR calibration with the \cite{zho_ne19} SFR calibration based \netwo, \nethree, and metallicity and found the ZHS model overestimates SFR by about twofold (0.31 dex) before correcting for IMF assumption. In Section~\ref{sec:corr_methods} we noted the difference in IMF assumption between the ZHS model and the other SFR calibrations considered in this study is known to account for about 0.18 dex of this offset \citep{murphy+11}. The remaining 0.13 dex offset may result from the assumed value for solar metallicity as the SFR from the ZHS model has a direct dependence on this value. Reducing our assumed KK04 solar metallicity by this 0.13 dex (from 12 + log$_{10}$[O/H]~=~8.8 to 8.67) removes the offset. The H~II regions considered in this work have KK04 values ranging from about 8.1 to 9.2, but regions from normal spiral galaxies like the Milky Way have metallicities between 8.7 and 9.2 in this calibration as shown in Figures \ref{fig:Ne3_ratios} and \ref{fig:PAH_ratios}. This indicates that the zero-point of the metallicity scale could be the cause of the offset in SFR from the ZHS model. Recent studies have also found that calibrations based on spectral energy distribution (SED) fitting of entire galaxies can overestimate SFR by a similar factor to that observed in Figure~\ref{fig:sfr_NeZ_T33}b \citep{leja}. We expect T33 to provide an ideal SFR that is independent of the assumptions in SED modelling, so the offset between \citet{zho_ne19} and our calibration may be evidence of this overestimate from SED-based SFR tracers, although after correcting for the IMF, we do not see this level of disagreement. We note that for distant, unresolved galaxies, it will be necessary to take into account neon emission lines powered by AGN rather than SF in some cases, as investigated by \citet{zho_ne19}.

Our comparisons between T33-based SFR tracers and others rely on an assumed constant H~II region electron temperature of 10$^4$~K. Changing this assumption by a factor of two results in a difference of $\sim37\%$ (0.14 dex). The temperature of H~II regions is metallicity dependent, so an additional correction term may be necessary to reduce the variations from this dependence, especially for observations at high redshifts.

The high-metallicity PAH band SFR tracer of \citet{shipley} was found to correlate well with SFR given by T33 measurements despite the weak correlations between PAH bands and T33 seen in Section~\ref{sec:results}. This demonstrates a key difference between calibrations for SFR from luminosities and calibrations for \sfrsig\ from surface brightnesses. We found the PAH 11.3~\micron\ and 7.7~\micron\ tracers underestimate SFR by about 60$\%$ (0.2 dex). A possible cause for this is the assumed limit that defines the high-metallicity range. Our high-metallicity group of regions is defined based on the median value of our dataset while that of \citet{shipley} is a fixed value based on a different metallicity calibration. It is also likely that differences in the PAH band ratios have a role in the offset. In Section~\ref{sec:bandratios} we found that the ratio of 11.3~\micron\ and 7.7~\micron\ PAH bands varies within star-forming regions compared to the immediate surrounding area. This implies a calibration such as \citet{shipley} that is based on emission from a galaxy will then represent an average of such variations between PAH bands. Both this difference in the relative strength of PAH band emission and our assumed high-metallicity definition may contribute to the observed offset between the \citet{shipley} and T33 SFR models.

\subsection{Metallicity Dependence of PAH Properties}

A number of previous studies have used photometric ratios of PAH-dominated 8~\micron\ emission relative to continuum at 24~\micron\ to trace the abundance of PAHs \citep{engel05,draine07} and to explore its metallicity dependence in galaxies. In our work, we can take a step further by measuring the integrated strength of individual PAH features normalized by neighboring continuum. Our results support the conclusion of these works that PAH abundance varies with metallicity. In addition, we found that normalizing PAH emission with 10~\micron\ continuum behaves similarly to 24~\micron\ as a function of metallicity. We have shown that all normalized PAH band correlations with metallicity have statistically equivalent Spearman coefficients $\rho_{Z}~\approx$~0.5.

As an additional investigation we studied ratios of ionized-to-neutral PAHs after subtracting the local average in an annulus of outer radius 5.5$\arcsec$ surrounding the 3.5$\arcsec$ region. We found no significant change in the ratio of 6.2~\micron\ to 7.7~\micron\ emission which is an indicator of the relative size of PAH molecules in some models \citep[e.g.][]{draine01}. This suggests that in terms of size, PAH molecules are distributed similarly in the surrounding diffuse area as they are in the star-forming region. The background subtraction decreases the 11.3~\micron\ to 7.7~\micron\ PAH ratio, indicating an increase in the ionization of PAH molecules closer to the star-forming region. Such an increase could be explained by the harder and more intense radiation fields near the star-forming region compared to the diffuse interstellar medium \citep{bakes01}.

\section{Conclusions}\label{sec:concl}
We have used resolved observations of thermal 33 GHz radio continuum, mid-IR emission from ionized neon and PAHs, and local metallicity to calibrate star formation rate tracers in nearby star-forming galaxies.
We summarize our main conclusions in the following:
\begin{itemize}
    \item All individual infrared emission features considered show significant metallicity-dependent offsets in correlations with T33. Known hybrid tracers of SFR such as \halpha\ and \twenmic\ are also found to have metallicity offsets with respect to T33.
    
    \item \sumneon\ alone is found to better correlate with T33 than the commonly used hybrid SFR tracer of \halpha\ and \twenmic\ emission. Residual scatter in the \sumneon-T33 correlation was found to be completely dependent on metallicity.
    
    \item A mid-infrared star-formation rate tracer with accuracy equivalent to thermal 33 GHz emission can be constructed using \sumneon\ and any PAH feature normalized by dust continuum. This accuracy is the same as using metallicity directly as the second parameter.
    
    \item All PAH correlations with T33 are statistically similar and very weak. Including a correction for metallicity improves the correlation but significant scatter remains.
    
    \item Continuum-normalized PAH emission is highly metallicity dependent and follows similar trends as 8/24~\micron\ ratios, which are suggested to trace PAH abundance.

    \item We found combining tracers of opposite metallicity dependence in T33 correlations leads to a significant improvement in correlation strength. This trend is evident in combinations such as [Ne~III]-24~\micron, \halpha-\twenmic, and [Ne~III]-PAH/10~\micron. We proposed this pairing of opposite metallicity-dependence features shows a novel pattern for constructing star-formation rate tracers wherein an indicator of ionized gas is corrected for variations due to metallicity by the additional indicator.
    
    \item We found subtracting local background emission from an H~II region decreases the 11.3~\micron\ to 7.7~\micron\ PAH ratio. This implies a decrease in the relative abundance of neutral PAH molecules closer to the star-forming region.
\end{itemize}

Our study is unique in using resolved thermal 33 GHz emission and local metallicity measurements for individual star-forming regions to calibrate SFR tracers. The mid-infrared SFR tracers presented in this study have wide applicability in analyzing future observations with the James Webb Space Telescope. The Mid-Infrared Instrument (MIRI) will detect \netwo\ out to redshift z~$\approx$~1.2 and PAH 7.7~\micron\ as far as z~$\approx$~2.5. In the 2030s, proposed telescopes such as the Space Infrared Telescope for Cosmology and Astrophysics (SPICA) and the Origins Space Telescope will have the potential to detect these features out to the epoch of reionization, enabling investigations throughout the full history of star formation.

\acknowledgements

We thank the anonymous referee and the statistical referee for useful comments that improved the paper. The authors thank Daniela Calzetti, J. D. T. Smith, and Adam Leroy for useful discussions.  C.W.\ and K.S.\ also thank Jeremy Chastenet, I-Da Chiang, and Petia Yanchulova Merica-Jones for their input over the years. We thank the SINGS team for their efforts in delivering processed data products for the community. 

C. W.\ and K. S.\ acknowledge funding support from NASA ADAP grants NNX16AF48G and
NNX17AF39G and National Science Foundation grant
No.\ 1615728.

This research has made use of the NASA/IPAC Extragalactic Database (NED) and NASA/IPAC Infrared Science Archive (IRSA) which are operated by the Jet Propulsion Laboratory, California Institute of Technology, under contract with the National Aeronautics and Space Administration. This work uses observations made with the {\em Spitzer Space
Telescope}, which is operated by the Jet Propulsion Laboratory,
California Institute of Technology, under a contract with NASA. The National Radio Astronomy Observatory is a facility of the National Science Foundation operated under cooperative agreement by Associated Universities, Inc. This research has made use of NASA's Astrophysics Data System.

\appendix

\section{Optimal Correlations Under Various Observational Restrictions} \label{sec:SFR_analysis}

We investigated which fits with Eq.~\ref{eq:triple} provided the most accurate model of observed T33 and subsequently the most accurate SFR under several data restrictions. This is useful because observations at higher redshift, for example, may exclude emission at longer wavelengths such as the \nethree\ or \netwo\ lines.

\begin{figure*}
\includegraphics[width=3.5in]{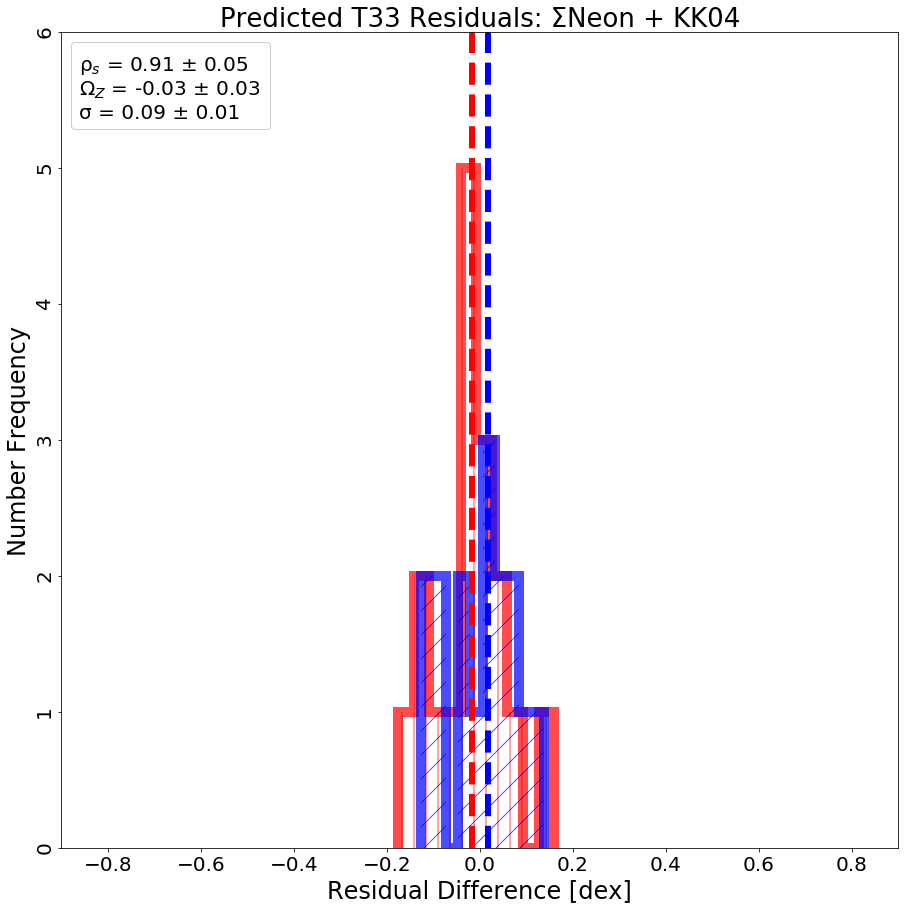}
\includegraphics[width=3.5in]{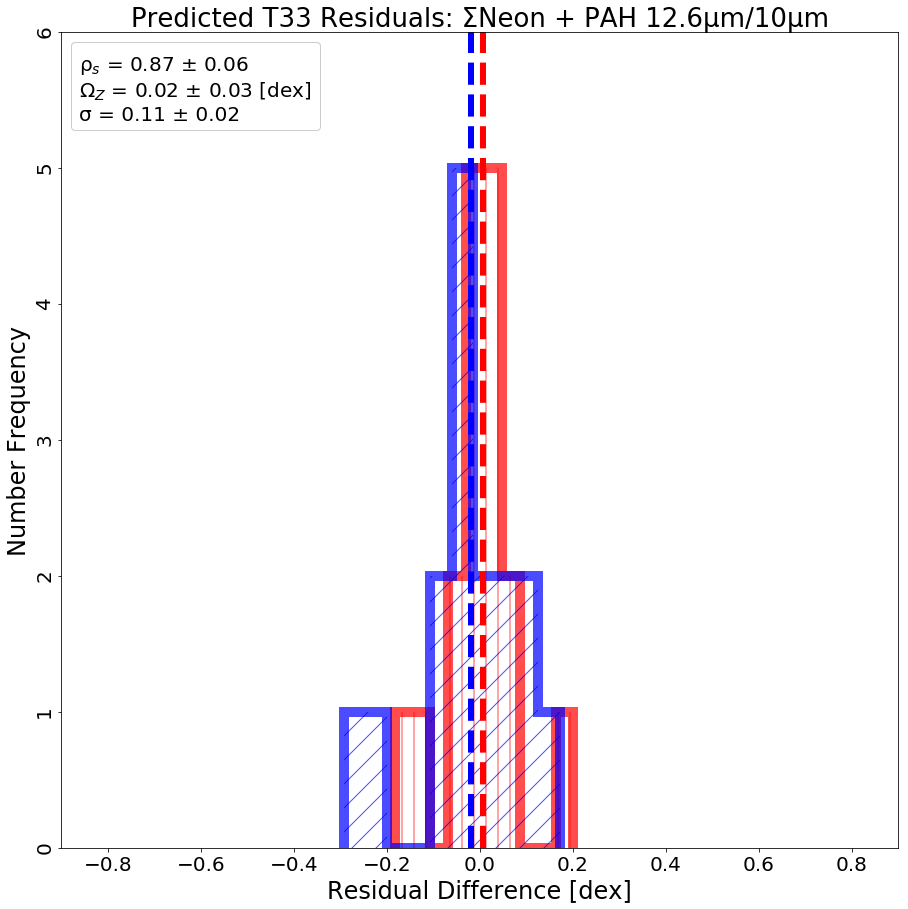}

\caption{Residual difference in predicted T33 using Eq.~\ref{eq:triple} based on \sumneon\ surface brightness and (a) KK04 metallicity values, (b) 10~\micron\ continuum-normalized PAH 12.6~\micron\ emission (12.6/10). All other normalized PAH features improve \rhos, \Zoff, and \tildel\ comparably to 12.6/10. }
\label{fig:totNe_pairs}
\end{figure*}

We find the best SFR tracer that can be made exclusively from emission features between 5.2 and 15.6~\micron\ uses \it{X}\rm~=~\sumneon\ and \it{Y}\rm~=~PAH 12.6/10~\micron. A histogram of the residual values for modelled T33 using this tracer is shown in Figure~\ref{fig:totNe_pairs}b. Other continuum-normalized PAH bands result in statistically similar correlations (\rhos\ within 1$\sigma$), so the 12.6~\micron\ feature is chosen for observational convenience. This feature is nearest to the [Ne~II] line at 12.81~\micron\ which must be measured to determine \sumneon\ and allows the range of necessary spectral observations to be reduced to about 6~\micron-wide from 10 to 16~\micron\ (including continuum to measure \nethree). Star-formation rates can be derived from this 6~\micron-wide region of the mid-infrared spectrum with negligible residual metallicity dependence. The Mid-Infrared Instrument (MIRI) on the upcoming James Webb Space Telescope will detect this range of mid-IR spectrum in galaxies to redshift z $\leq$ 0.8. 

Observations at higher redshifts may exclude the \nethree\ line so we investigated the best tracer that can be made with emission features between 5.2 and 13.6~\micron\ (z $\leq$ 1.12 for MIRI). We found in Section~\ref{sec:T33_corr_1} that the best-correlated individual observable in this range is the \netwo\ line. We find the best tracer in this range is made by combining [Ne~II] as the basis and including the nearby 12.6~\micron\ PAH feature and its continuum-normalized form PAH 12.6/10~\micron\ continuum, as shown in Table~\ref{tab:model_quality}. This three parameter tracer based on [Ne~II], PAH 12.6~\micron, and 10~\micron\ continuum has \rhos\ equivalent to the commonly used \halpha\ and \twenmic\ combination, about 0.78, and has a statistically insignificant \Zoff, as shown in Figure~\ref{fig:offsetspear_2}.

\begin{figure*}
\includegraphics[width=2.3in]{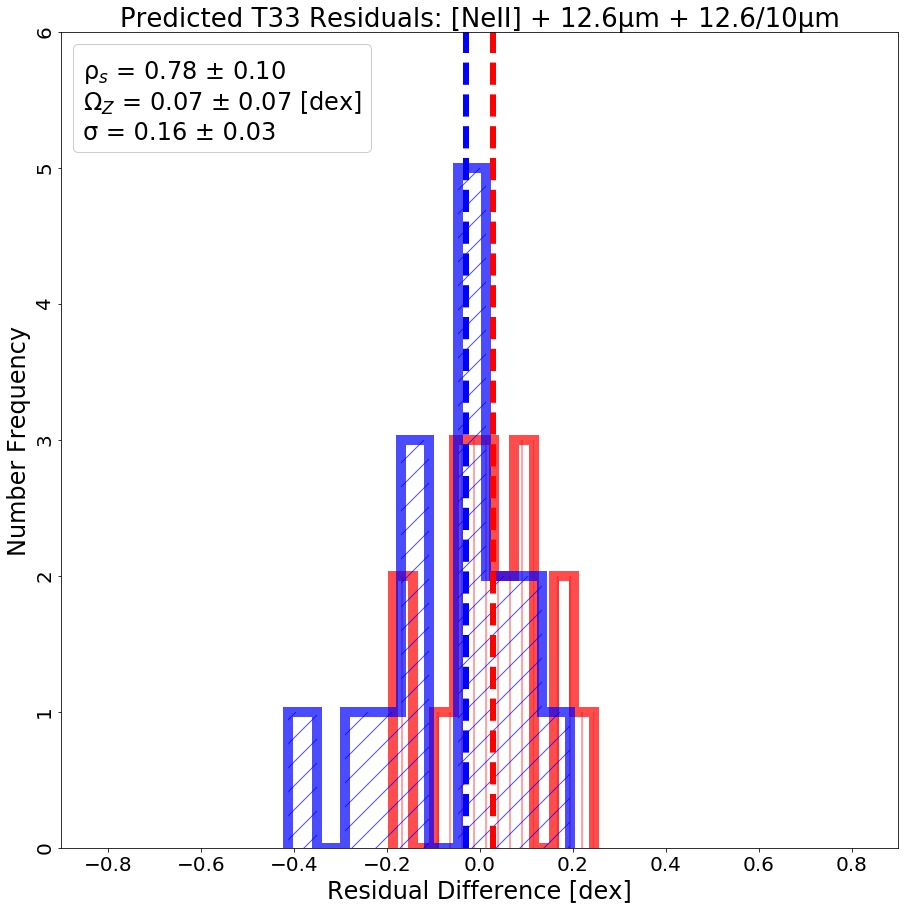}
\includegraphics[width=2.3in]{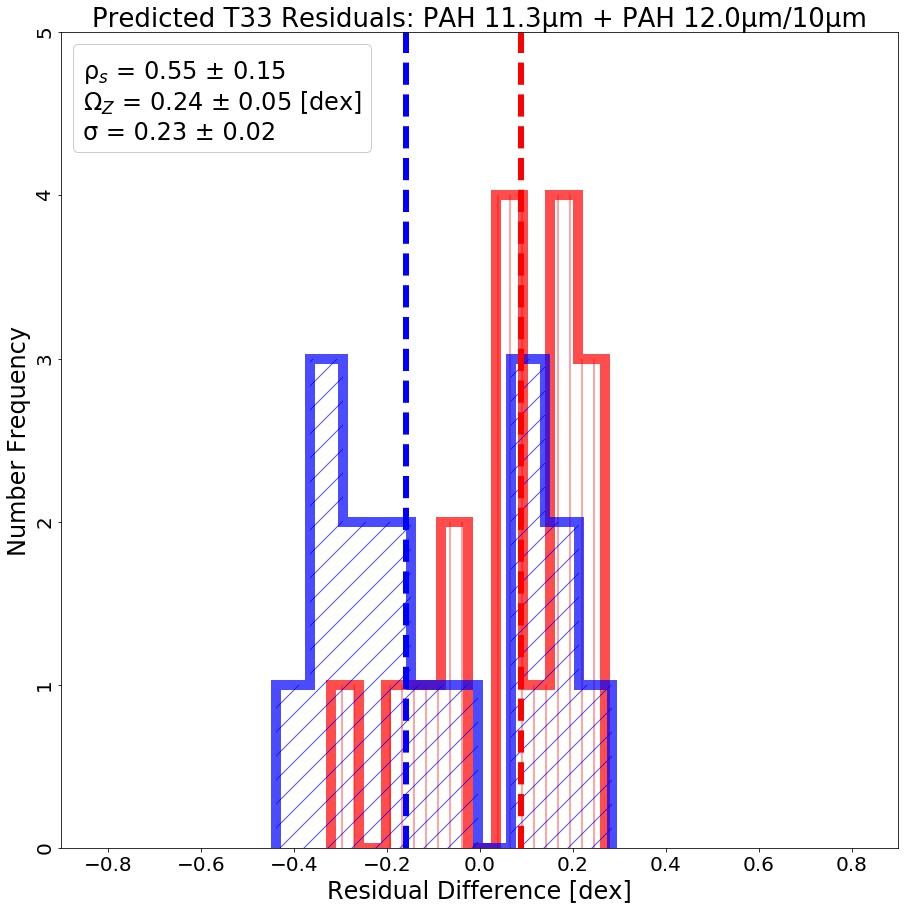}
\includegraphics[width=2.3in]{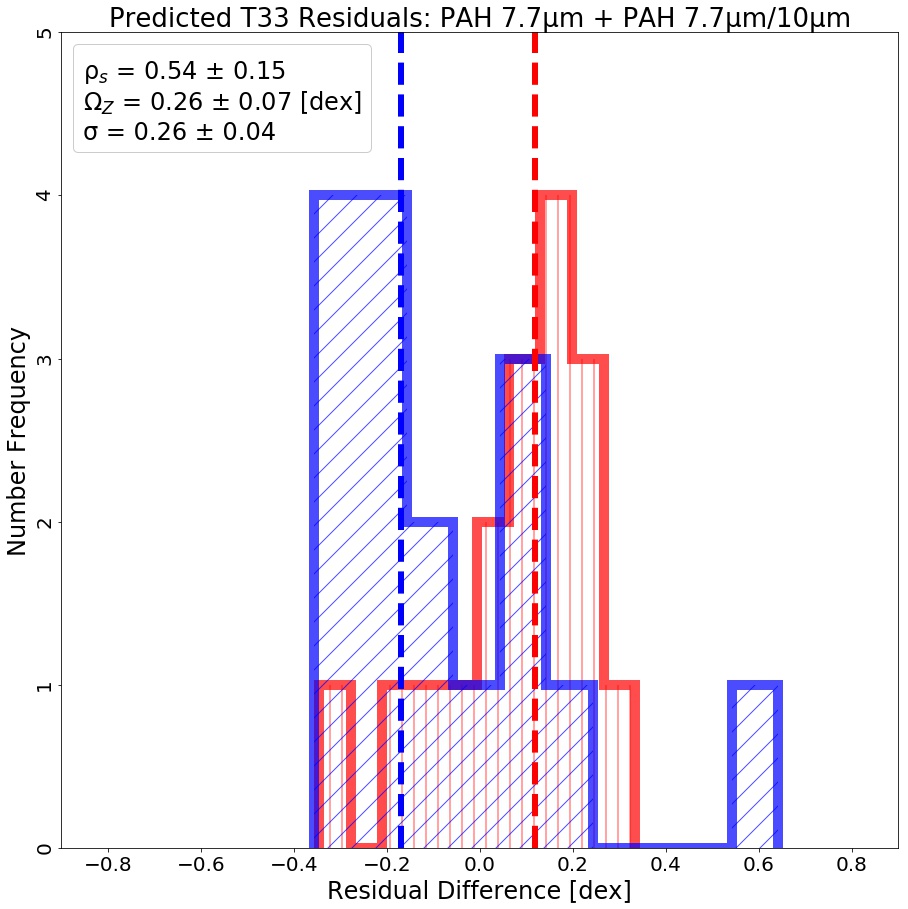}
\caption{Residual difference in predicted T33 using Eq.~\ref{eq:triple} based on: (a) \netwo, PAH 12.6~\micron, and 12.6/10~\micron\ continuum,
(b) PAH 11.3~\micron\ and PAH 12.0/10~\micron\ continuum, and
(c) PAH 7.7~\micron\ and PAH 7.7/10~\micron\ continuum.}
\label{fig:obs_hist}
\end{figure*}

We continued restricting our possible observables by excluding \netwo\ to investigate the best tracer that can be made with emission features between 5.2 and 12.6~\micron. In this range we are left with only PAH bands, of which there is no single feature that is better correlated with T33 than the others. For this reason we search for the best improvements to all PAH band correlations but with a focus on the strongest features such as 7.7 and 11.3~\micron. We find the best multi-feature correlations in this range involve a bright PAH feature paired with a 10~\micron\ continuum normalized PAH band. However, the maximum value of \rhos\ that can be achieved with these features is about 0.55 and the large \Zoff\ of the individual PAH band tracers remains unaffected as shown in Figures \ref{fig:obs_hist}b and c. This is seen when comparing the individual tracer based on PAH 7.7~\micron\ shown in Figure~\ref{fig:PAH77}b with the PAH 7.7~\micron\ and 7.7/10~\micron\ pair tracer shown in Figure~\ref{fig:obs_hist}c. 

In Section~\ref{sec:T33_corr_1} we noted a trend where high-metallicity regions are significantly better correlated with T33 when considered separately from low-metallicity regions. We find that this trend is preserved when a normalized PAH feature is included as seen in Figure~\ref{fig:obs_hist}c. This figure shows the metallicity offset is similar to 7.7~\micron\ alone (Figure~\ref{fig:PAH77}b), but the overlap between high- and low-metallicity points is notably decreased. We find that the correlation coefficient with T33 for high-metallicity points separately from low reaches about 0.6, and likewise for low-metallicity regions separately from high.

\section{Correlations of Mid-IR Emission Feature Ratios with Metallicity}\label{sec:Z_corr}

We investigated the potential of a mid-IR metallicity tracer with a similar procedure to that of Section~\ref{sec:corr_methods}. We modify the general form of Eq.~\ref{eq:triple} to solve for KK04 metallicities in terms of unitless [O/H] in place of T33 in units of mJy sr$^{-1}$. We also restrict our possible parameters to unitless quantities such as band ratios and continuum-normalized bands. These correlations use the set of 46 regions with high-confidence metallicity measurements as described in Section~\ref{sec:Z_data}.

\begin{figure*}
\includegraphics[width=3.5in]{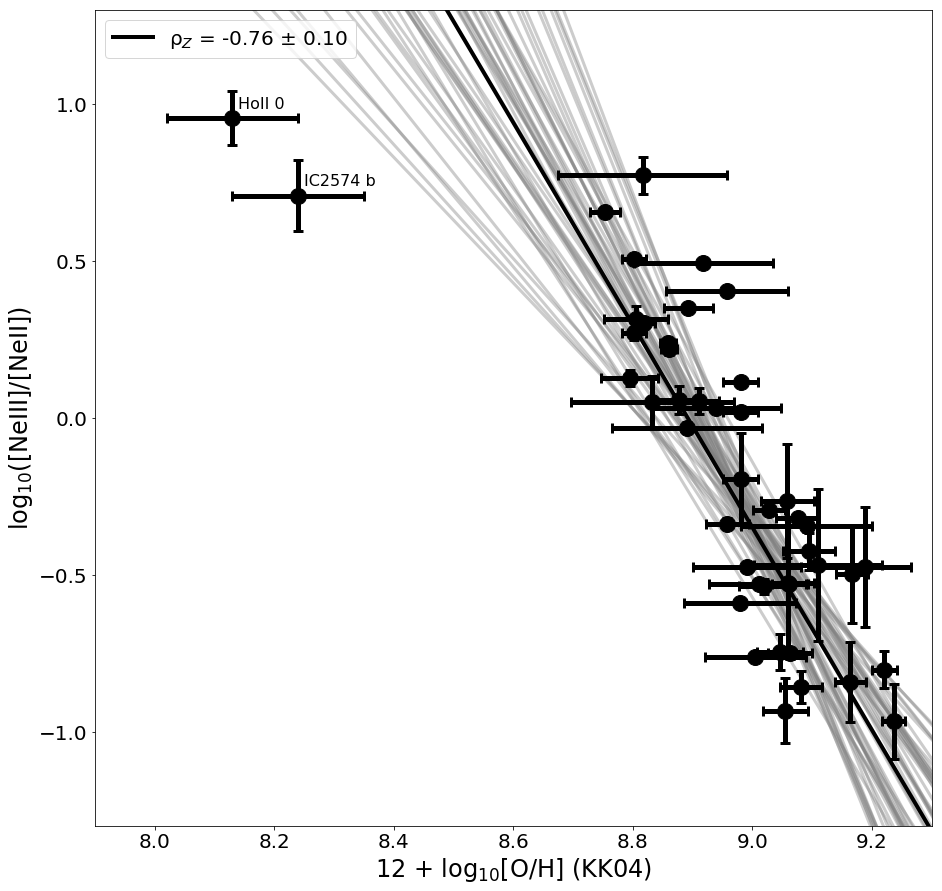}
\includegraphics[width=3.5in]{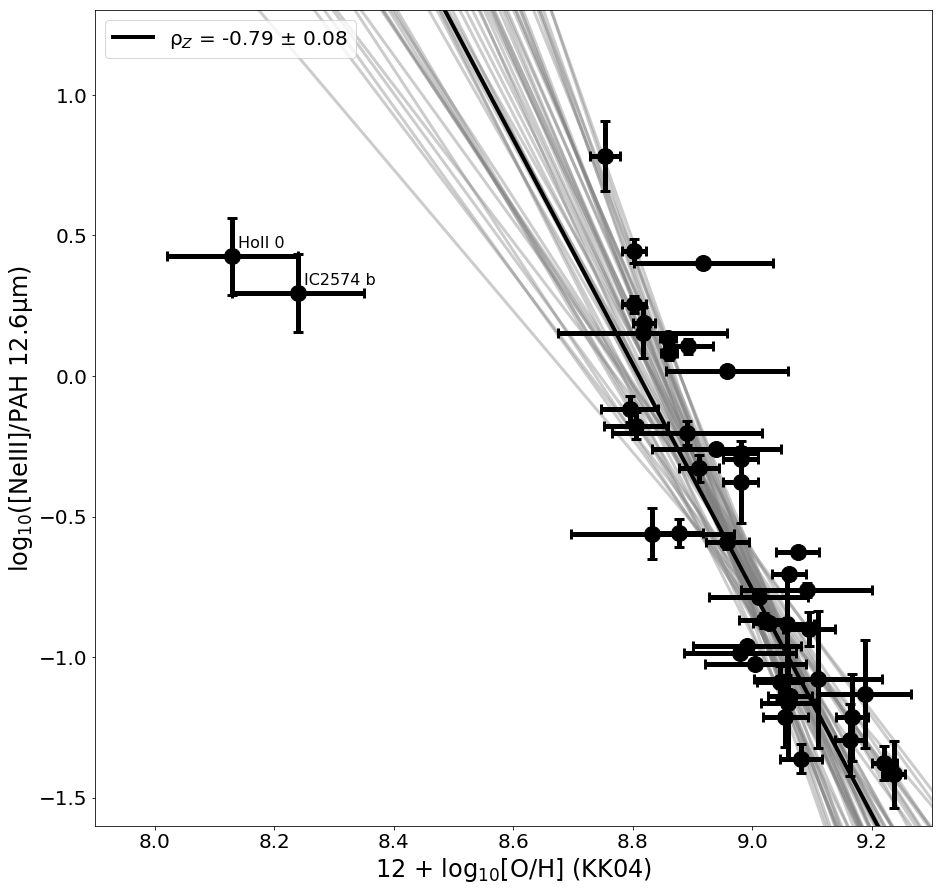}

\caption{KK04 metallicity as a function of \nethree\ ratios: (a) [Ne~III]/\netwo\ and (b) [Ne~III]/PAH 12.6~\micron.}
\label{fig:Ne3_ratios}
\end{figure*}

We find strong correlations between metallicity and ratios involving \nethree\ emission. Figure~\ref{fig:Ne3_ratios} shows KK04 metallicities as a function of [Ne~III] ratios with the 12.6~\micron\ PAH band and the \netwo\ line. We find [Ne~III] ratios with [Ne~II] and all PAH bands are statistically equivalent with one another based on the Spearman correlation coefficient with metallicity $\rho_Z \sim -0.76$ for each. Figure~\ref{fig:Ne3_ratios} also shows these models using ratios of \nethree\ fail to accurately reproduce values for the lowest-metallicity regions (12 + log[O/H] $<$ 8.7$_{KK04}$ or $<$ 8.2$_{PT05}$). Our sample contains two regions below this metallicity that are located in the dwarf galaxies Holmberg-II and IC2754. These results for KK04 correlations are found to be true for correlations with PT05 metallicities as well. We find including a second parameter \textit{Y} does not improve correlations between [Ne~III] ratios and metallicity. 

\begin{figure*}
\includegraphics[width=2.3in]{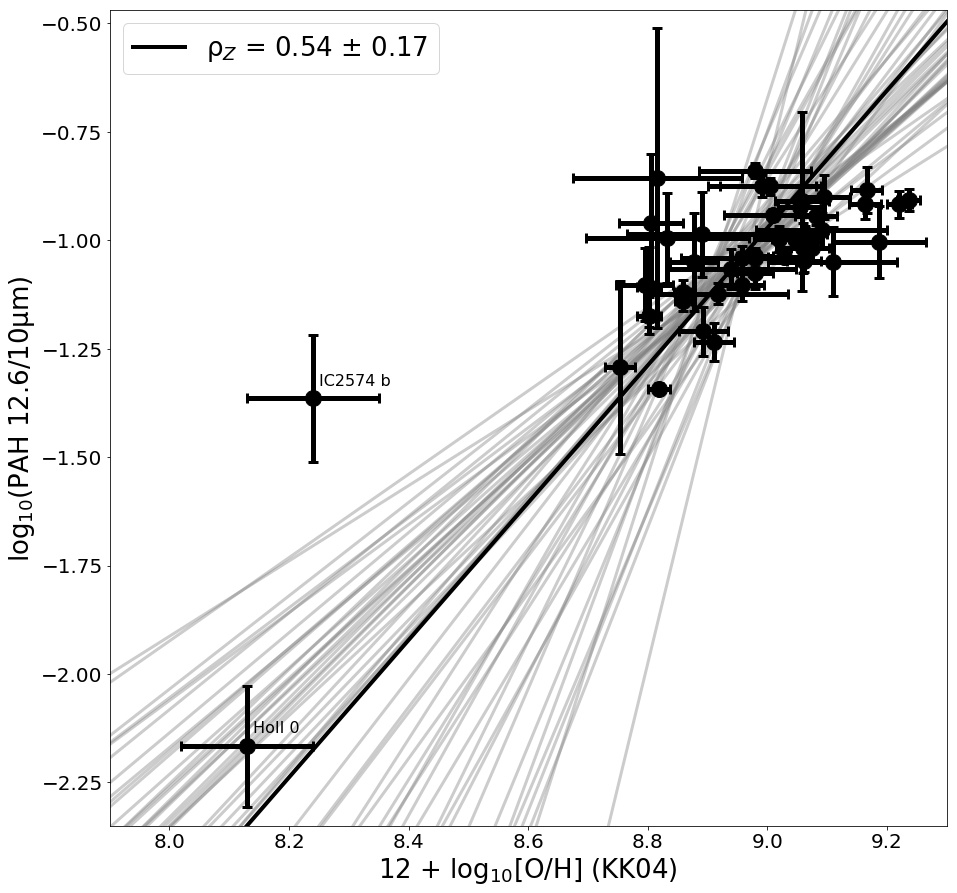}
\includegraphics[width=2.3in]{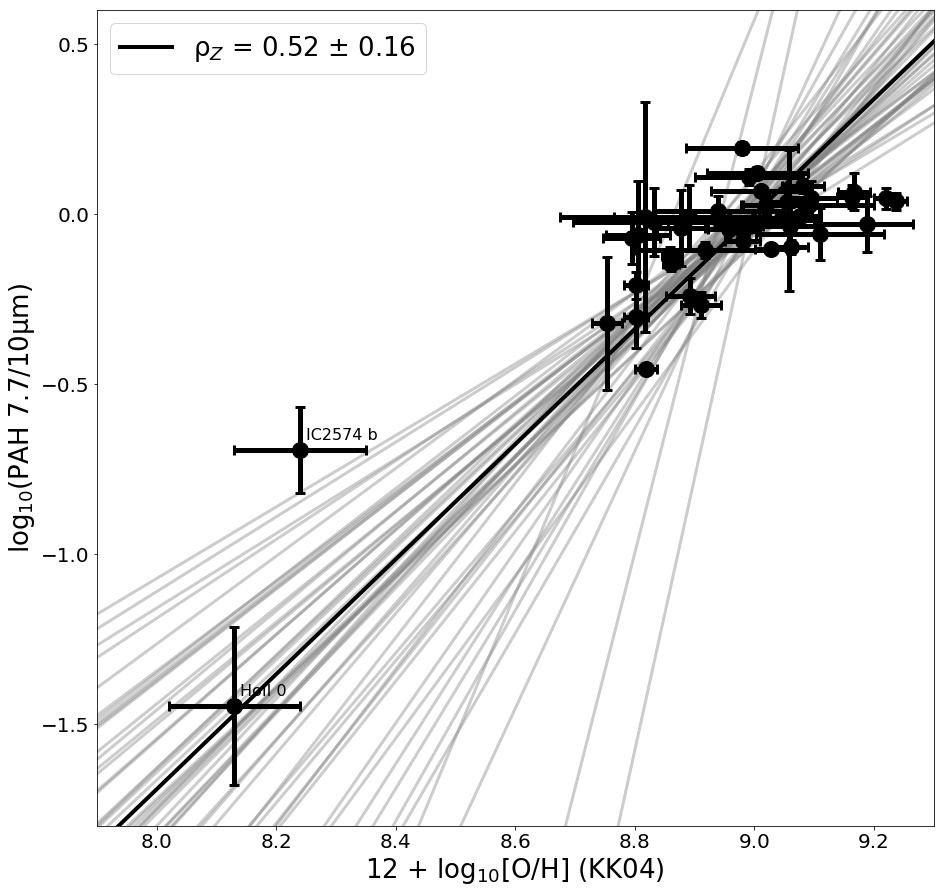}
\includegraphics[width=2.3in]{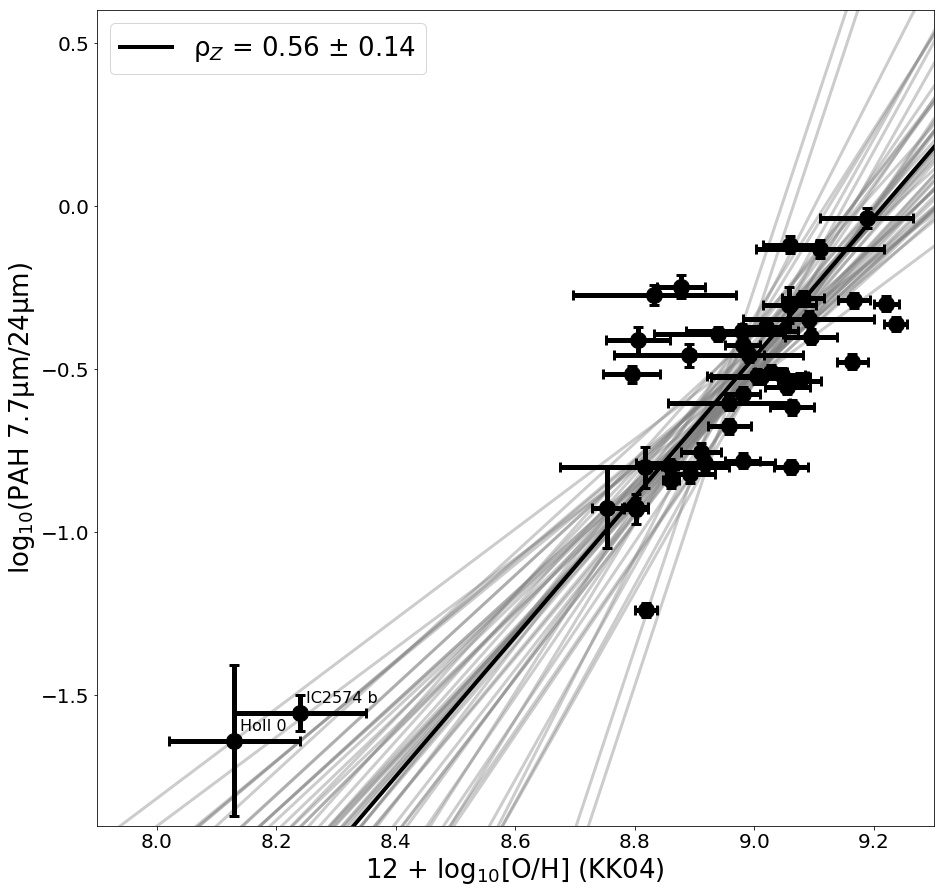}

\caption{KK04 metallicity as a function of 10~\micron\ continuum-normalized PAH bands: 
(a) PAH 12.6/10~\micron\ and (b) PAH 7.7/10~\micron;
as well as (c) PAH 7.7~\micron\ band normalized by 24~\micron\ continuum.}
\label{fig:PAH_ratios}
\end{figure*}

We also find significant correlations between metallicity and continuum-normalized PAH bands. Figure~\ref{fig:PAH_ratios} shows three of these normalized PAH correlations with metallicity: PAH 12.6/10~\micron, PAH 7.7/10~\micron, and PAH 7.7/24~\micron. These normalized bands can be used to improve correlations with T33 and reduce the metallicity offset of single feature tracers, as seen in Table~\ref{tab:model_quality}. We find normalizing by the continuum at 10~\micron\ is equivalent to that of 24~\micron\ in metallicity correlations with $\rho_Z \sim 0.54$ within $1\sigma$. Compared to the \nethree\ ratio metallicity correlations in Figure~\ref{fig:Ne3_ratios}, these continuum-normalized PAH bands show a weaker correlation but they more accurately predict the lowest-metallicity points. Similar to [Ne~III] ratios, we find no improvement in correlations with metallicity by adding a second parameter \textit{Y}. A tracer with \textit{X}~=~[Ne~III]/12.6 and \textit{Y}~=~7.7/10~\micron, for example, improves their individual $\rho_Z$ values but not by greater than the 1$\sigma$ uncertainty.

\section{Individual Correlation Histograms}\label{sec:hists}

\begin{figure*}
\centering
\includegraphics[width=3in]{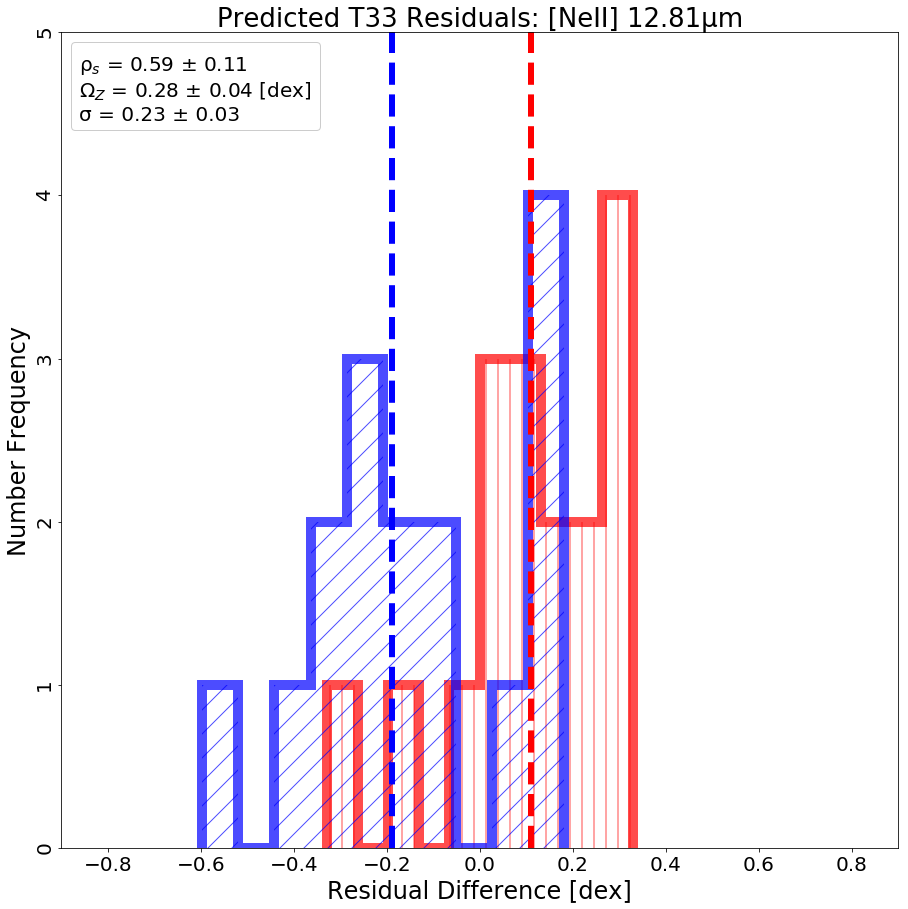} \\
\includegraphics[width=3in]{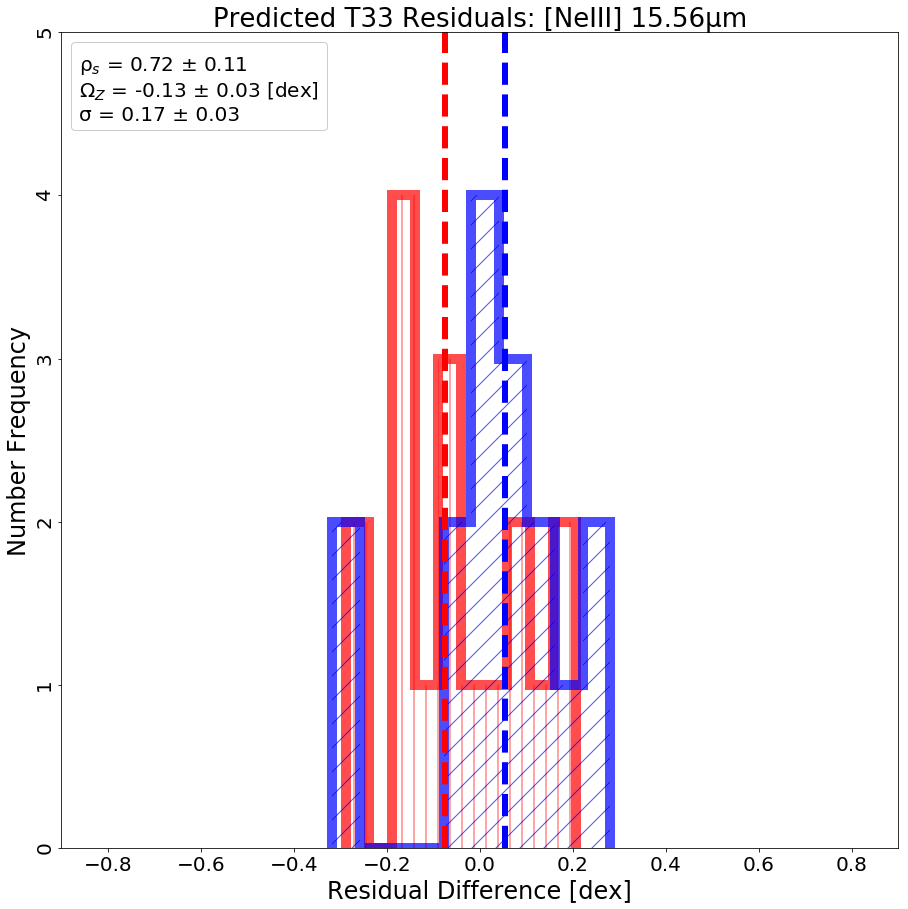}\\
\includegraphics[width=3in]{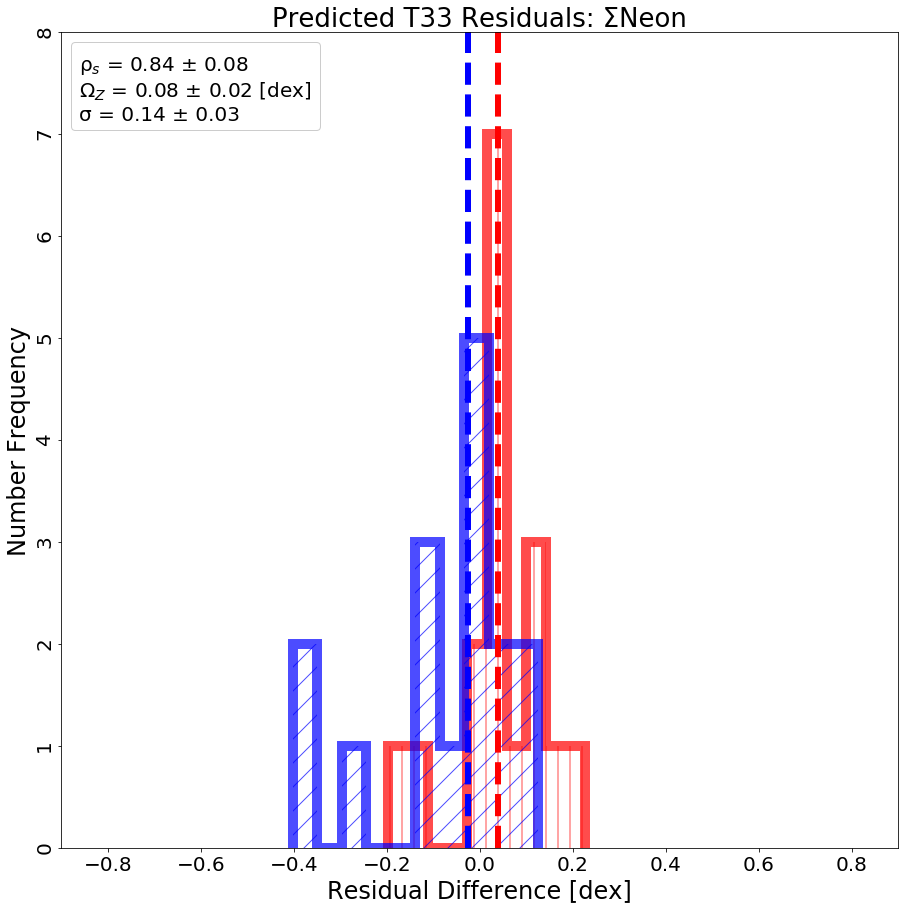}
\caption{Residual difference in predicted T33 for calibrations with:\\(a) \netwo, (b) \nethree, and (c) \sumneon.}
\label{fig:neon_summ}
\end{figure*}
\begin{figure*}
\includegraphics[width=3.5in]{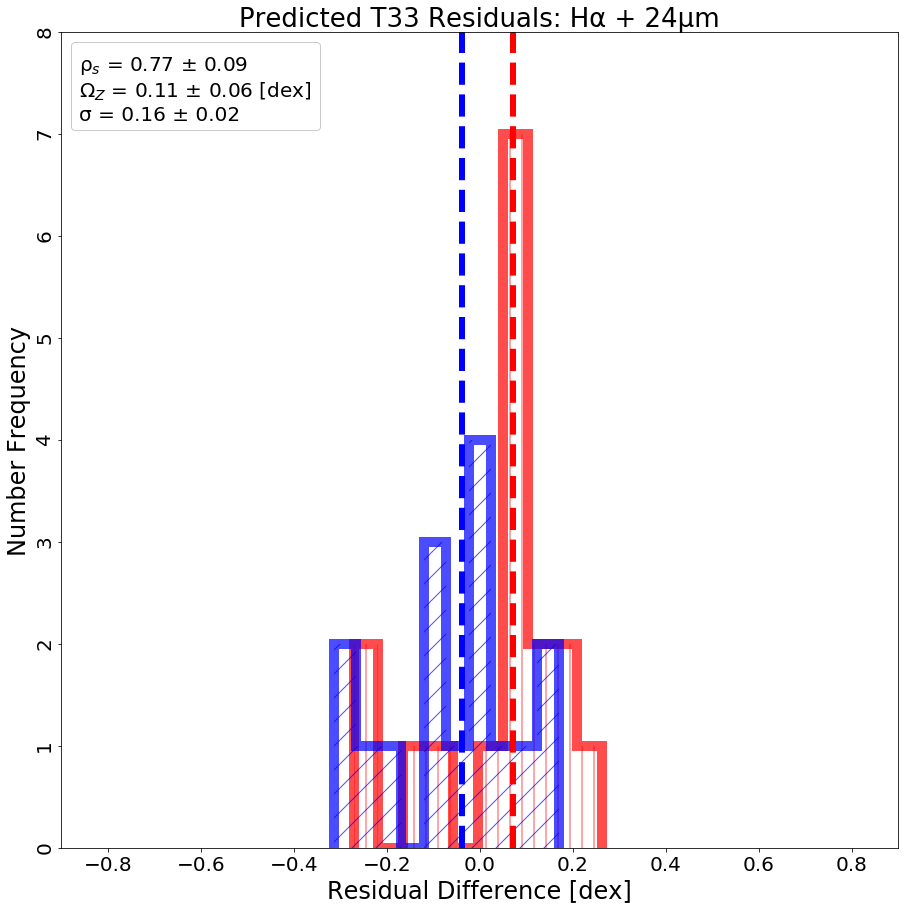}
\includegraphics[width=3.5in]{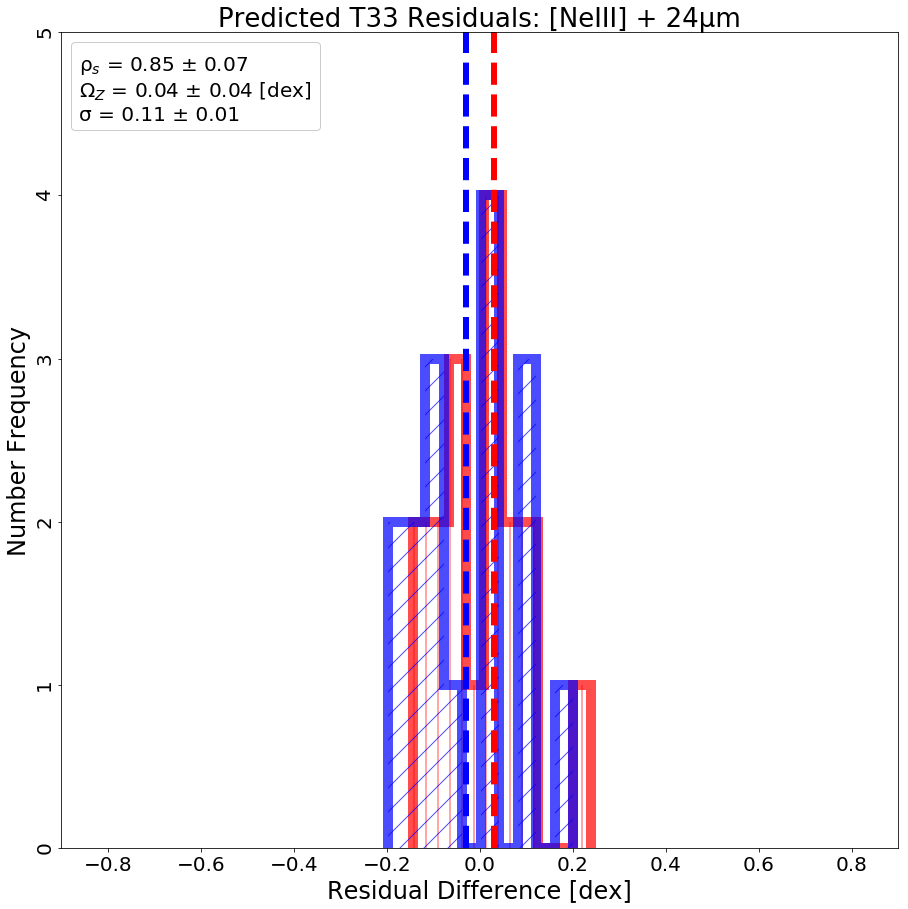}
\caption{Residual difference in predicted T33 based on \twenmic\ emission and (a) \halpha\ and (b) \nethree\ emission.}
\label{fig:Ha_Ne3+24}
\end{figure*}
\begin{figure*}
\includegraphics[width=3.5in]{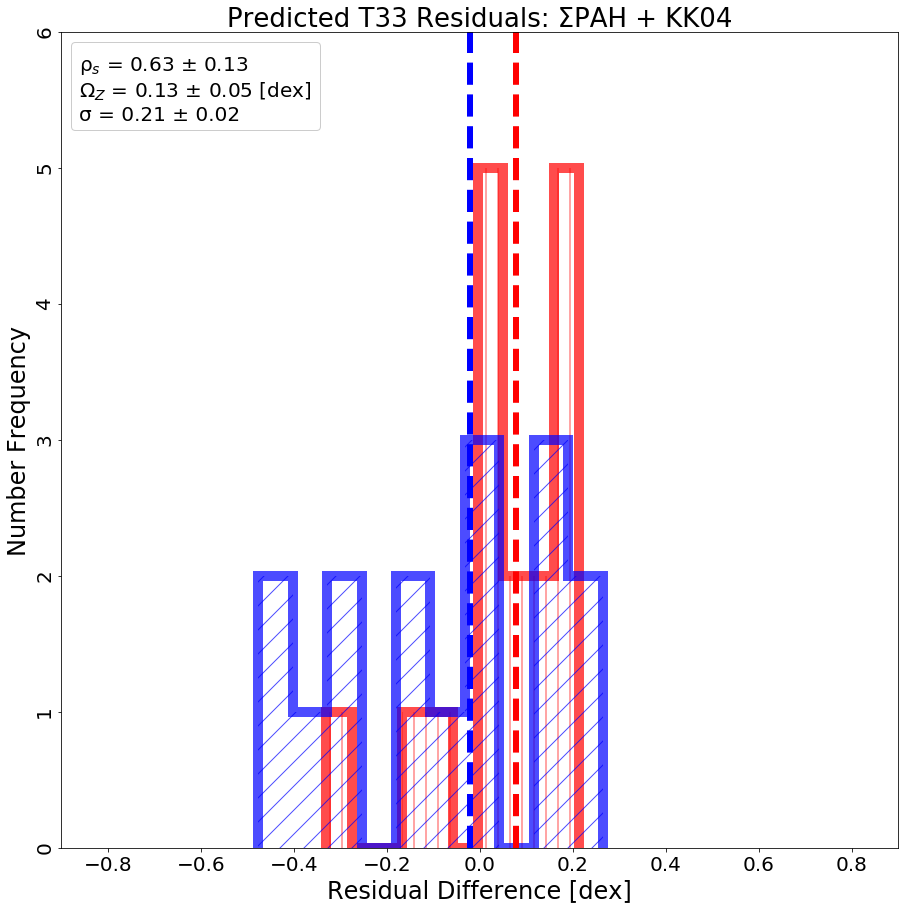}
\includegraphics[width=3.5in]{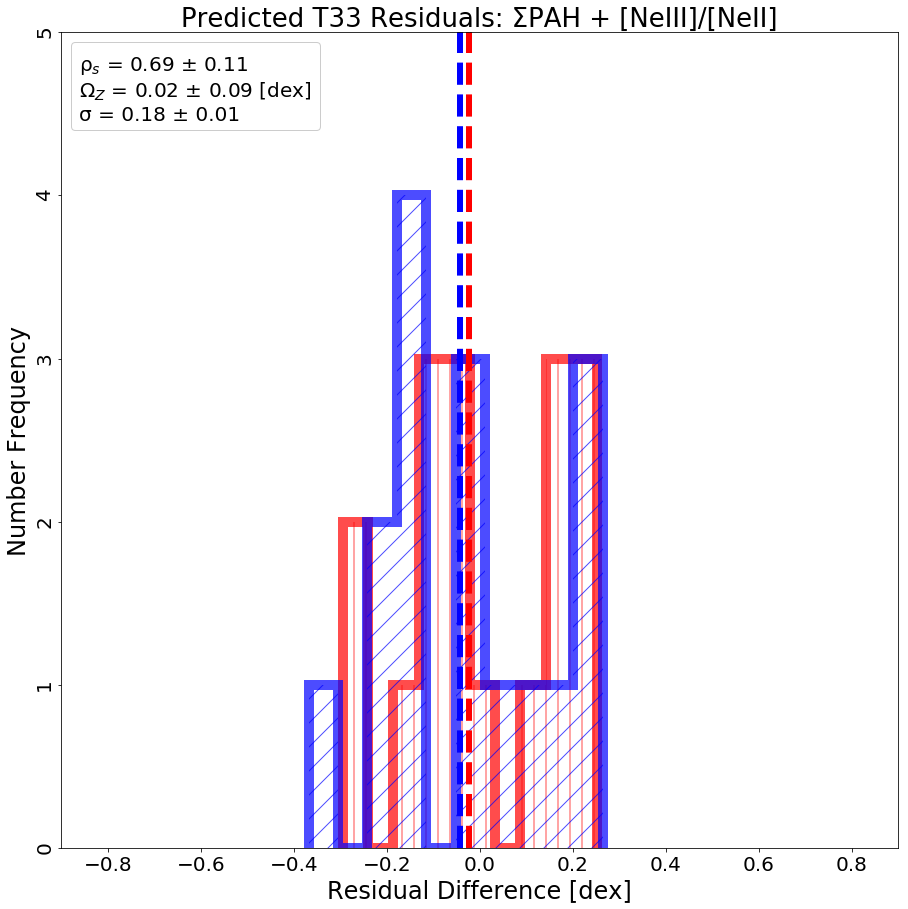}
\caption{Residual difference in predicted T33 based on $\Sigma$PAH emission and (a) KK04 metallicity and (b) the ratio of \nethree\ to \netwo.}
\label{fig:PAH_Z}
\end{figure*}

%%%%%%%%%%%%%%
% APPENDIX %
%%%%%%%%%%%%%%

%%%%%%%%%%%%%%
% REFERENCES %
%%%%%%%%%%%%%%

\bibliographystyle{aasjournal}
\bibliography{references}

%\clearpage

\end{document}